\documentclass[preprint2]{aastex}

 \def\gtsima{$\; \buildrel > \over \sim \;$}
 \def\simgt{\lower.5ex\hbox{\gtsima}}

\shorttitle{Metal distributions of four galaxy groups}
\shortauthors{Sasaki et al.}

\begin{document}


\title{Metal distributions out to 0.5~$r_{180}$ in the intracluster medium of four galaxy groups observed with Suzaku}


\author{
Toru \textsc{sasaki}\altaffilmark{1}, 
Kyoko \textsc{matsushita}\altaffilmark{1},
and Kosuke \textsc{sato}\altaffilmark{1}}



\altaffiltext{1}{Department of Physics, Tokyo University of Science, 
1-3 Kagurazaka, Shinjuku-ku, Tokyo 162-8601; j1213703@ed.tus.ac.jp; matusita@rs.kagu.tus.ac.jp}


\begin{abstract}
We studied the distributions of metal abundances and metal-mass-to-light 
ratios in the intracluster medium (ICM) of four galaxy groups, MKW~4, 
HCG~62, the NGC~1550 group, and the NGC~5044 group, out to $\sim$ 0.5~$r_{180}$ 
observed with Suzaku.  The Fe abundance decreases with radius, and 
about 0.2--0.4 solar beyond 0.1~$r_{180}$.  At a given radius in units 
of $r_{180}$, the Fe abundance in the ICM of the four galaxy groups were 
consistent or smaller than those of clusters of galaxies.  The Mg/Fe 
and Si/Fe ratios in the ICM are nearly constant at the solar ratio 
out to 0.5~$r_{180}$.  We also studied systematic uncertainties in 
the derived metal abundances comparing the results from two versions 
of atomic data for astrophysicists (ATOMDB) and single- and 
two-temperature model fits.  Since the metals have been synthesized 
in galaxies, we collected $K$-band luminosities of galaxies from Two 
Micron All Sky Survey catalogue (2MASS) and calculated the integrated
iron-mass-to-light-ratios (IMLR), or the ratios of the iron mass in 
the ICM to light from stars in galaxies.  The groups  with smaller 
gas mass to light ratios have smaller IMLR values and the IMLR 
inversely correlated with the entropy excess.  Based on these 
abundance features, we discussed the past history of metal enrichment 
process in groups of galaxies.
\end{abstract}



\section{Introduction}
\setcounter{footnote}{1}

Clusters and groups of galaxies are the best laboratories for study 
of their thermal and chemical evolution history governed by baryons.  
The observations of metals in the intracluster medium (ICM), 
synthesized by supernovae (SNe) in galaxies give an important clue 
in studying the evolution of galaxies.

The ratio of the metal mass in the ICM to the total light from galaxies 
in clusters, the metal-mass-to-light-ratio, is a key parameter in 
investigating star-formation history in these systems. Using ASCA data, 
\citet{Makishima2001} summarized iron-mass-to-light-ratios (IMLRs),  
with the $B$-band luminosity for various objects as a function of 
their plasma temperature and found that IMLRs in groups are systematically  
smaller than those in clusters.  Using Chandra data, \citet{Rasmussen2009} 
reported that IMLRs 
of galaxy groups within $\sim 0.5~r_{180}$
\footnote{$r_{180}$ is the radius within which the mean density of
the cluster is 180 times the critical density of the universe. 
In this paper, we adopted $r_{180}$ as the virial radius.}
show a positive correlation with total mass of groups. 

Groups of galaxies also differ from richer systems in stellar and 
gas mass fractions within $r_{500}$
\citep{Lin2004, Sun2009, Giodini2009, Zhang2011}.
This dependence on the system mass has been sometimes interpreted 
that the star-formation efficiency depends on the system mass.
The gas-density profiles in the central regions of groups and poor 
clusters are observed to be shallower and the relative entropy level 
is higher than the oretical  predictions \citep{Ponman1999, Ponman2003}.
The entropy profile of the ICM is more expressive characterization 
of the rmodynamical history than temperature.  The excess entropies 
in poor systems are thought to be caused by non-gravitational processes, 
likely results of the preheating of central AGN, SNe, and so on, 
injected the energy to the ICM.  The metal distribution in the ICM 
can be a useful tracer of the history of gas heating in the early 
epoch, because the relative timing of metal enrichment and heating 
should affect the present amount and distribution of the metals 
in the ICM.

The abundance ratios of O, Mg and Si to Fe reflect the relative
enrichment from SNe Ia and core-collapse SNe (hereafter SNecc).
Using Chandra data, \citet{Rasmussen2007} reported that the Si/Fe 
ratio in the ICM of groups of galaxies increases with radius and 
the SNecc contribution dominates at $r_{500}$.  With Suzaku 
observations, we can derive the metal abundances of O and Mg beyond 
cool cores of groups and clusters of galaxies, because the X-ray 
Imaging Spectrometer (XIS) instrument \citep{Koyama2007}
has a lower background level and higher spectral sensitivity below 
1 keV\@.  
With Suzaku, the abundance profiles of O, Mg, Si and Fe were 
derived for several groups and clusters out to 
$\sim$0.3--0.4~$r_{\rm 180}$ \citep{Matsushita2007, Komiyama2009, Sato2007, 
Sato2008, Sato2009a, Sato2009b, Sato2010, Murakami2011, Sakuma2011}, 
and, with XMM, Si/Fe ratios were derived out to 0.5 $r_{\rm 180}$ 
of the Perseus cluster \citep{Matsushita2013Perseus} and the Coma 
cluster \citep{Matsushita2013Coma}.  The radial abundance profiles 
of Mg, Si and S were mostly similar to that of Fe.  The number ratio 
of SNecc and SNe Ia to enrich the ICM is estimated to be $\sim$3 
\citep{Sato2009b, Sato2010}.

In this paper, we describe our study of four galaxy groups, MKW~4, 
HCG~62, the NGC~1550 group, and the NGC~5044 group, observed out 
to $\sim0.5~r_{180}$ with Suzaku.  
MKW~4, the NGC~1550 group, and the NGC~5044 group have
central dominant giant elliptical or S0 galaxies.
In contrast, the central region ($r < 1.1'$) of HCG~62 is dominated by three galaxies.
All groups have nearly symmetric spatial distributions in the X-ray band
\citep{dellAntonio1995, Kawaharada2003, David1995}.
Among these systems, results of 
Suzaku data of HCG~62, the NGC~1550 group, and the NGC~5044 group
have already been published out to $0.2~r_{180}$ (1 pointing), 
$0.5~r_{180}$ (2 pointings), and $0.3~r_{180}$ (3 pointings),
respectively \citep{Tokoi2008, Sato2010, Komiyama2009}.
MKW~4 has been observed with XMM and metal distribution 
in cool core is reported by \citet{OSullivan2003}. 
In this paper, we included six new pointing observations 
(2 for MKW~4, 1 for HGC~62, 1 for the NGC~1550 group, and 2 for the 
NGC~5044 group) with Suzaku.  In section \ref{sec:obs}, we summarize 
the observations and data preparation. Section \ref{sec:analysis} 
describes our analysis of the data. In section \ref{sec:results}, we 
summarize the distributions of temperature, the Fe abundance,
abundance ratios to Fe, and the IMLR.
We discuss our results in section \ref{sec:discussion}.

Recently, atomic data for astrophysicists (ATOMDB), version 2.0.1 
which includes major updates of the atomic data and new calculation of
lines  has been released \citep{Foster2010}.
These differences in the atomic data are critical to study metal 
abundances in the ICM of galaxy groups where Fe-L lines dominate 
the spectra. In order to compare ATOMDB versions between version 
1.3.1 and 2.0.1, we reanalyzed the regions which have been reported 
with Suzaku using the latest responses and background.


We use Hubble constant $H_{0}=$70 km s$^{-1}$ Mpc$^{-1}$ in this paper.
The solar abundance table is given by \citet{Lodders2003}.
In this paper, errors are quoted at a 90\% confidence level for 
a single parameter of interest.


\section{Observations and Data Reduction}
\label{sec:obs}

\begin{deluxetable}{llllllll}
\tabletypesize{\scriptsize}
\tablewidth{0pt}
\tablecaption{
Suzaku observation logs and basic properties 
for four galaxy groups, MKW~4, HCG~62, the NGC~1550 group, and 
the NGC~5044 group. 
\label{tb:obs_log}
}
\tablehead{
 \colhead{Group} &  \colhead{Field name} &  \multicolumn{2}{l}{Sequence number} & \multicolumn{2}{l}{Exposure time} & \multicolumn{2}{l}{(RA, DEC)$^{a}$}  \\
\colhead{} & \colhead{} & \colhead{} & \colhead{} &  \multicolumn{2}{l}{ksec} & \multicolumn{2}{l}{J2000.0} }
\startdata
MKW~4      & East         & \multicolumn{2}{l}{805081010} &  \multicolumn{2}{l}{61.8} & \multicolumn{2}{l}{($12^{\rm h}04^{\rm m}31^{\rm s}.3, +02^{\circ}13'44.0''$)} \\
                   & North        & \multicolumn{2}{l}{805082010} &  \multicolumn{2}{l}{65.0} & \multicolumn{2}{l}{($12^{\rm h}05^{\rm m}43^{\rm s}.0, +01^{\circ}54'20.9''$)}  \\
HCG~62     & Center      & \multicolumn{2}{l}{800013020} &  \multicolumn{2}{l}{93.8} & \multicolumn{2}{l}{($12^{\rm h}53^{\rm m}05^{\rm s}.8, -09^{\circ}12'07.9''$)}   \\
                   & West         &  \multicolumn{2}{l}{805031010}&  \multicolumn{2}{l}{55.9}& \multicolumn{2}{l}{($12^{\rm h}51^{\rm m}59^{\rm s}.4, -09^{\circ}05'01.3''$)}   \\
NGC~1550 & Center      & \multicolumn{2}{l}{803017010} &  \multicolumn{2}{l}{73.8}& \multicolumn{2}{l}{($04^{\rm h}19^{\rm m}47^{\rm s}.7, +02^{\circ}24'37.8''$)}  \\
                   & East          & \multicolumn{2}{l}{803018010} &  \multicolumn{2}{l}{28.1} & \multicolumn{2}{l}{($04^{\rm h}20^{\rm m}59^{\rm s}.5, +02^{\circ}24'31.0''$)} \\
                   & Northeast &  \multicolumn{2}{l}{803046010} &  \multicolumn{2}{l}{61.9} & \multicolumn{2}{l}{($04^{\rm h}20^{\rm m}35^{\rm s}.8, +02^{\circ}36'31.0''$)}  \\
NGC~5044 & Center      & \multicolumn{2}{l}{801046010}  &  \multicolumn{2}{l}{19.7} & \multicolumn{2}{l}{($13^{\rm h}15^{\rm m}24^{\rm s}.1, -16^{\circ}23'23.6''$)}  \\
                   & 15$'$North&  \multicolumn{2}{l}{801047010}&  \multicolumn{2}{l}{54.6} & \multicolumn{2}{l}{($13^{\rm h}15^{\rm m}24^{\rm s}.1, -16^{\circ}08'34.8''$)}  \\
                   & 15$'$East  &  \multicolumn{2}{l}{801048010}&  \multicolumn{2}{l}{62.4} & \multicolumn{2}{l}{($13^{\rm h}16^{\rm m}26^{\rm s}.8, -16^{\circ}23'16.8''$)}  \\
                   & 30$'$North&  \multicolumn{2}{l}{804013010} &  \multicolumn{2}{l}{56.6} & \multicolumn{2}{l}{($13^{\rm h}16^{\rm m}23^{\rm s}.8, -15^{\circ}53'38.8''$)}  \\
                   & 30$'$South& \multicolumn{2}{l}{804014010}  & \multicolumn{2}{l}{53.3} & \multicolumn{2}{l}{($13^{\rm h}14^{\rm m}59^{\rm s}.6, -16^{\circ}51'46.4''$)}   \\ \tableline
Group & $z$ & $ \left<kT\right>^{b} $ & $N_{H}^{c}$ & $D_{L}^{d}$ &$r_{180}^{e}$ & 1 arcmin$^{d}$ & (RA, Dec)$^{f}$ \\
 &  & keV & $10^{20}~{\rm cm}^{-2}$ & Mpc & Mpc & kpc & J2000.0 \\\tableline
MKW~4 & 0.020 & 1.8 & 1.76 & 87.0 & 1.18 & 24.3 &  ($12^{\rm h}04^{\rm m}27^{\rm s}.0, +01^{\circ}53'45.0''$)\\
HCG~62 &0.0145  & 1.5 & 3.31  & 64.3 & 1.08  & 17.8 & ($12^{\rm h}53^{\rm m}05^{\rm s}.3, -09^{\circ}12'01.2'$) \\
NGC~1550 & 0.0124 & 1.2 & 10.2 & 53.6 & 0.97 & 15.2 & ($04^{\rm h}19^{\rm m}37^{\rm s}.9, +02^{\circ}24'36.0''$) \\ 
NGC~5044 &  0.0093 & 1.0 &4.87 & 40.0 &0.88 & 11.5 &  ($13^{\rm h}15^{\rm m}24^{\rm s}.0, -16^{\circ}23'07.8''$)  \\
\enddata
\tablenotetext{a}{Average pointing direction of the XIS, written in the RA\_NOM and DEC\_NOM keywords of the event FITS files.}
\tablenotetext{b}{
Average temperature as used in \citet{Rasmussen2007} for MKW~4, \citet{Tokoi2008} for HCG~62, 
\citet{Sato2010} for the NGC~1550 group, and \citet{Komiyama2009} for the NGC~5044 group.}
\tablenotetext{c}{The Galactic hydrogen column density \citep{Kalberla2005}.}
\tablenotetext{d}{NASA Extragalactic database(http://ned.ipac.caltech.edu/).}
\tablenotetext{e}{The virial radius derived from the equation with the average temperature as described in \citet{Markevitch1998, Evrard1996}.}
\tablenotetext{f}{The central coordinate for the annuals regions, which
 corresponds to the X-ray peak for each group.}
\end{deluxetable}

\begin{figure*}[!htpd]
\begin{center}
\includegraphics[width=0.37\textwidth,angle=0,clip]{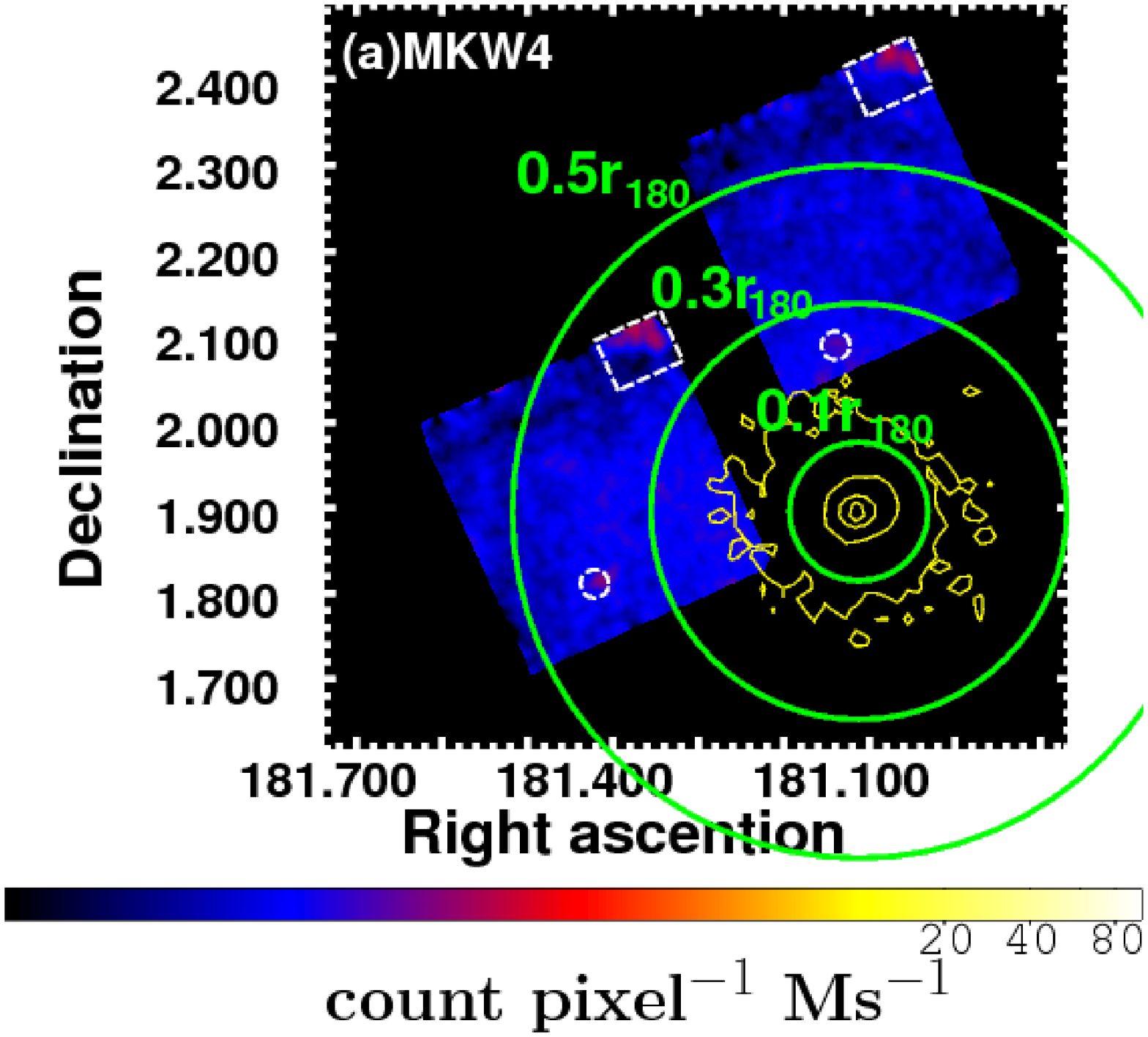}
\includegraphics[width=0.37\textwidth,angle=0,clip]{all_HCG62_05_20_4bin_4ga_paper_cps_v2.eps}
\includegraphics[width=0.45\textwidth,angle=0,clip]{all_NGC1550_05_20_4bin_4ga_paper_cps_v2.eps}
\includegraphics[width=0.3\textwidth,angle=0,clip]{all_NGC5044_05_20_4bin_4ga_paper_ds9_cps_v2.eps}
\caption{
Combined XIS images of (a) MKW~4, (b) HCG~62, (c) the NGC~1550 group,
and (d) the NGC~5044 groups in 0.5--2.0 keV\@.
The exposure time was corrected, though vignetting was not corrected.
The instrumental backgrounds (NXB) were subtracted. The images 
were smoothed by a Gaussian of $\sigma=$16 pixels $\approx 17''$.  
The green solid circles correspond to  0.1, 0.3, and 0.5~$r_{180}$\@.
The white dashed circles and boxes show point-like sources and 
hot pixels, respectively, which were excluded in our analysis.
For MKW~4, the X-ray contour map observed with XMM-Newton in 
logarithm scale from 0.5--2.0 keV are overlaid in yellow.
}
\label{fig:image}
\end{center}
\end{figure*}

We selected four galaxy groups, MKW~4, HCG~62, the NGC~1550 group, 
and the NGC~5044 group, which have been observed up to $0.5~r_{180}$ 
with Suzaku.  The observation logs are given in table \ref{tb:obs_log}, 
and the X-ray images in 0.5--2.0 keV are shown in figure
\ref{fig:image}.  The average temperature and the Galactic 
hydrogen column density for each group are also given in 
table \ref{tb:obs_log}.  The virial radius for each group is 
calculated from the following equation with the average temperature, 
$r_{180}=1.95 (H_{0}/100)^{-1} \sqrt{\langle kT \rangle/10 \  \rm keV}$
Mpc \citep{Markevitch1998, Evrard1996}, and shown in table 
\ref{tb:obs_log}.

In this study, we used only the Suzaku XIS data.  The XIS 
instrument consists of four sets of X-ray CCDs (XIS~0, 1, 2, 
and 3).  XIS~1 is a back-illuminated (BI) sensor, while 
XIS~0, 2 and 3 are front-illuminated (FI).  Because XIS~2
has not been available since 2006 November, all the observations 
except for the HCG~62 Center and NGC~5044 Center/15$'$ East/15$'$ North
were carried out by XIS~0, 1, and 3\@.  The instruments were 
operated in the normal clocking mode (8 s exposure per frame) 
with standard $5\times5$ or $3\times3$ editing mode.  We used 
the standard selection data\footnote{http://www.astro.isas.ac.jp/suzaku/process/v2changes/\\criteria\_xis.html}.
The analysis was performed by HEAsoft version 6.11 and 
XSPEC 12.7.0\@.  In the new atomic data (ATOMDB version 
2.0.1), the emission line properties have been updated, 
especially for Fe-L and Ne-K complex lines.  These changes 
would give new results for the temperature and metal abundances 
in the spectral analysis for the lower temperature system, such 
as galaxy groups.  We, therefore, examined the spectral fits 
by changing the old (version 1.3.1) or new (version 2.0.1) ATOMDB 
versions for the same spectra and responses to compare with 
the previous works.  In this paper, the results with the ATOMDB 
version 2.0.1 are shown unless noted otherwise.  We also discuss 
the difference between the old and new ATOMDB versions in subsection 
\ref{sec:atomdb} in detail.

We generated Ancillary Response Files (ARFs) by 
``xissimarfgen'' Ftools task \citep{Ishisaki2007}, assumed a 
uniform sky of 20$'$ radius.  
Because the surface brightness profile of the ICM is far from uniform, 
an ARF file can be generated for an assumed spatial distribution
(e.g., $\beta$-model surface brightness profile) on the sky \citep{Ishisaki2007}. 
However, particularly for offset observations, surface brightness profiles,
such as a $\beta$-model profile, which derived from the shape of the central 
region in the clusters, might not have a good agreement with those 
in the outer regions.  We, therefore, adopted the uniform sky ARFs
in our analysis, and estimated the differences of the fit parameters 
with the $\beta$-model or uniform ARFs as the systematic errors. 
The resultant fit parameters such as temperatures and abundances, 
except for the normalizations with the two kinds of  ARFs agreed well. 
As for the normalizations, excluding the innermost region where
the effect of the point spread function (PSF) is severe,
the difference was within statistical errors.
Consequently, those differences were negligible for our results, 
particularly for the temperature, abundances, and the associated 
parameters in the outer regions.

The effect of degrading energy resolution by radiation damage 
was included in the redistribution matrix files by ``xisrmfgen'' 
Ftools task.  
The effect of contaminations on 
the optical blocking filter (OBF) of the XISs was included in 
the spectral fits as the photoelectric absorption model 
($\it varabs$ model).  The C/O ratio of the contaminant was 
fixed to be 6.0 in number ratio.
We employed the night Earth database generated by the 
``xisnxbgen'' Ftools task for the same detector area to subtract 
the non-X-ray background (NXB).  

We searched for point-like sources with ``wavdetect'' tool in 
CIAO\footnote{http://cxc.harvard.edu/ciao/} in 0.5--2.0 keV\@.
As shown by white circles in figure \ref{fig:image}, we 
subtracted the point-like sources with 1$'$/1.5$'$/2$'$ radii.
We also excluded the area around the hot pixels\footnote{http://www.astro.isas.ac.jp/suzaku/doc/suzakumemo/suzakumemo-2010-01.pdf} 
as shown in figure \ref{fig:image}.
The flux level of the faintest source was about 5$\times10^{-14}$
erg s$^{-1}$ cm$^{-2}$ in 0.5--2.0 keV with a power-law model 
of a fixed photon index, $\Gamma=$1.4, and the contribution from 
unresolved sources were taken into account when we subtracted 
the Cosmic X-ray Background (CXB) spectrum.

\section{Spectral Analysis}
\label{sec:analysis}

\begin{figure*}
\begin{center}
  \includegraphics[width=0.310\textwidth,angle=0,clip]{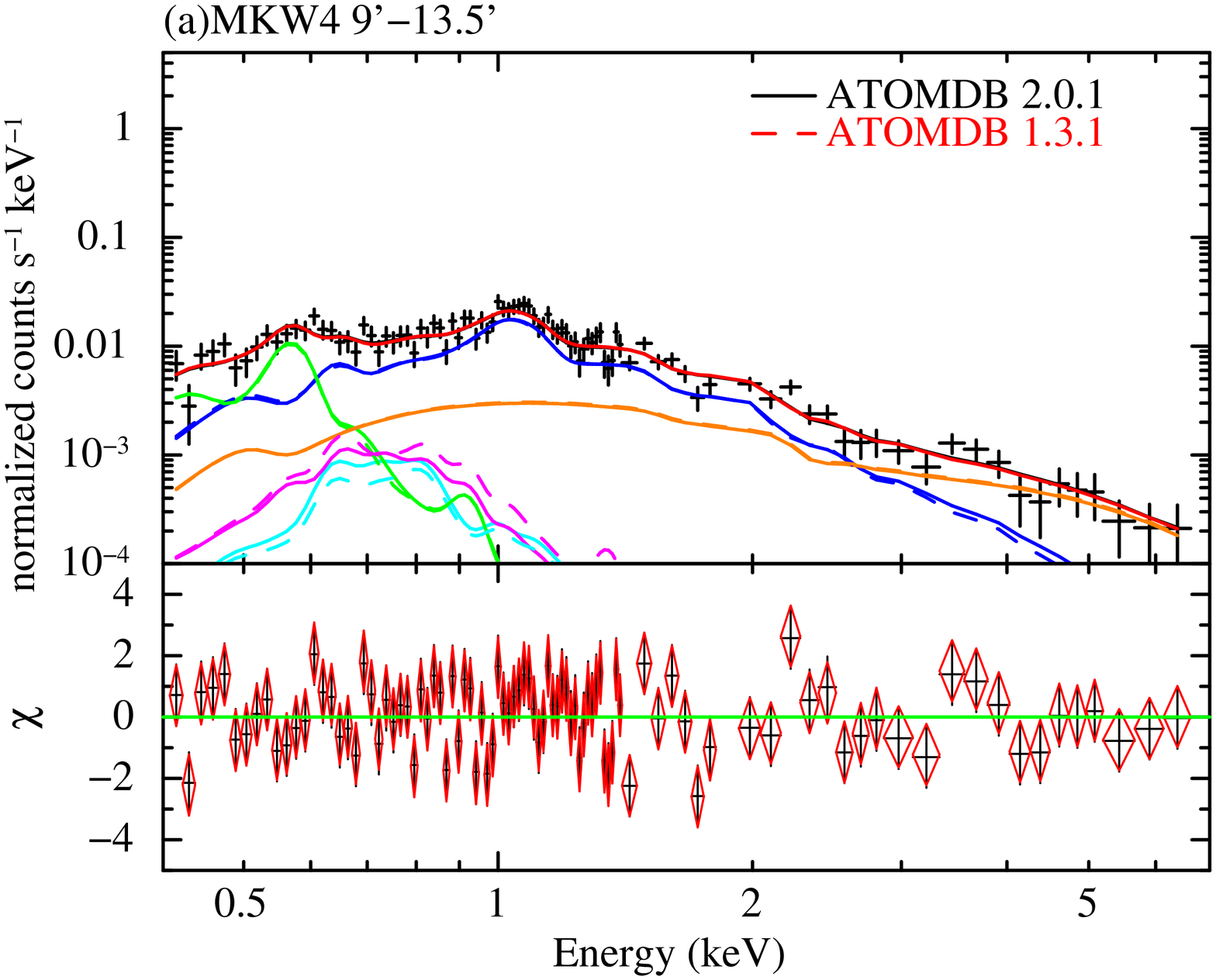}
  \includegraphics[width=0.310\textwidth,angle=0,clip]{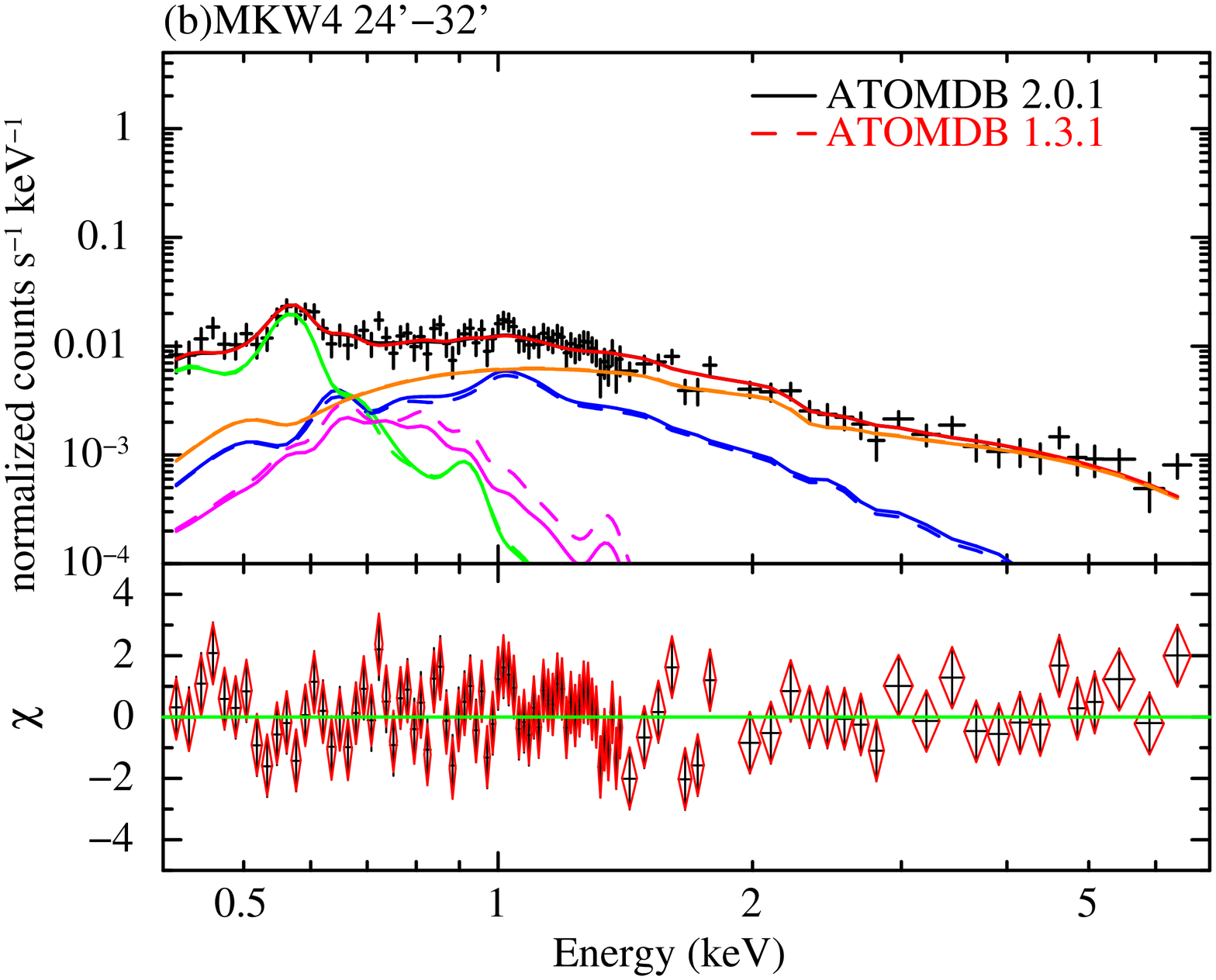}
  \includegraphics[width=0.310\textwidth,angle=0,clip]{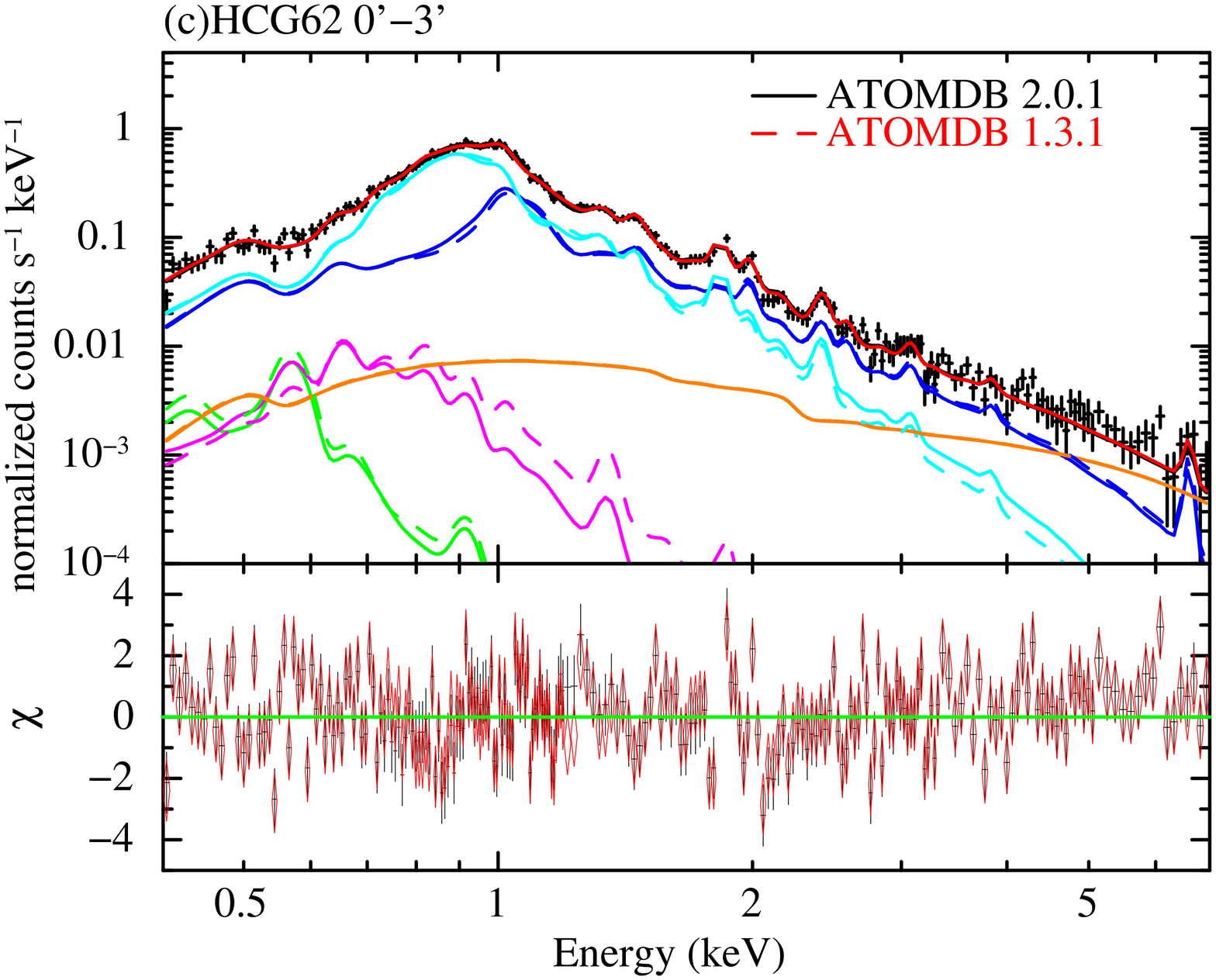}
  \includegraphics[width=0.310\textwidth,angle=0,clip]{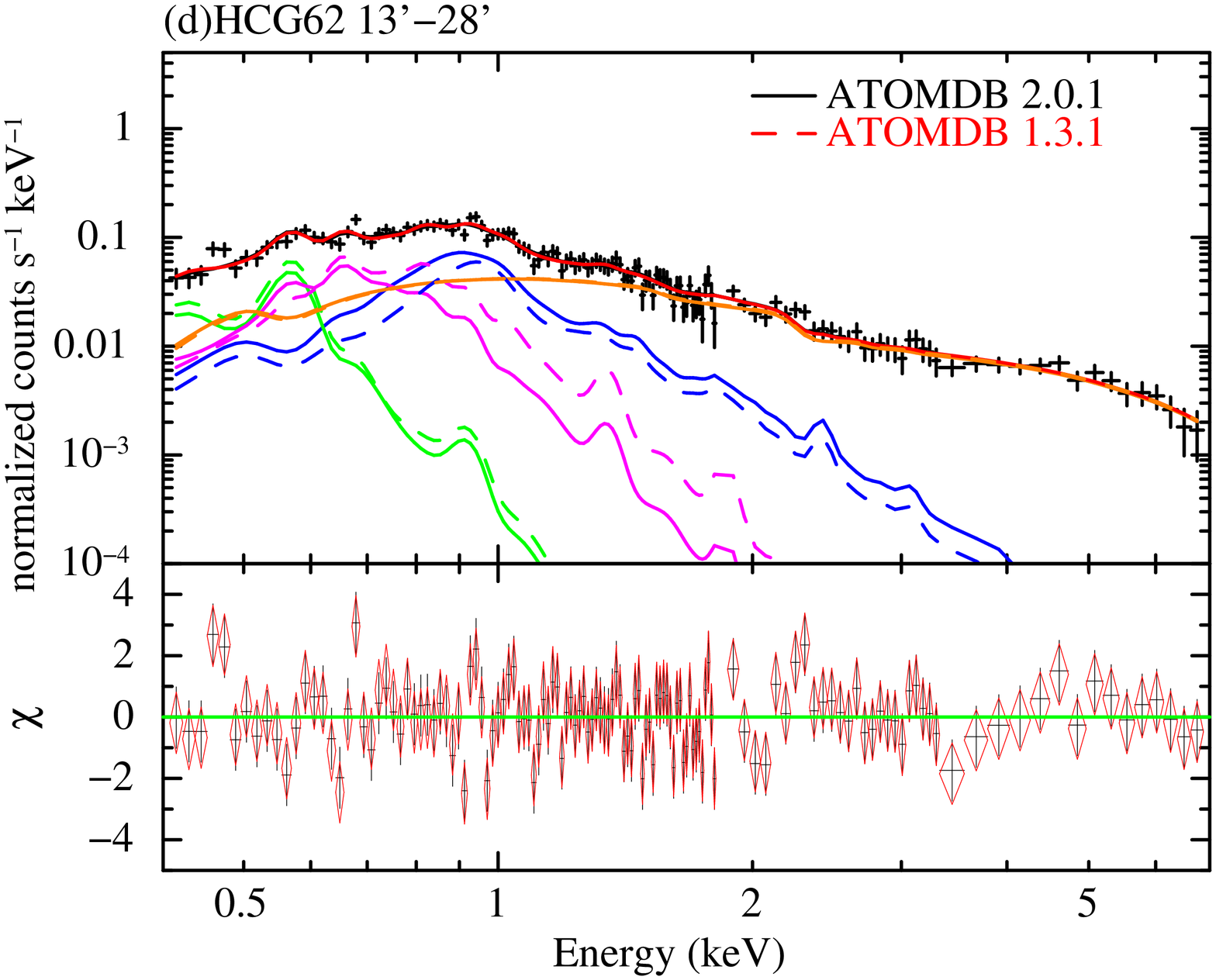}
  \includegraphics[width=0.310\textwidth,angle=0,clip]{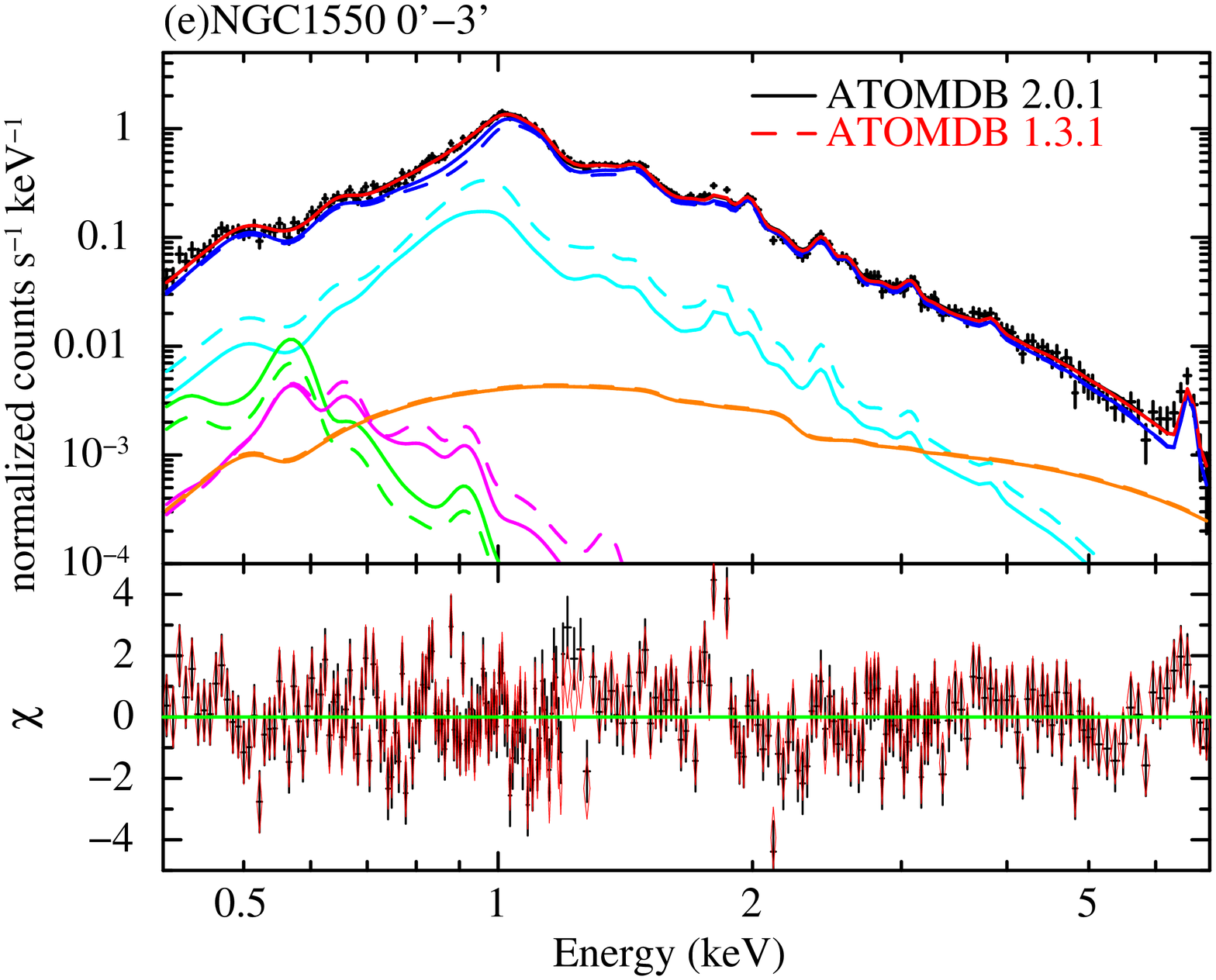}
  \includegraphics[width=0.310\textwidth,angle=0,clip]{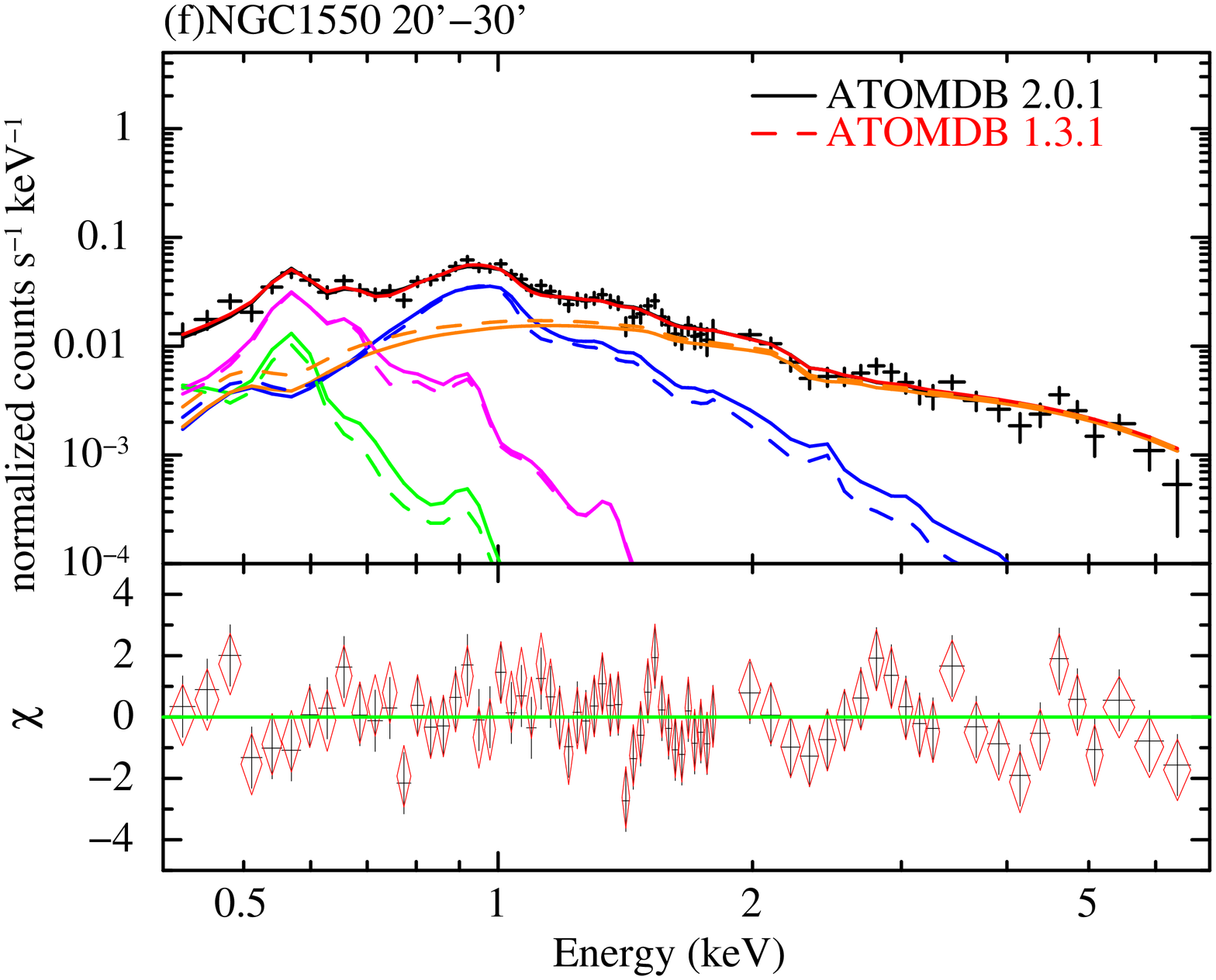}
  \includegraphics[width=0.310\textwidth,angle=0,clip]{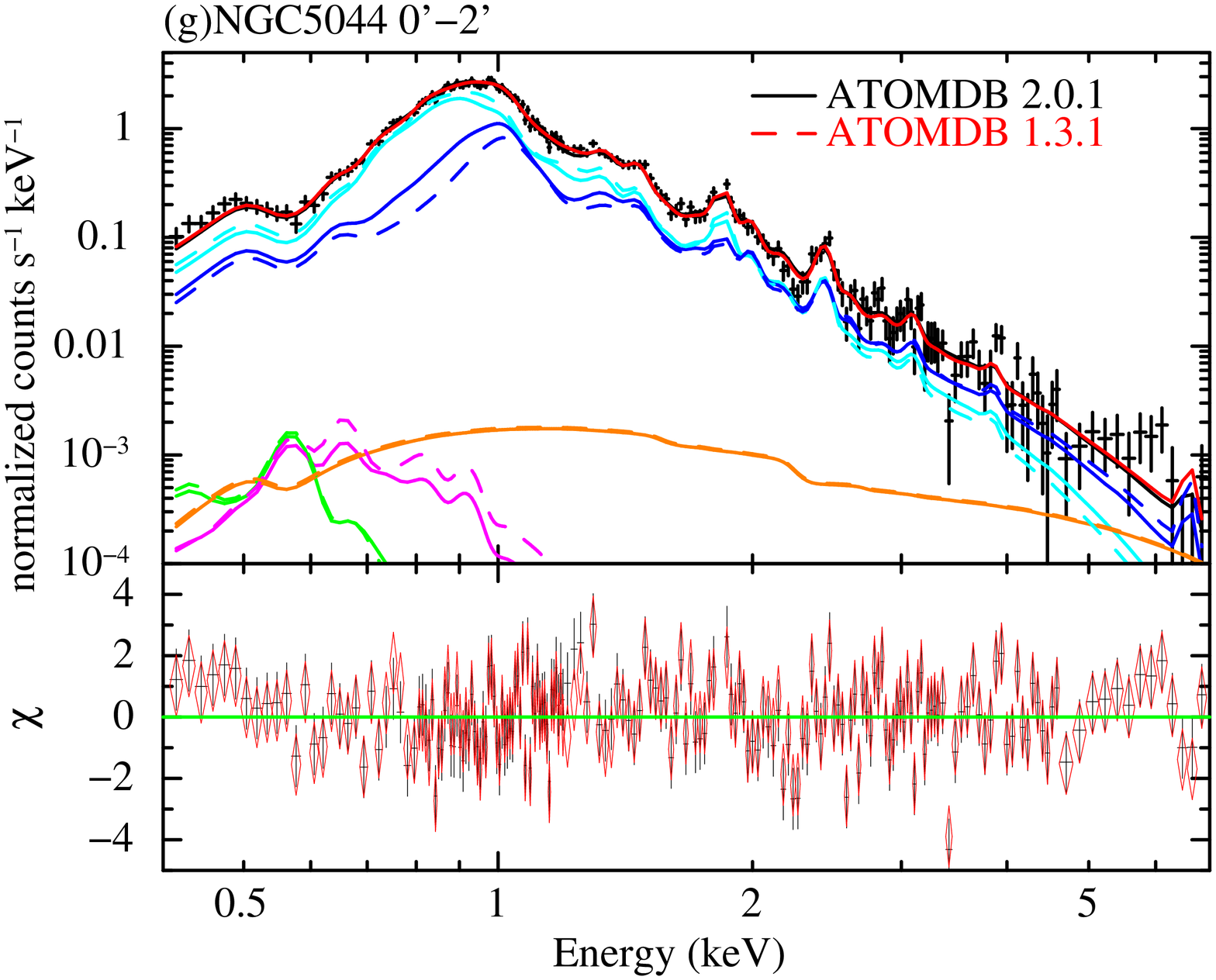}
  \includegraphics[width=0.310\textwidth,angle=0,clip]{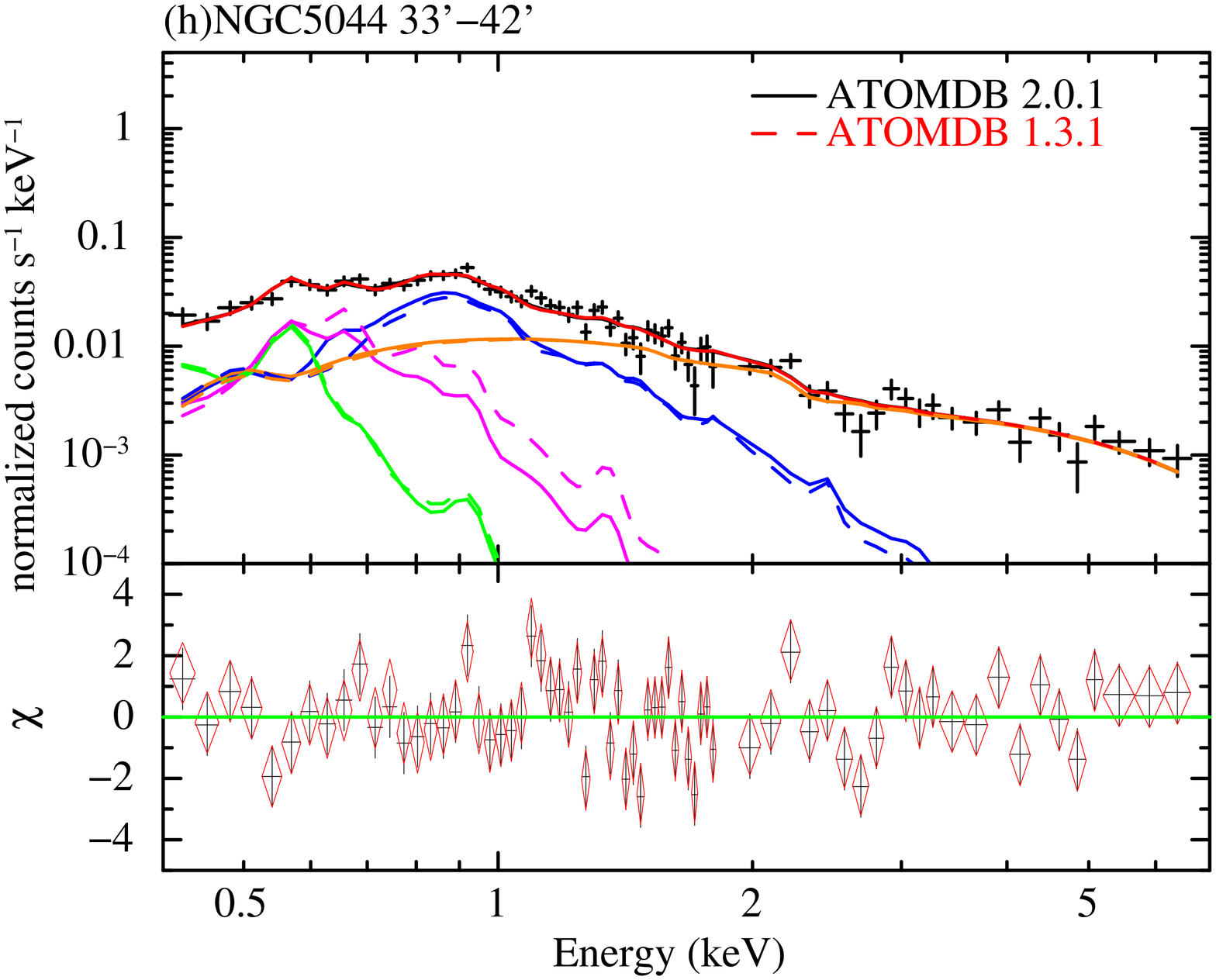}
\end{center}
\caption{
Upper panels show the observed XIS~1 spectra by black crosses. 
(a) and (b) show the spectra extracted from 9$'$--13.5$'$ and 
24$'$--32$'$, respectively, for MKW~4.  (c) and (d) show the 
spectra from 0$'$--3$'$ and 13$'$--28$'$, respectively, for HCG~62.
(e) and (f) show the spectra extracted from 0$'$--3$'$ and 
20$'$--30$'$ for the NGC~1550 group, respectively. (g) and (h) show 
the spectra from 0$'$--2$'$ and 33$'$--42$'$ for the NGC~5044 group, 
respectively.
Red and black lines indicate the best-fit model with the ATOMDB 
version 1.3.1 and 2.0.1, respectively.  
Note that black lines for the best-fit model are completely overlapped by the red lines.
The estimated NXB component
was subtracted.  The hotter and cooler ICM components for the 2T model 
fits are shown by blue and cyan for the inner regions, and the ICM component 
for the 1T model fits is indicated by blue for the outermost regions.
The CXB, LHB, and MWH components are indicated by orange, green, and 
magenta lines, respectively.  Dashed and solid lines show the best-fit 
models with the ATOMDB version 1.3.1 and 2.0.1, respectively.  
The energy band around the Si-K edge (1.82-1.84 keV) is excluded 
in the spectral fits.  The lower panels show the fit residuals in 
units of $\sigma$.}
\label{fig:spectrum}
\end{figure*}

We extracted the spectra from annular regions
(9$'$--13.5$'$, 13.5$'$--18$'$, 18$'$--24$'$, and 24$'$--32$'$ for MKW~4; 
0$'$--3$'$, 3$'$--6$'$, 6$'$--13$'$, and 13$'$--28$'$ for HCG~62;  
0$'$--3$'$, 3$'$--6$'$, 6$'$--12$'$, 12$'$--20$'$, and 20$'$--30$'$ 
for the NGC~1550 group;
0$'$--2$'$, 2$'$--4$'$, 4$'$--6$'$, 6$'$--9$'$, 9$'$--15$'$, 15$'$--25$'$, 
25$'$--33$'$, and 33$'$--42$'$ for the NGC~5044 group)
as shown in figure \ref{fig:spectrum},
centered on the coordinates as shown in table 
\ref{tb:obs_log}.  
Each spectrum was binned carefully to observe 
details in metal lines, especially below ~1 keV as shown in figure 
\ref{fig:spectrum}\@.  We used the energy ranges, 0.4--7.0 keV and 
0.5--7.0 keV for the BI and FI detectors, respectively.  We excluded 
the energy band around the Si-K edge (1.82--1.84 keV) because its 
response was not modeled correctly.  In the spectral fits of BI 
and FI data, only the normalization parameter was allowed to take 
different values between them.  It is important to estimate the 
Galactic and CXB emissions accurately, because the spectra, 
particularly in outer regions of groups, suffer from the Galactic 
and CXB emissions strongly.  We summarizes the background 
estimations in Appendix \ref{sec:bgd}.

We assumed that the ICM emission was represented by a single-- 
(hereafter 1T) or two--temperature (hereafter 2T) {\it vapec} model
\citep{Smith2001}\@.  The metal abundances of He, C, N, and 
Al were fixed to be a solar value.  We divided other metals into 
six group; O, Ne, Mg, Si, S=Ar=Ca, Fe=Ni, and allowed them to vary. 
For the 2T model for the ICM, the metal abundances of the two 
components were assumed to have the same value.  We fitted all 
the spectra with the 1T and 2T models for the ICM with the ATOMDB 
version 1.3.1 and 2.0.1 for estimating the best-fit model.  
For HCG~62, the NGC~1550 group, and the NGC~5044 group, we fitted the spectra 
for each annular region and the outermost region simultaneously 
for the background constraints as described in Appendix \ref{sec:bgd}.  
The temperature and 
normalization for each region were allowed to be free, but the
abundance of each element was assumed to have the same value. 
For MKW~4, the spectra from all the annulus regions were fitted
simultaneously for constraining the background well.

\section{Results}
\label{sec:results}


We summarize results of the spectral fits with the ATOMDB version 
2.0.1 in subsection \ref{sec:fittingresults}.  In this subsection, 
we show radial profiles of the temperature, normalizations, 
abundances, and metals to Fe ratios derived from the {\it vapec} 
model.  The differences of our results with the old and new ATOMDB 
versions are mentioned in subsection \ref{sec:atomdb}.  
In subsection \ref{sec:xmmchandra}, our results are compared with 
the previous results.
Subsection \ref{sec:sys} describes uncertainties by systematic errors for the 
background estimations.  We derive $K$-band luminosity profiles of 
the member galaxy for each group from galaxy catalogues in 
subsection \ref{sec:K-band}.  In subsection \ref{sec:imlr}, we 
calculate gas mass, iron mass, gas-mass-to-light-ratios (GMLRs), 
and IMLR profiles.

\subsection{Results of spectral fits}
\label{sec:fittingresults}

\begin{figure}[!t]
  \begin{center}
    \includegraphics[width=0.4\textwidth,angle=0,clip]{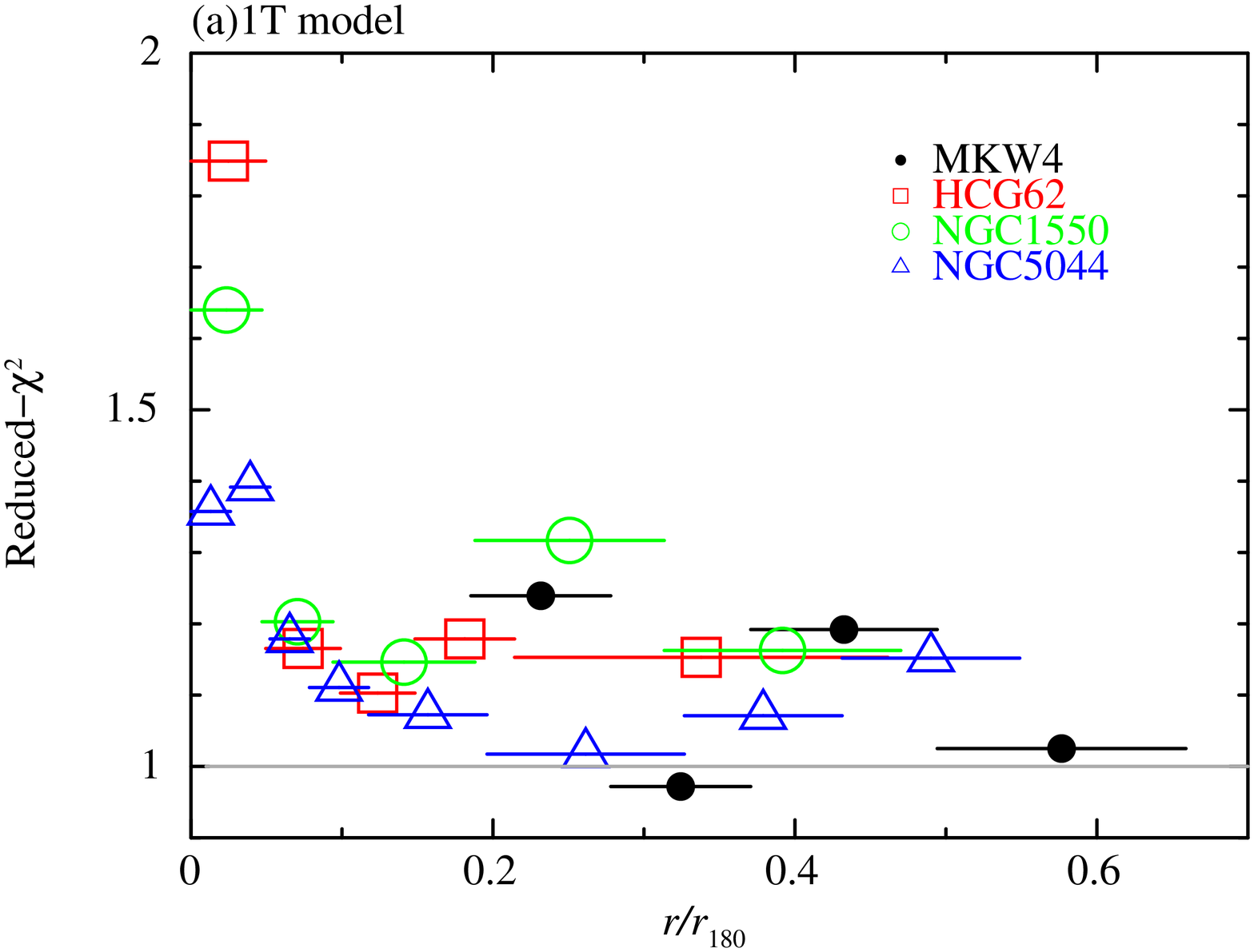}
    \includegraphics[width=0.4\textwidth,angle=0,clip]{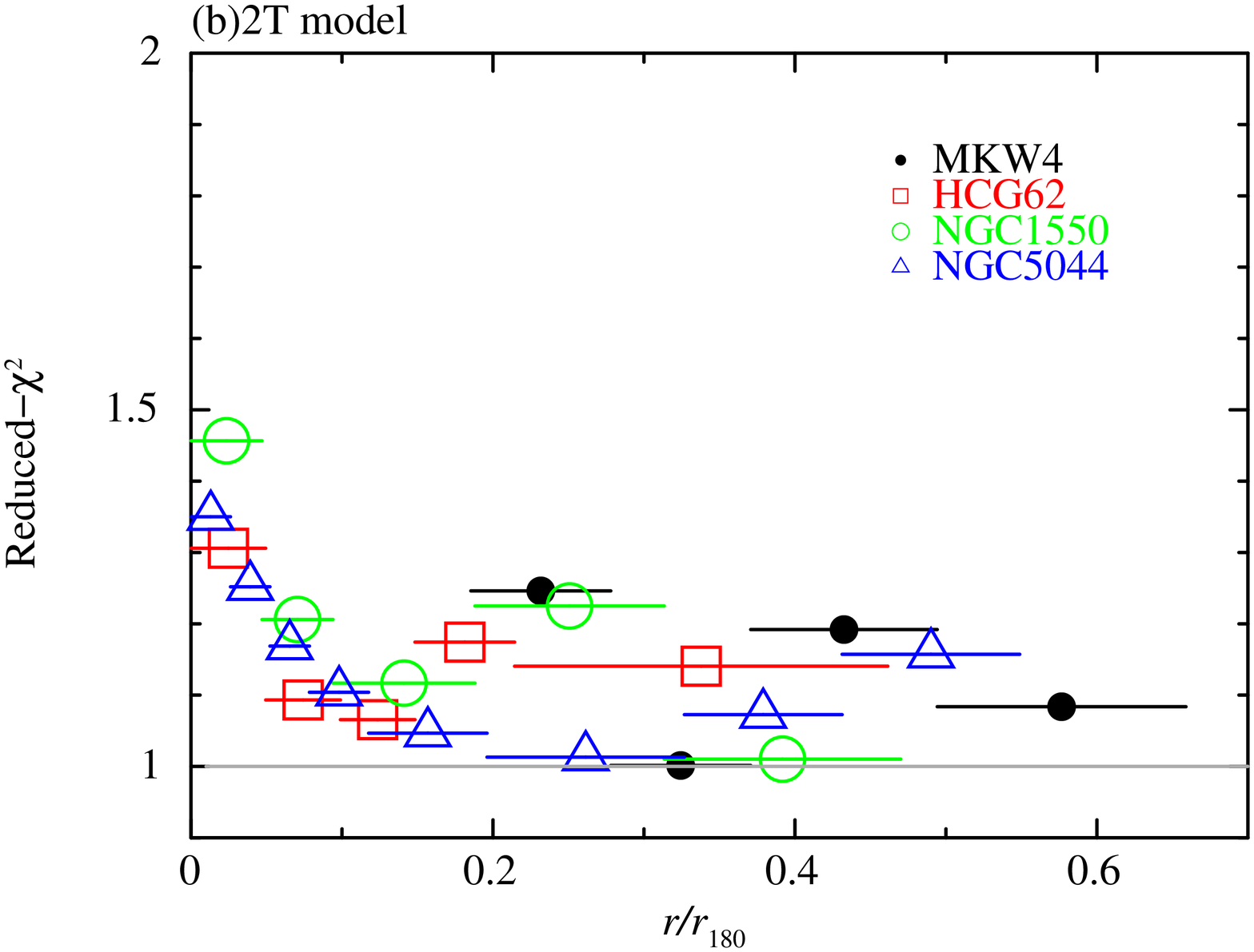}
  \end{center}
  \caption{
(a) Radial profiles of the reduced $\chi^{2}$ derived from the 1T model 
fits with the ATOMDB version 2.0.1. (b) The same figure as (a), but 
from the 2T model fits.
}
\label{radial_chi}
\end{figure}

We fitted the NXB-subtracted spectra with the 1T and 2T models.
All the spectra were well-represented by the 1T or 2T model as 
shown in figure \ref{fig:spectrum}. 
 Figure \ref{radial_chi} shows radial reduced $\chi^{2}$ profiles 
for each spectral fit.  
The fit statistics in $\chi^2$ test significantly favored 
the 2T model rather than the 1T model within $0.05~r_{180}$.  In the 
$r>0.05~r_{180}$ region, however, the differences in the reduced 
$\chi^2$ between the 1T and 2T model fits were only several \%, and 
the 1T model represented the observed spectra fairly well.  
Although the reduced $\chi^{2}$ profiles 
decreased with radius up to $\sim 0.2~r_{180}$, the reduced $\chi^{2}$ 
in the $r > 0.2~r_{180}$ had a flatter slope than those in the 
central region.  In the $F$-test probabilities, the spectral fits up 
to $\sim 0.1~r_{180}$ were clearly improved by the 2T model rather
than the 1T model.  On the other hand, in the $r > 0.2~r_{180}$, 
the improvements were not significant.


\begin{figure}[!t]
  \begin{center}
           \includegraphics[width=0.4\textwidth,angle=0,clip]{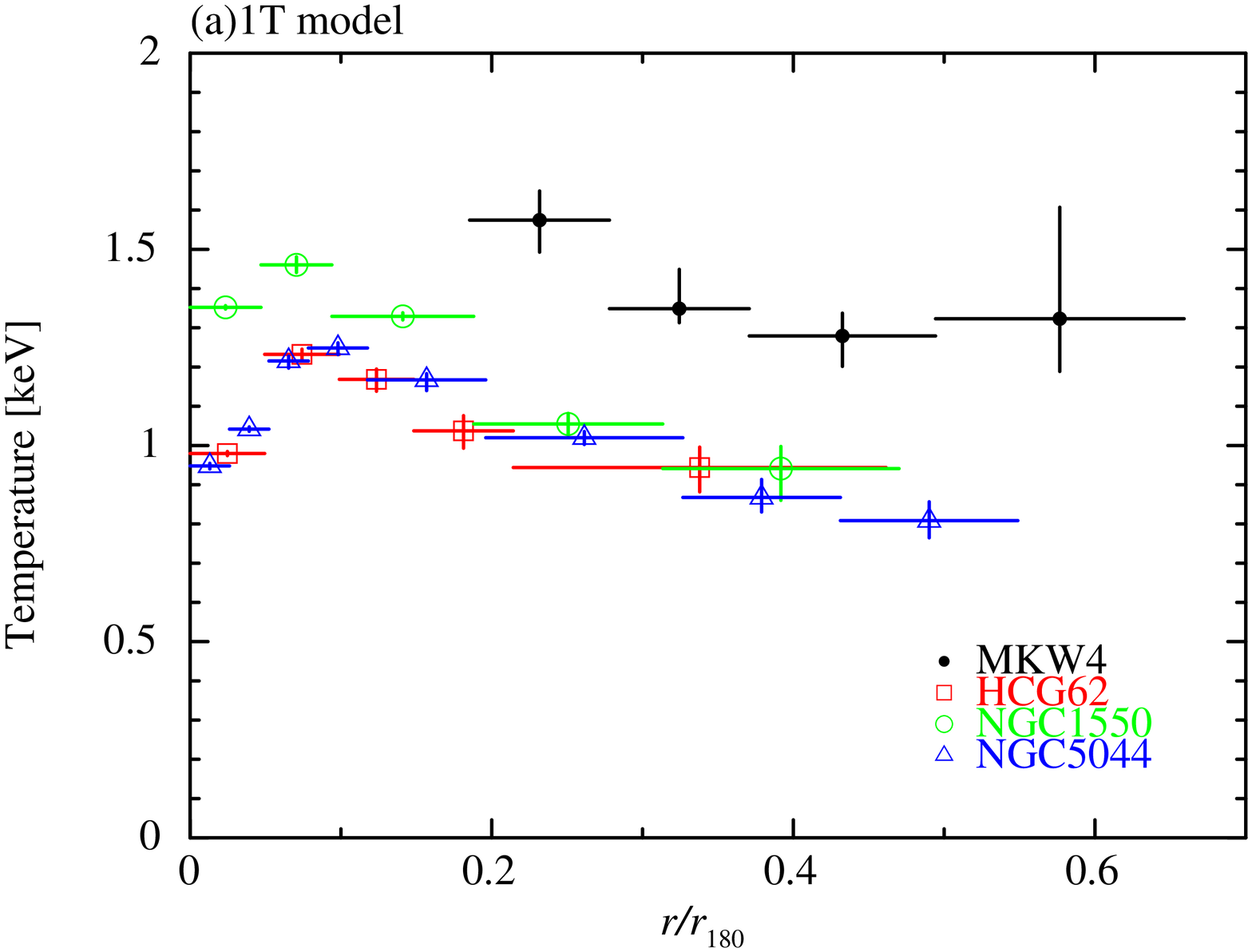}
           \includegraphics[width=0.4\textwidth,angle=0,clip]{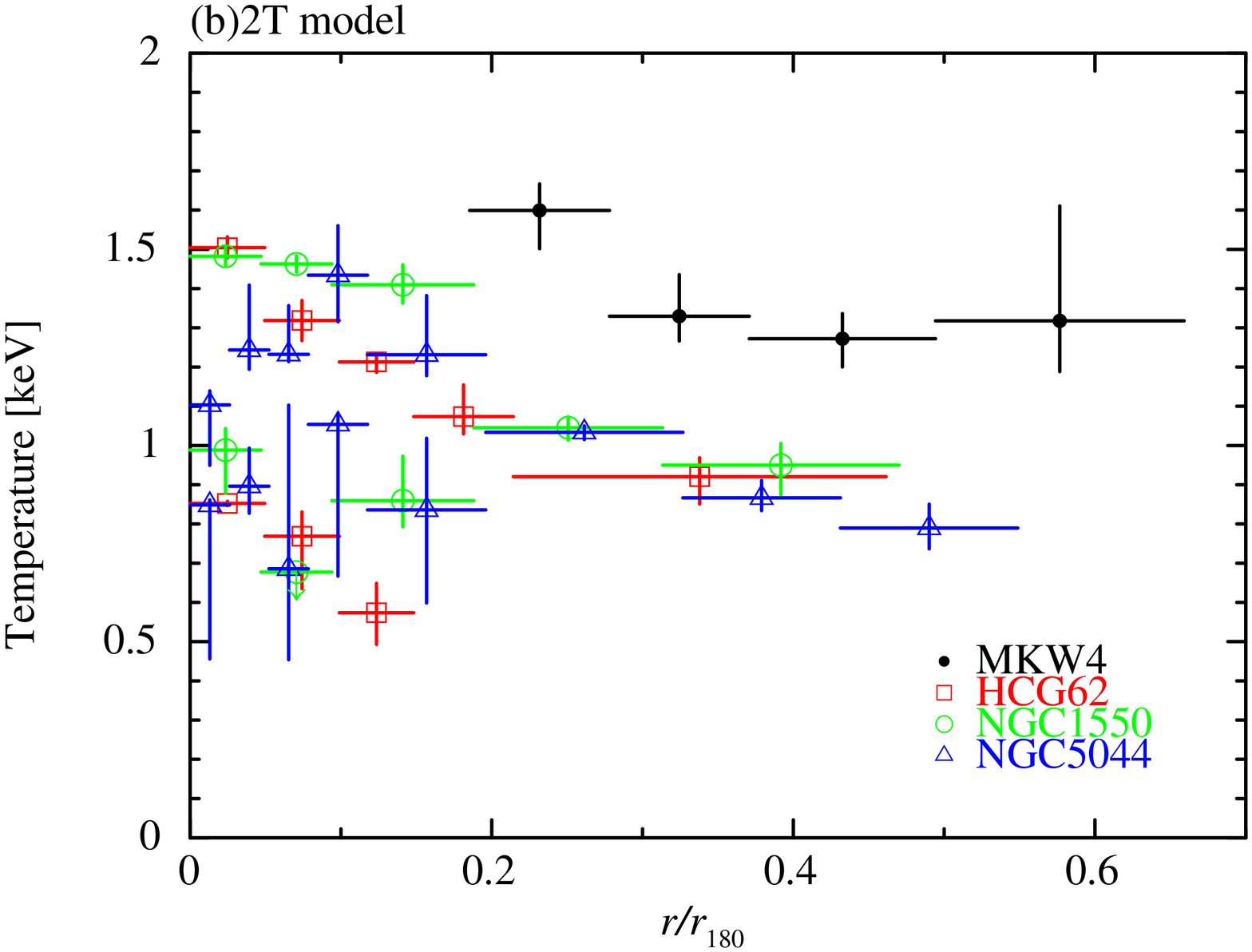}  
             \end{center}
  \caption{
Radial temperature profiles derived from the 1T (left) and 
2T (right) model fits. Black, red, green, and blue correspond to the 
resultant temperatures for MKW~4, HCG~62, the NGC~1550 group, and 
the NGC~5044 group, respectively. 
In figure (b), the cooler component in 3$'$--6$'$ of the NGC1550 group 
was derived only upper limit.
}
\label{fig:kTprofile}
\end{figure}

The resultant radial temperature profiles from the 1T and 2T model 
fits with the ATOMDB version 2.0.1 are shown in figure 
\ref{fig:kTprofile}.   With the 1T model, the temperatures slightly 
increased with radius up to $0.1~r_{180}$, and it declined to about 
a half of the peak temperature at $0.5~r_{180}$\@.  With the 2T model, 
within $0.2~r_{180}$, the resultant temperatures of both the cooler 
and hotter components for HCG~62 and the NGC~1550 group decreased 
with radius.  Although the temperature profiles for the NGC~5044 group 
with the 2T model had larger error bars, the hotter component profile 
looks similar to that with the 1T model.  Beyond $0.2~r_{\rm 180}$, 
we were not able to constrain the temperature of the minor component 
for the 2T model, and the temperatures of the major component were 
close to those from the 1T model fits.  

\begin{figure*}
  \begin{center}
          \includegraphics[width=0.4\textwidth,angle=0,clip]{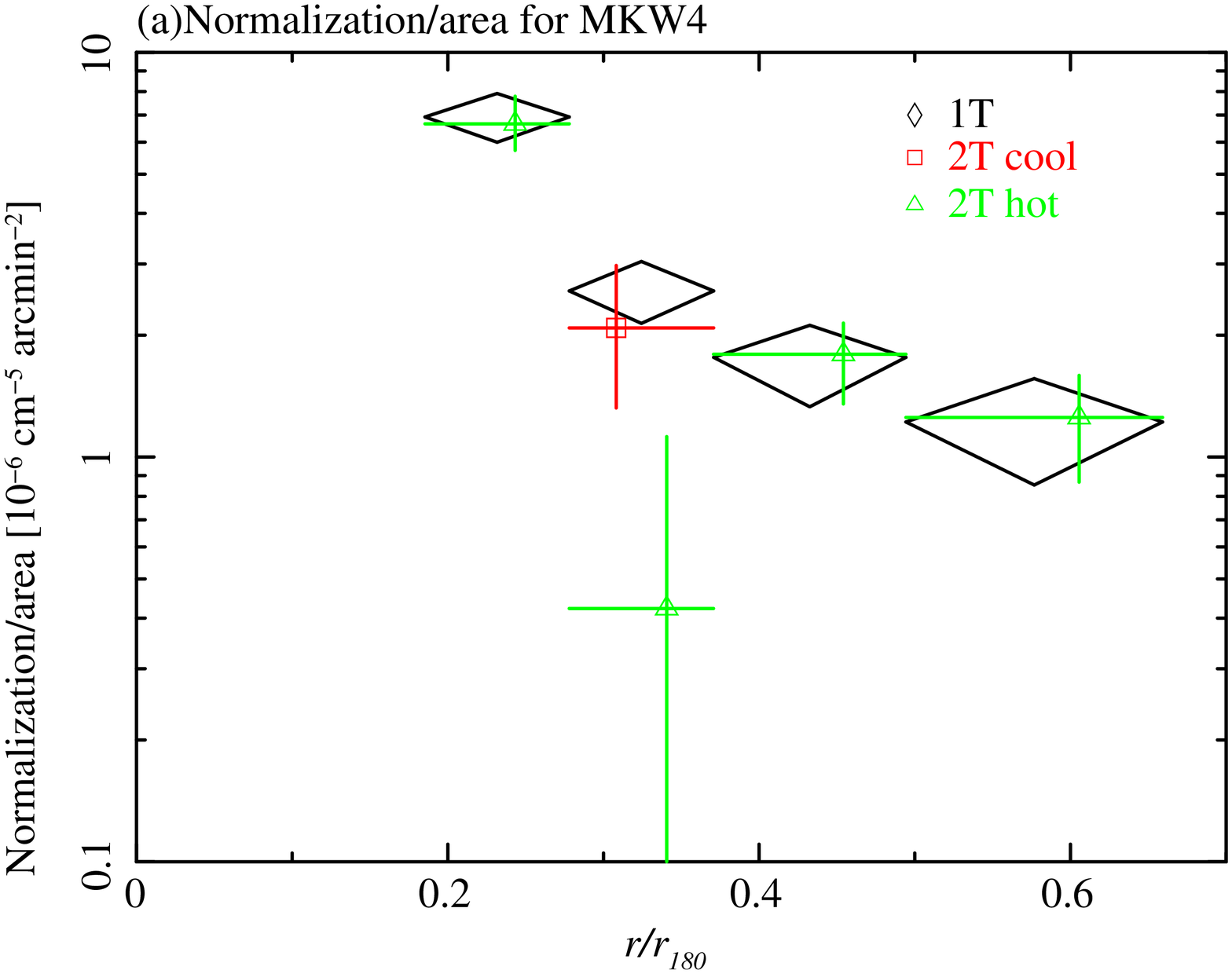}
          \includegraphics[width=0.4\textwidth,angle=0,clip]{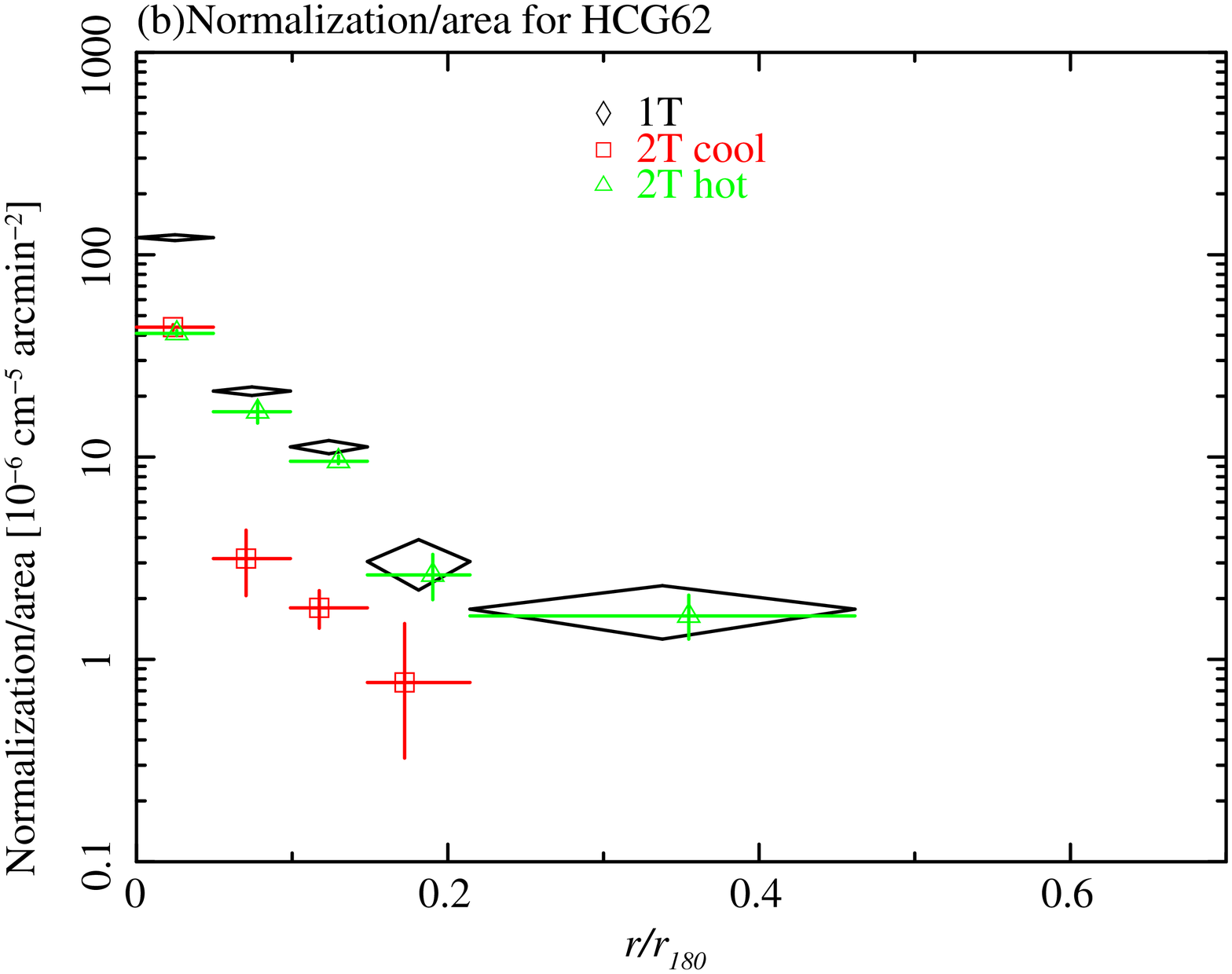}
          \includegraphics[width=0.4\textwidth,angle=0,clip]{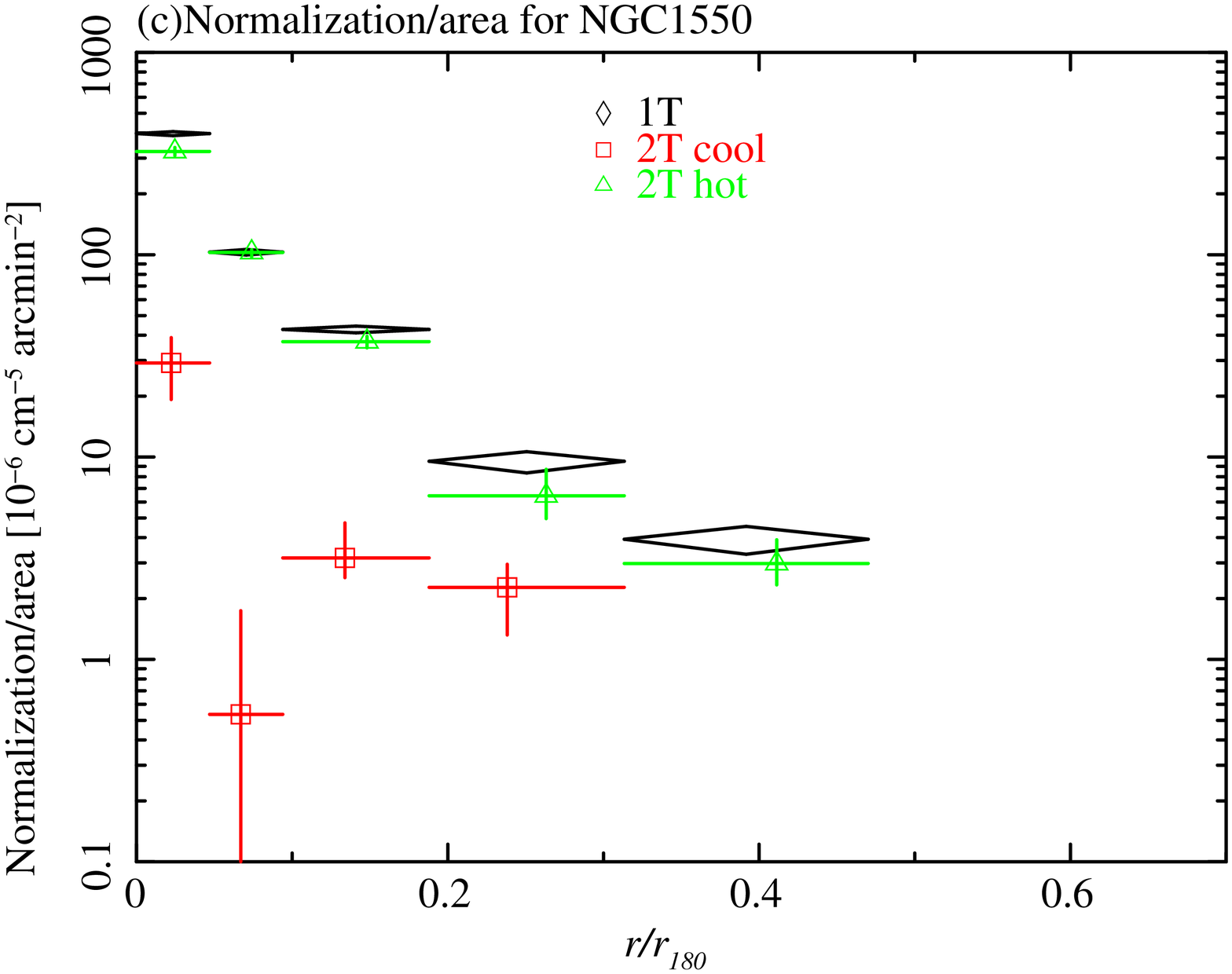}
          \includegraphics[width=0.4\textwidth,angle=0,clip]{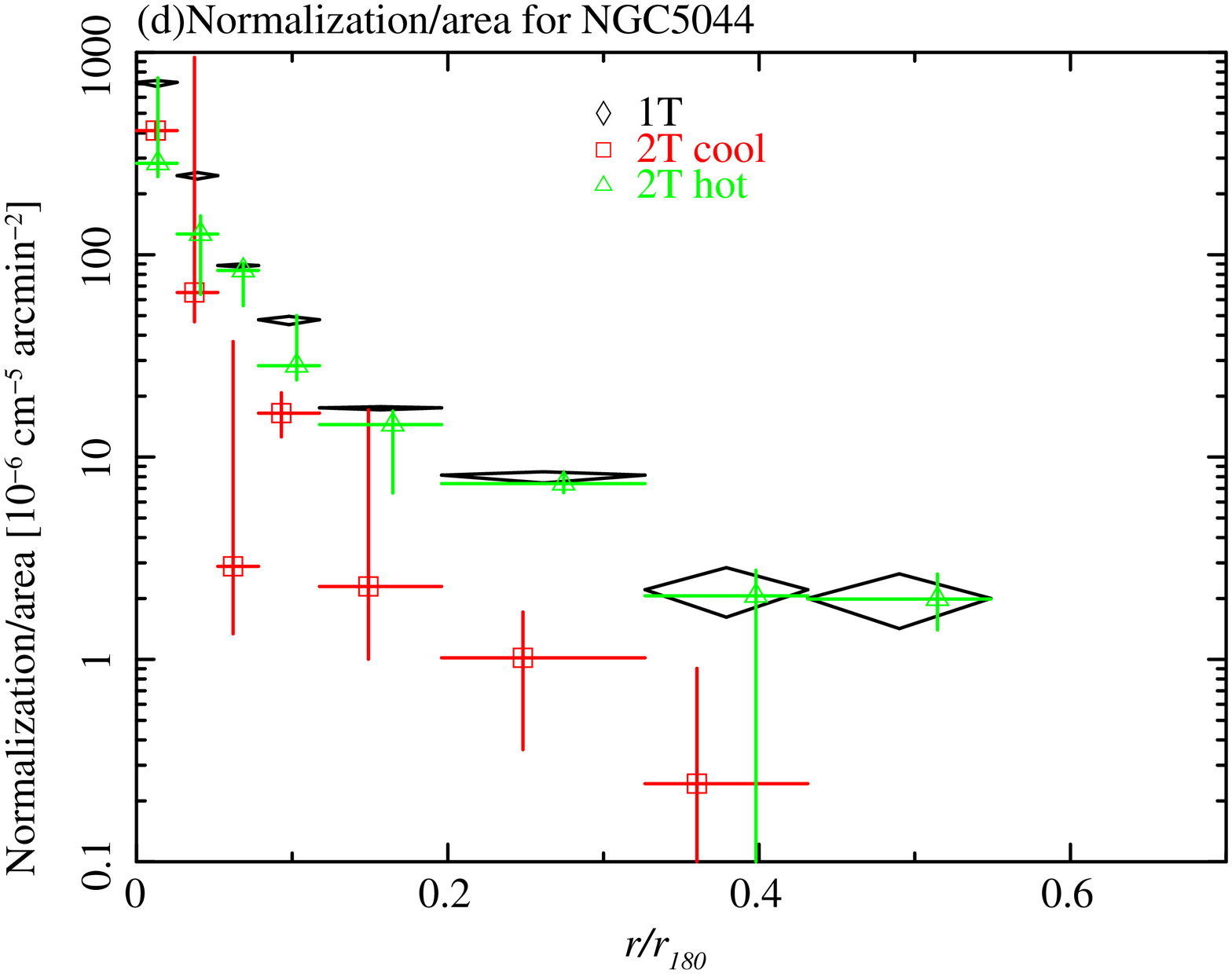}
  \end{center}
  \caption{
Radial normalization profiles divided by the area for each group. 
Black diamonds indicate normalization profiles divided by the 
area with the 1T model.  Red squares and green triangles correspond 
to the normalizations divided by the area of the cooler and hotter 
components with the 2T model, respectively.
}
\label{fig:normprofile}
\end{figure*}

Figure \ref{fig:normprofile} shows radial profiles of the resultant
normalizations, which are derived from the spectral fits with 
the 1T and 2T models, divided by the area from which each spectrum 
is extracted.  The normalizations divided by the area, which 
corresponded to the surface brightness profile, decreased with radius 
for both the 1T and 2T model fits.  The normalization profiles for 
the 1T model fits were close to the sum of those of the cooler and 
hotter components for the 2T model fits.

\begin{figure*}[!t]
      \includegraphics[width=0.32\textwidth,angle=0,clip]{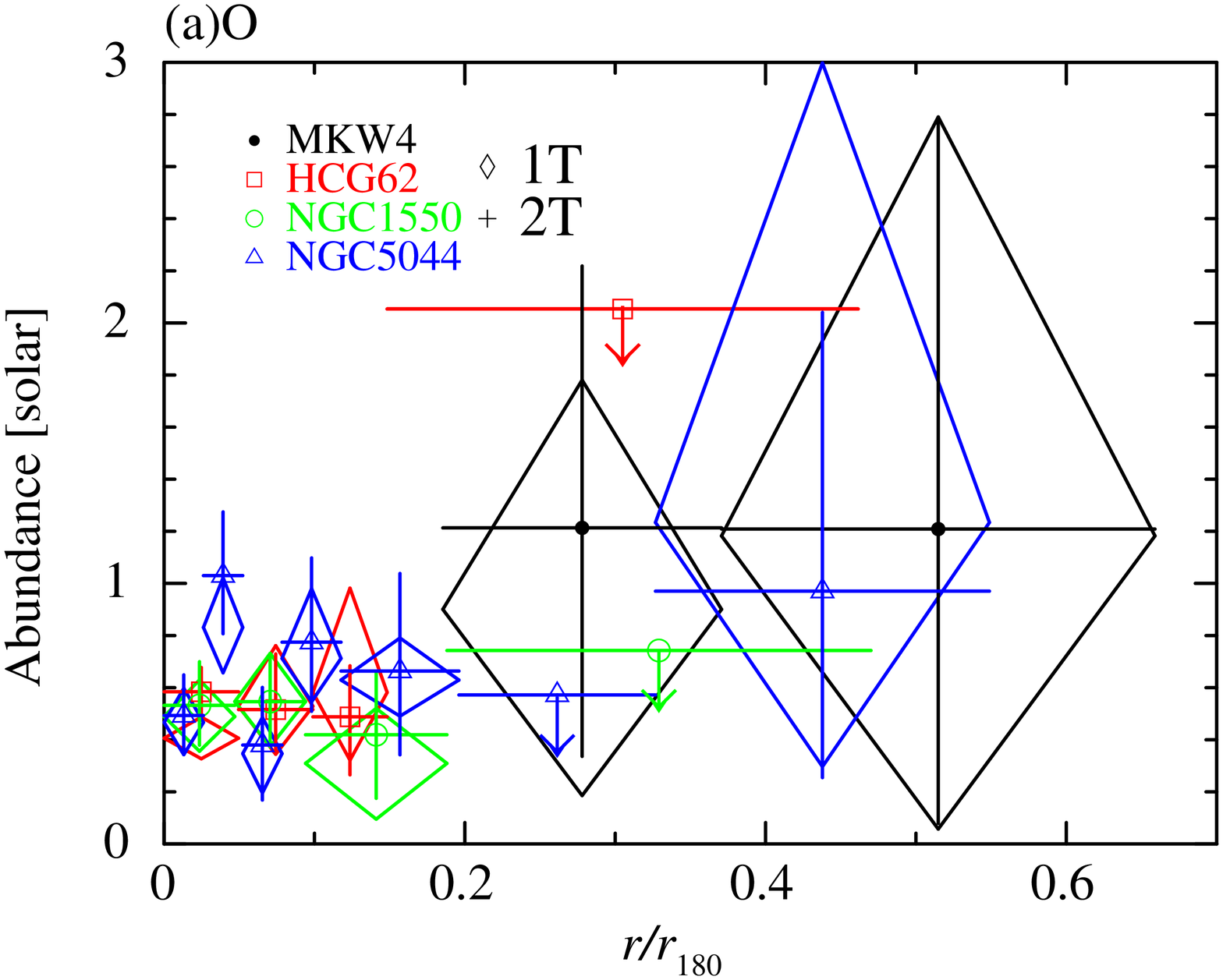}    
      \includegraphics[width=0.32\textwidth,angle=0,clip]{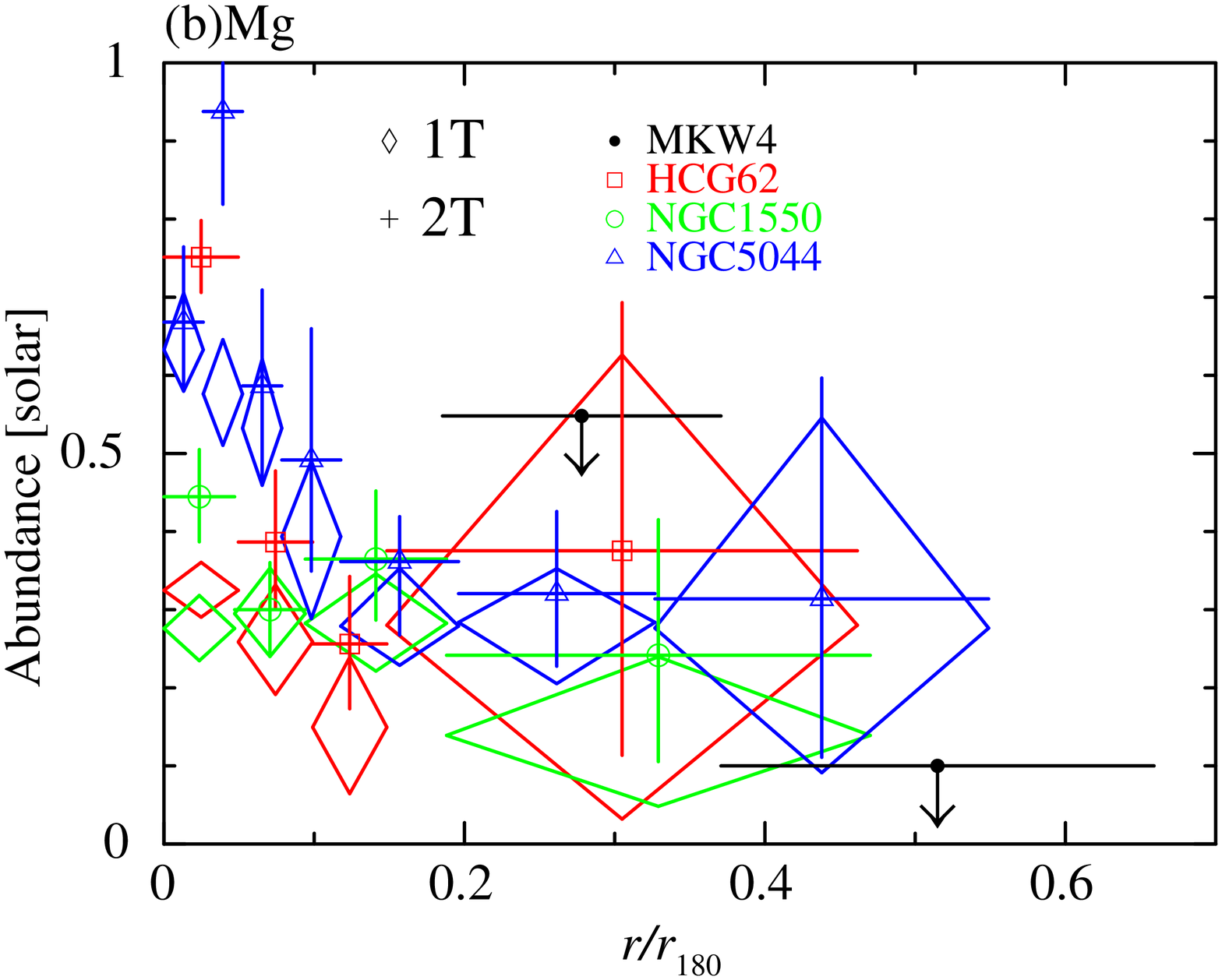}    
      \includegraphics[width=0.32\textwidth,angle=0,clip]{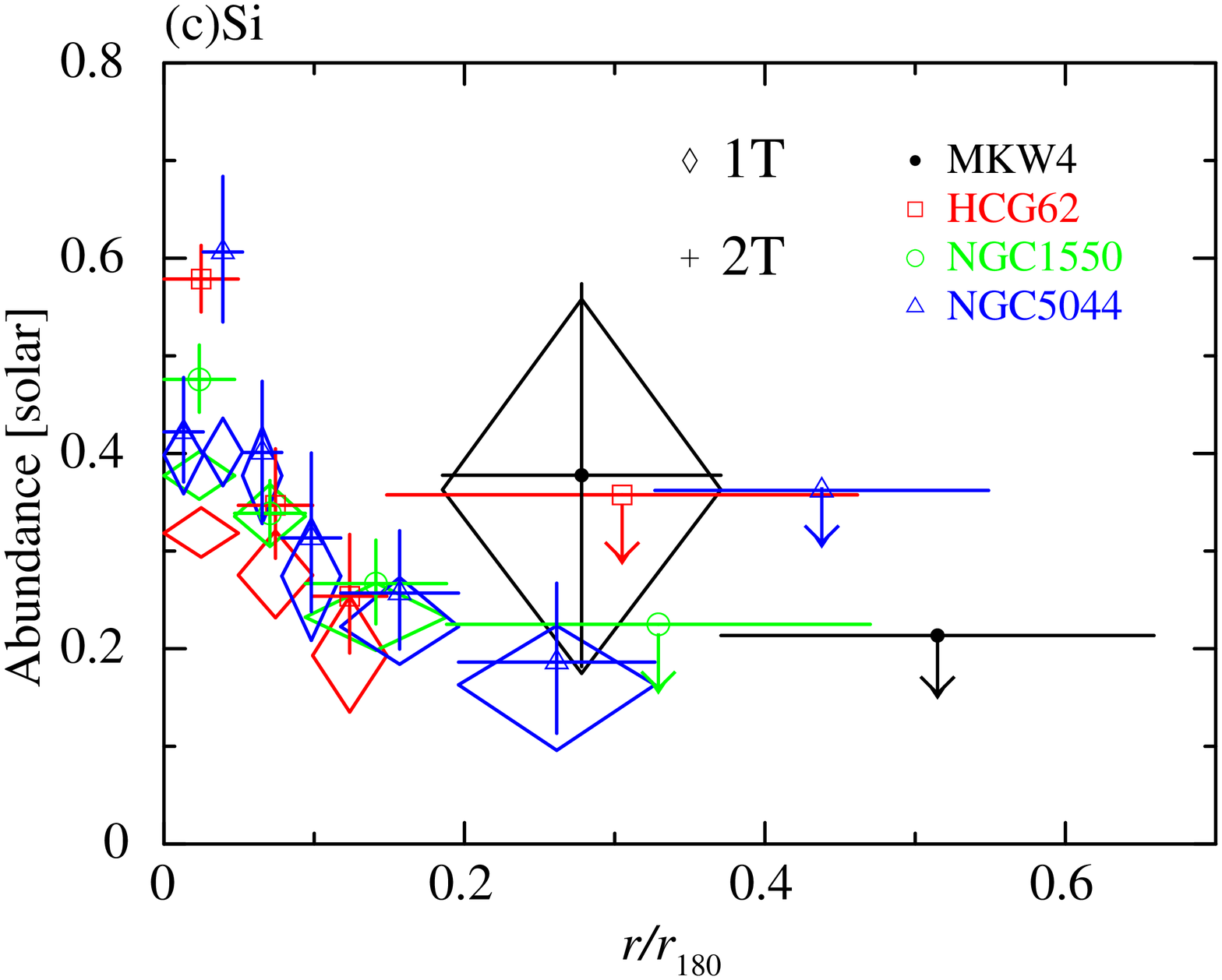}    
    \includegraphics[width=0.32\textwidth,angle=0,clip]{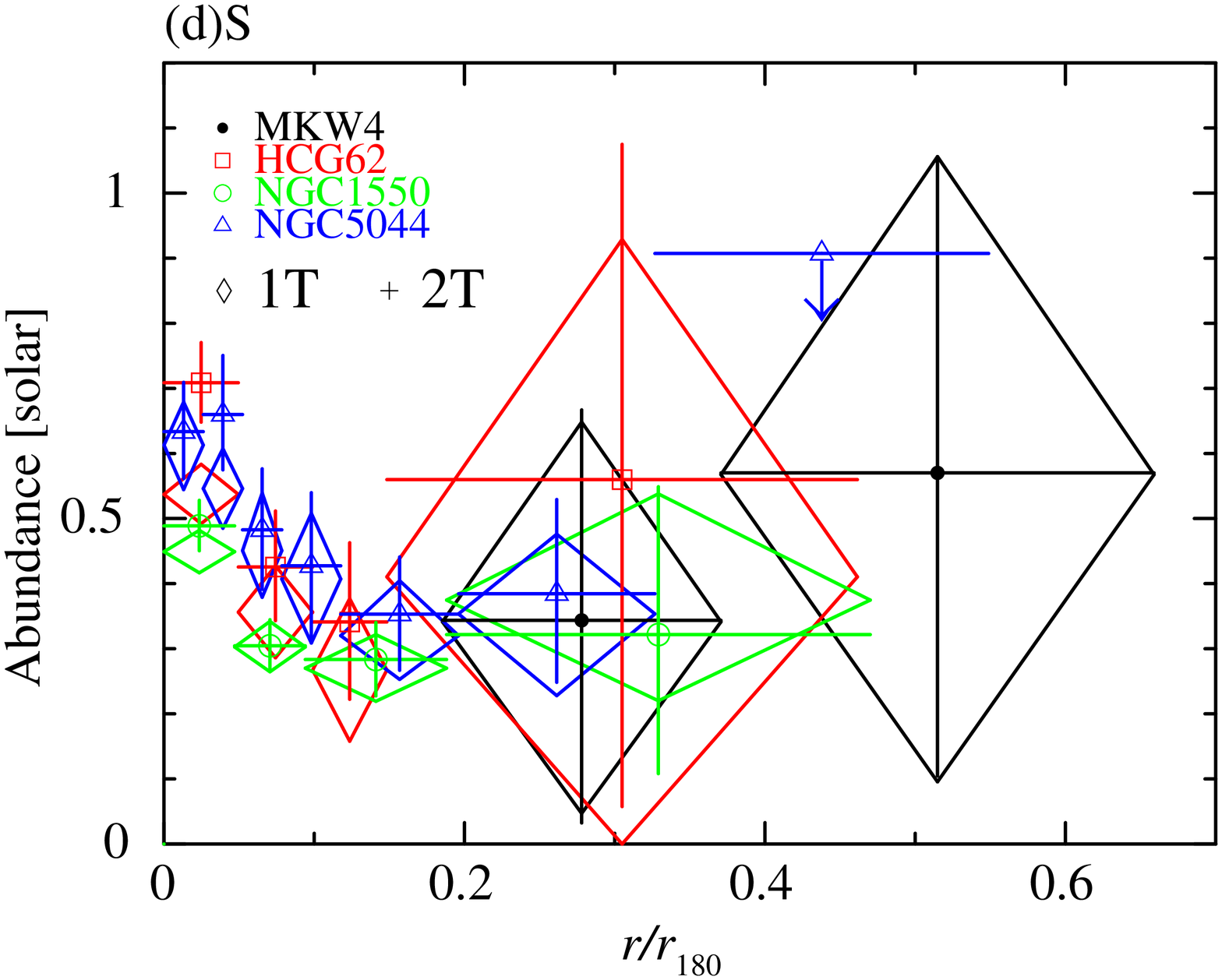}    
    \includegraphics[width=0.32\textwidth,angle=0,clip]{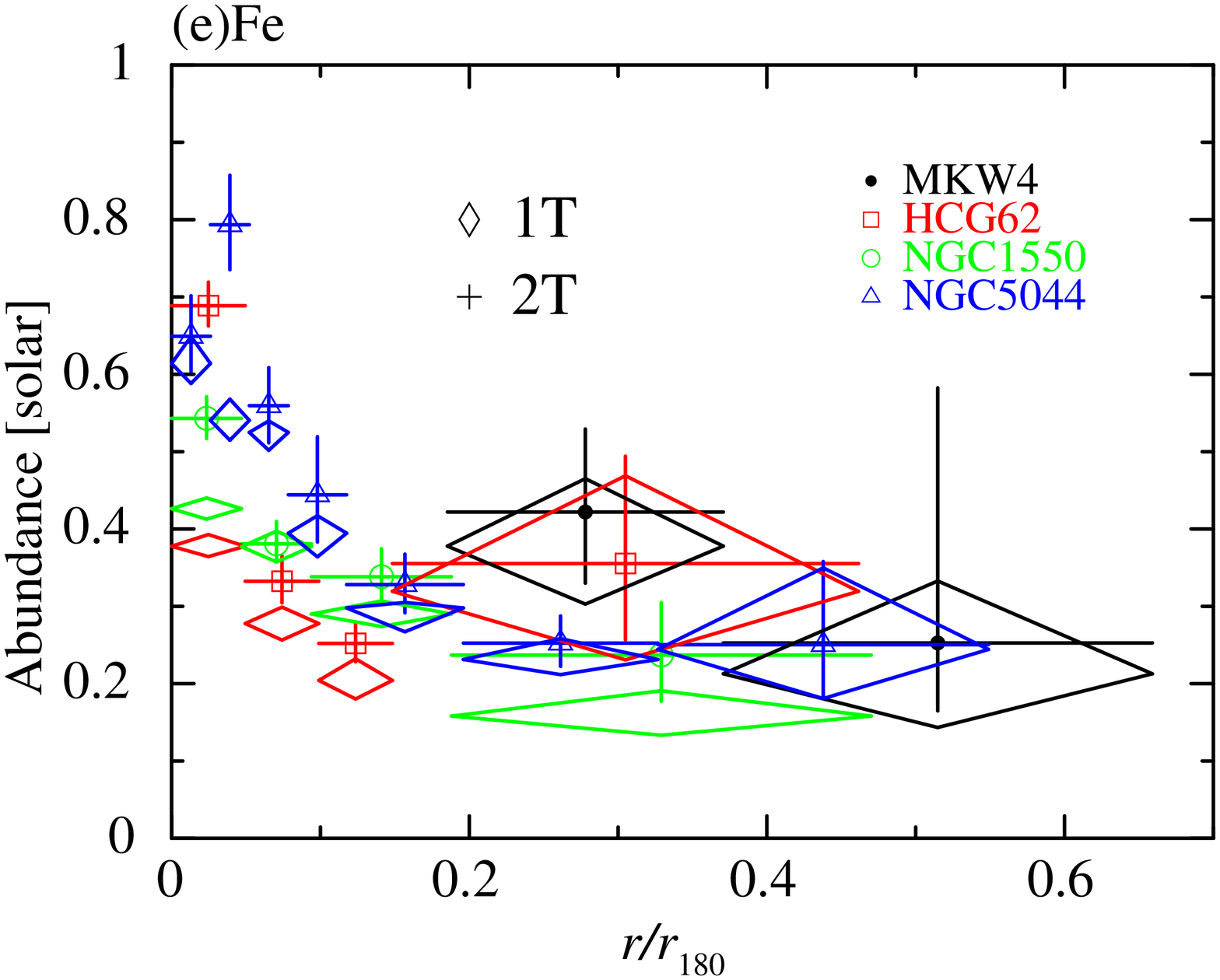}    
  \caption{
Abundance profiles of (a) O, (b) Mg, (c) Si, (d) S, and (e) Fe in 
the ICM with the ATOMDB version 2.0.1\@. The diamonds and crosses
indicate the abundances derived from the 1T and 2T model fits,  
respectively.  The notations of colors are same as in figure 
\ref{radial_chi}.  As for the abundances which were not 
constrained significantly, here we plotted only the larger upper 
limit derived from the 1T or 2T model.
}
\label{fig:feabundance}
\end{figure*}

Radial abundance profiles of O, Si, and S derived from the 1T 
and 2T model fits with the ATOMDB version 2.0.1 are summarized 
in figure \ref{fig:feabundance}.  The Mg and Fe abundances within 
$0.1~r_{180}$ had large scatter.  Beyond $0.1~r_{180}$, the results 
of all the abundances except for O with both the 1T and 2T model had 
similar profiles.  The O abundance profile showed no radial gradient 
although the statistical error were fairly large.  The Fe abundances 
decreased with radius, and reached $\sim 0.3$ solar at $0.5~r_{180}$.

\begin{figure*}[!th]
  \begin{center}
    \includegraphics[width=0.4\textwidth,angle=0,clip]{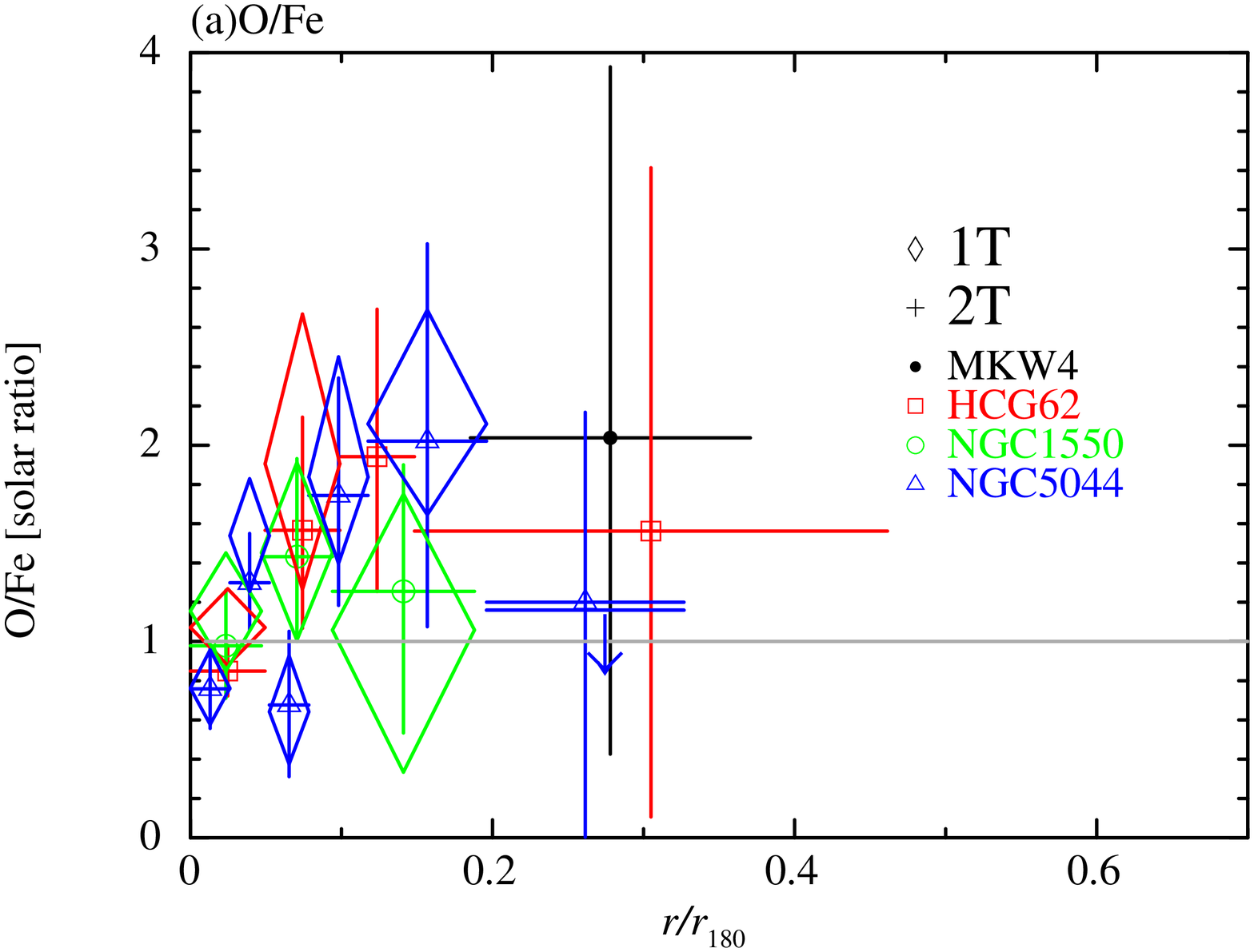}
    \includegraphics[width=0.4\textwidth,angle=0,clip]{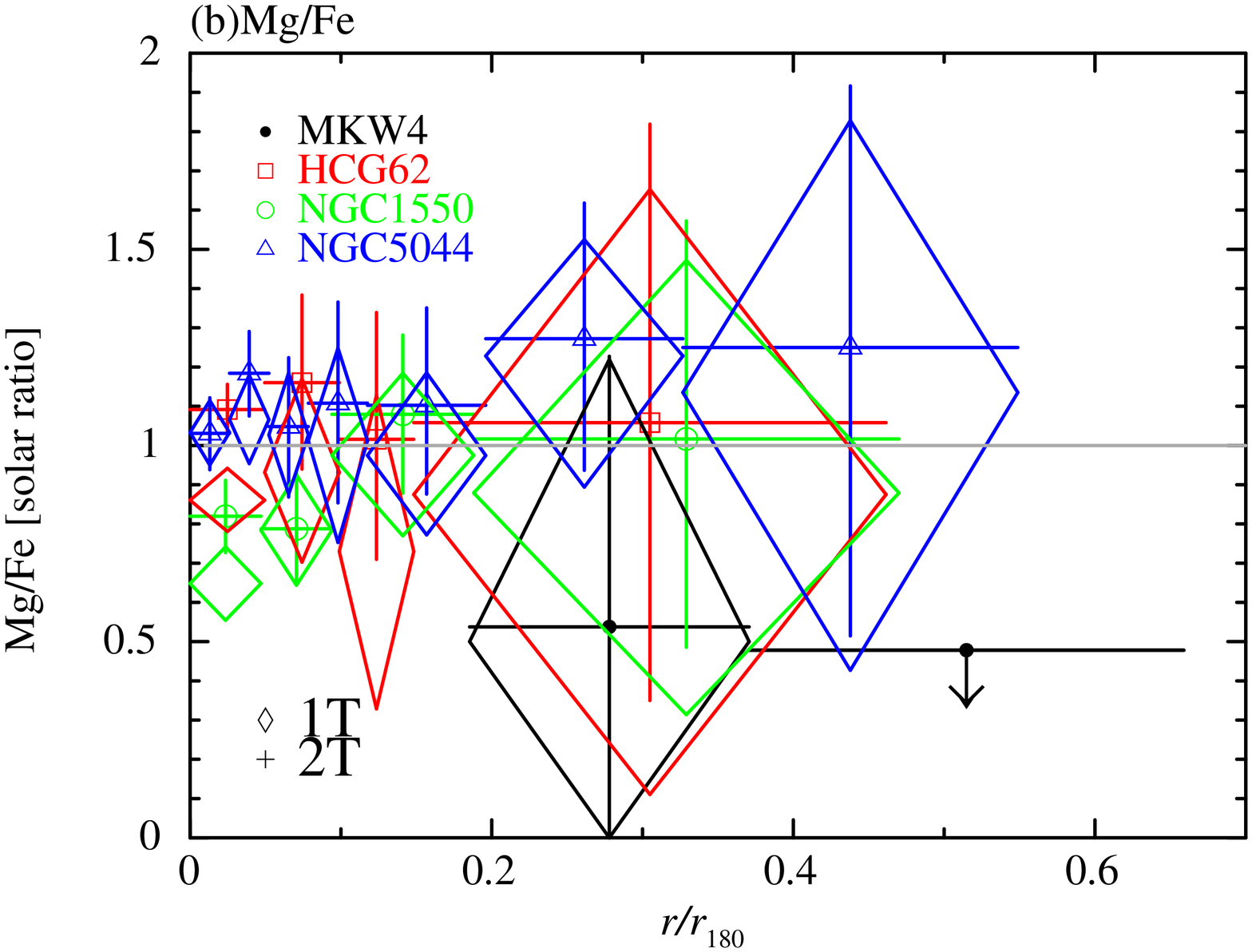}
    \includegraphics[width=0.4\textwidth,angle=0,clip]{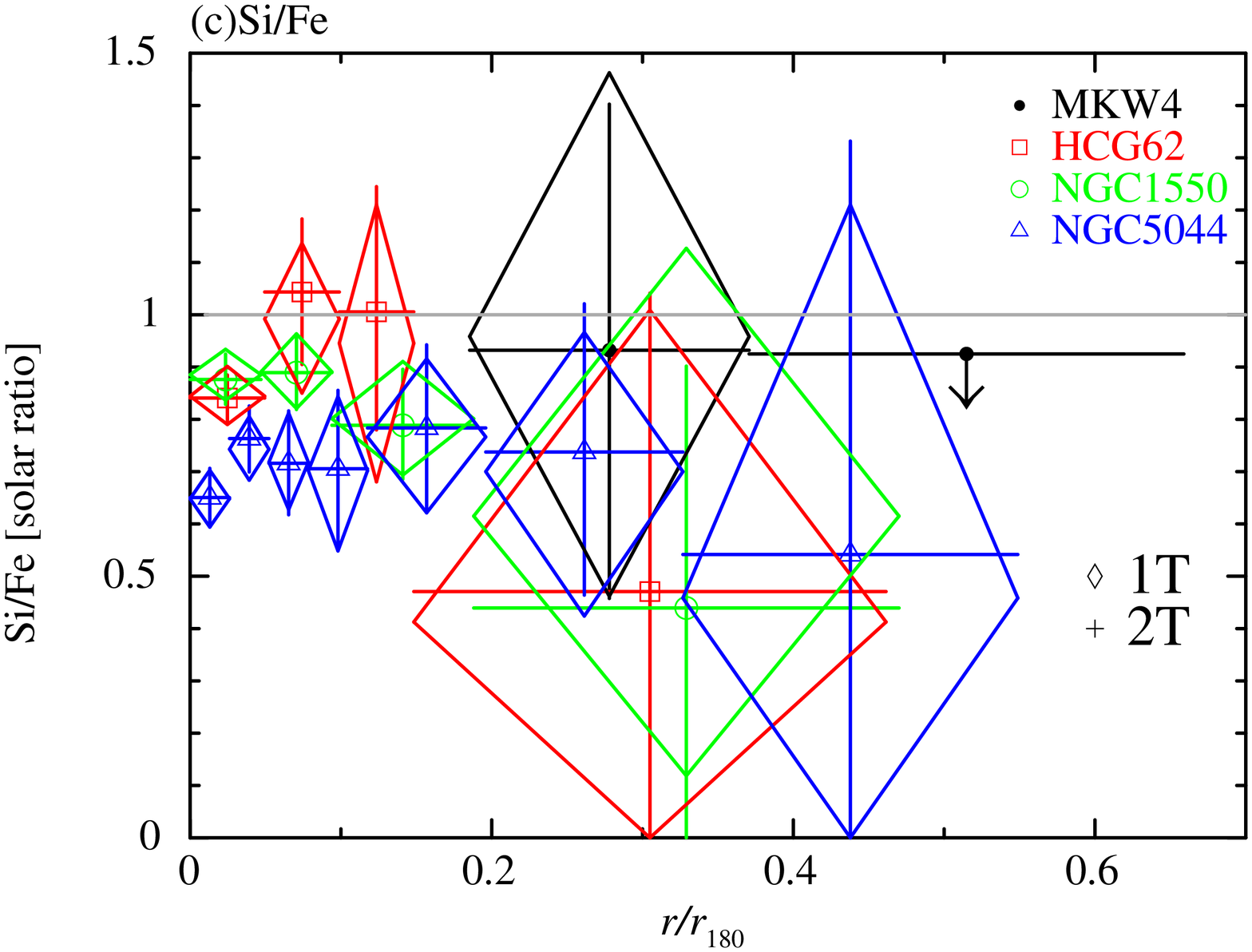}
    \includegraphics[width=0.4\textwidth,angle=0,clip]{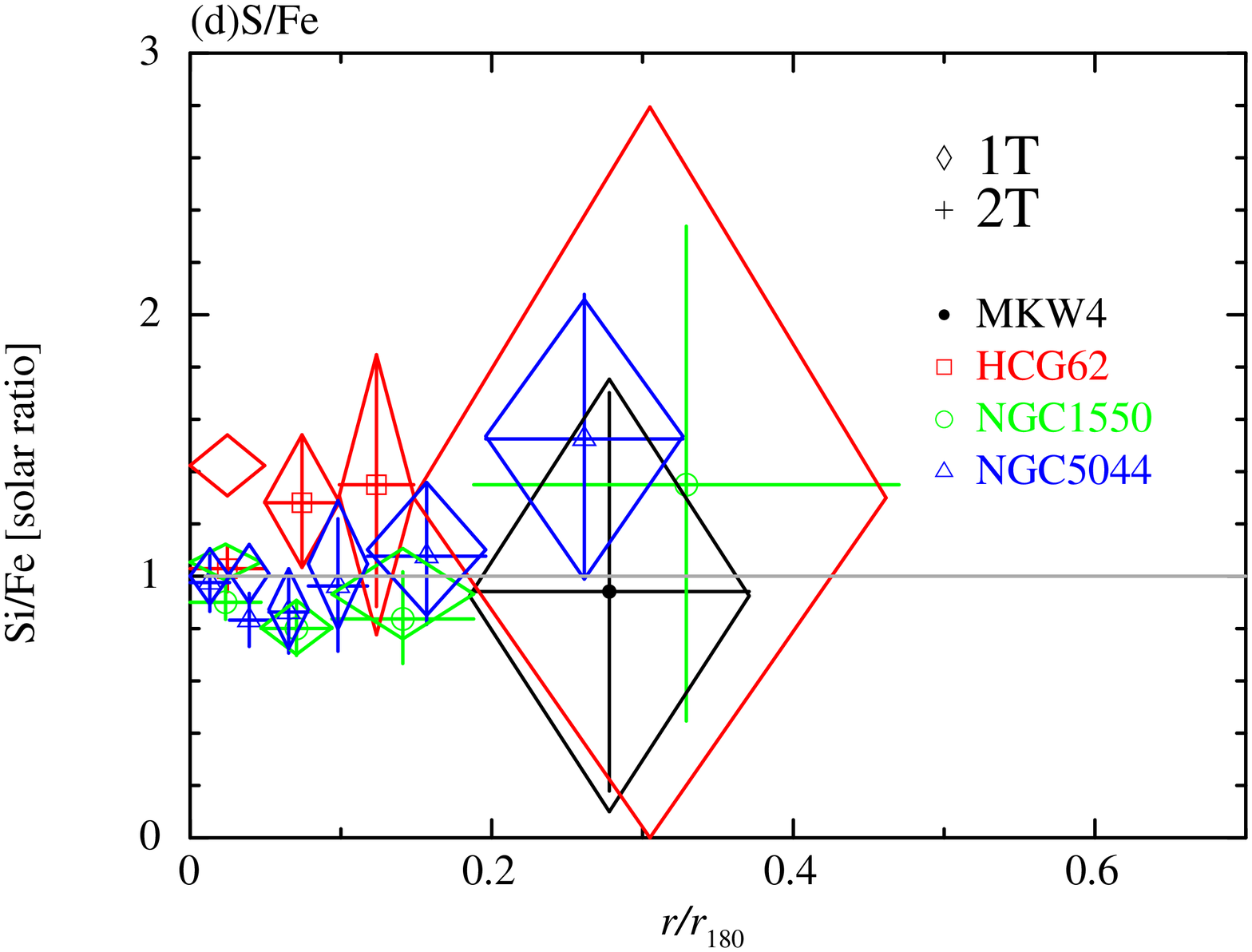}
   \end{center}
  \caption{
Radial profiles of (a) O/Fe, (b) Mg/Fe, (c) Si/Fe, and (d) S/Fe 
derived from the spectral fits with the 1T and 2T models with the 
ATOMDB version 2.0.1\@. The colors and marks are the same as in figure 
\ref{radial_chi}.
}
\label{fig:abundance}
\end{figure*}

We examined radial profiles of the abundance ratios, O/Fe, Mg/Fe, 
Si/Fe, and S/Fe as shown in figure \ref{fig:abundance}.  We do not 
show Ne/Fe ratio profiles because of large systematic uncertainties 
for the overlap with strong Fe-L lines.  
For estimating the abundance ratios rather 
than the absolute values, we calculated confidence contours between 
the metal abundances (O, Ne, Mg, and Si) and the Fe abundance. 
The derived abundance ratios of Mg/Fe, Si/Fe, and S/Fe were consistent to be a 
constant value of around $\sim 1$, although the error bars were fairly 
large, particularly in the outer region.  Both the 1T and 2T model 
fits gave similar abundance ratios.

\begin{deluxetable}{llcccccccc}
\tabletypesize{\scriptsize}
%
\tablewidth{0pt}
\tablecaption{
Weighted averages of the abundance ratios in units of the 
solar ratios with ATOMDB version 2.0.1.\label{tb:weight201}}
\tablehead{
\colhead{group} & \colhead{region} & \multicolumn{2}{c}{O/Fe [solar ratio]}  & \multicolumn{2}{c}{Mg/Fe [solar ratio]} & \multicolumn{2}{c}{Si/Fe [solar ratio]} & \multicolumn{2}{c}{S/Fe [solar ratio]}  \\
\colhead{} & \colhead{} & \colhead{1T} & \colhead{2T} & \colhead{1T} & \colhead{2T} & \colhead{1T} & \colhead{2T} & \colhead{1T} & \colhead{2T} \\ 
}
\startdata
MKW~4      & All$^{a}$ &$2.04^{+1.89}_{-1.61}$ &$2.04^{+3.10}_{-2.04}$  &$0.14^{+0.63}_{-0.14}$ &$0.14^{+0.63}_{-0.14}$  &$0.76^{+0.74}_{-0.73}$  &$0.74^{+0.69}_{-0.69}$ &$1.08^{+1.29}_{-1.08}$ &$1.05^{+1.19}_{-1.05}$ \\
HCG~62 & $< 0.1~r_{180}$ &$1.12^{+0.31}_{-0.30}$ &$0.88^{+0.21}_{-0.20}$  &$0.87^{+0.13}_{-0.12}$ &$1.10^{+0.10}_{-0.10}$  &$0.86^{+0.09}_{-0.08}$ &$0.86^{+0.07}_{-0.07}$  &$1.40^{+0.17}_{-0.17}$ &$1.06^{+0.14}_{-0.14}$ \\
 & $> 0.1~r_{180}$ &$2.60^{+2.61}_{-1.55}$ &$1.89^{+1.14}_{-1.02}$  &$0.76^{+0.58}_{-0.58}$ &$1.02^{+0.49}_{-0.46}$  &$0.86^{+0.40}_{-0.37}$ &$0.93^{+0.36}_{-0.33}$  &$1.31^{+0.83}_{-0.81}$ &$1.36^{+0.78}_{-0.73}$ \\
 & All$^{a}$ &$1.14^{+0.31}_{-0.30}$ &$0.92^{+0.21}_{-0.20}$  &$0.86^{+0.12}_{-0.12}$ &$1.09^{+0.10}_{-0.10}$  &$0.86^{+0.08}_{-0.08}$ &$0.86^{+0.07}_{-0.07}$  &$1.40^{+0.17}_{-0.17}$ &$1.07^{+0.13}_{-0.13}$ \\
NGC~1550 & $< 0.1~r_{180}$ &$1.25^{+0.41}_{-0.41}$ &$1.09^{+0.40}_{-0.37}$  &$0.69^{+0.13}_{-0.13}$ &$0.81^{+0.13}_{-0.13}$  &$0.89^{+0.07}_{-0.07}$ &$0.88^{+0.07}_{-0.06}$  &$0.98^{+0.09}_{-0.09}$ &$0.87^{+0.09}_{-0.09}$ \\
 & $> 0.1~r_{180}$ &$1.04^{+1.12}_{-0.55}$ &$1.21^{+1.04}_{-0.42}$  &$0.96^{+0.32}_{-0.32}$ &$1.07^{+0.31}_{-0.31}$  &$0.79^{+0.18}_{-0.17}$ &$0.77^{+0.17}_{-0.16}$  &$0.95^{+0.28}_{-0.28}$ &$0.85^{+0.29}_{-0.27}$ \\
 & All$^{a}$ &$1.22^{+0.38}_{-0.33}$ &$1.10^{+0.38}_{-0.28}$  &$0.73^{+0.12}_{-0.12}$ &$0.85^{+0.12}_{-0.12}$  &$0.88^{+0.06}_{-0.06}$ &$0.87^{+0.06}_{-0.06}$  &$0.98^{+0.09}_{-0.09}$ &$0.87^{+0.09}_{-0.09}$ \\
NGC~5044 & $< 0.1~r_{180}$ &$0.96^{+0.22}_{-0.21}$ &$0.98^{+0.23}_{-0.23}$  &$1.04^{+0.10}_{-0.10}$ &$1.09^{+0.10}_{-0.10}$  &$0.69^{+0.06}_{-0.06}$ &$0.70^{+0.06}_{-0.06}$  &$0.98^{+0.11}_{-0.11}$ &$0.90^{+0.11}_{-0.11}$ \\
 & $> 0.1~r_{180}$ &$1.73^{+0.84}_{-0.70}$ &$1.82^{+1.11}_{-1.17}$  &$1.05^{+0.27}_{-0.28}$ &$1.17^{+0.32}_{-0.30}$  &$0.74^{+0.21}_{-0.20}$ &$0.77^{+0.22}_{-0.22}$  &$1.18^{+0.38}_{-0.37}$ &$1.17^{+0.39}_{-0.38}$ \\
 & All$^{a}$ &$1.03^{+0.22}_{-0.20}$ &$1.01^{+0.23}_{-0.22}$  &$1.04^{+0.09}_{-0.09}$ &$1.10^{+0.10}_{-0.10}$  &$0.70^{+0.06}_{-0.06}$ &$0.71^{+0.06}_{-0.06}$  &$1.00^{+0.10}_{-0.10}$ &$0.91^{+0.10}_{-0.10}$ \\ 
\enddata
\tablenotetext{a}{
All the regions observed with Suzaku.}
\end{deluxetable}

We calculated the weighted averages of the abundance ratios derived 
from both the ATOMDB versions for the $r < 0.1~r_{180}$, $r > 0.1~r_{180}$, 
and whole regions, and summarized in table \ref{tb:weight201}. 
On the average, the abundance ratios between the $r<0.1~r_{180}$ and 
$r > 0.1~r_{180}$ regions were almost consistent. The weighted averages 
of the Mg/Fe, Si/Fe, and S/Fe ratios for the whole regions were 
consistent to be a solar ratio, although those of the O/Fe ratios had 
large scatter in the outer regions.  The ratios from the 1T and 2T 
model fits were similar to each other.  In addition, the abundance 
ratios with the ATOMDB version 2.0.1 were consistent with those with 
the version 1.3.1(table \ref{tb:weight}), except for the O/Fe ratios.

\subsection{Comparisons of the results with the ATOMDB version 1.3.1 
and 2.0.1\@}
\label{sec:atomdb}

In this subsection, we summarize comparisons of the results
 from the spectral fits  with the ATOMDB version 1.3.1 and 2.0.1\@.
Tables and figures are shown in the Appendix \ref{sec:apend_atomdb}.

In the outer regions, the spectral fits with both the ATOMDB versions 
gave almost same reduced $\chi^2$ (figure \ref{fig:chi_201131}).  
For the 1T model fits, the $\chi^{2}$ values 
with the ATOMDB version 2.0.1 were smaller than those with the 
version 1.3.1 within $0.05~r_{180}$\@. On the other hand, the fit 
statistics in the reduced $\chi^2$ favor the ATOMDB version 1.3.1 
than the version 2.0.1 with the 2T model within $0.05~r_{180}$.

We compared the 
temperatures derived from the spectral fits with both of the ATOMDB 
versions (figure \ref{fig:201131_temp} a--c).  
The resultant temperatures from the 1T model fits and 
the cooler component from the 2T model fits with the version 2.0.1 
were systematically higher than those with the version 1.3.1 by 
$\sim 0.1$ keV\@.  On the other hand, as for the hotter component of 
the 2T model fits, the spectral fits with either version gave 
similar temperatures.  

Radial profiles of the normalizations divided by the area from which 
each spectrum were extracted with the 1T and 2T models were also compared 
with both the ATOMDB versions (figure \ref{fig:201131_temp} d--f ).
As in the temperatures, the resultant normalizations for the 1T 
model fits and the cooler component for the 2T model fits with the 
version 2.0.1 were systematically smaller than those with the version 
1.3.1\@.  On the other hand, as for the normalizations of the hotter 
component with the 2T model, the spectral fits with either version 
also gave similar values.

\begin{figure*}
  \begin{center} 
    \includegraphics[width=0.4\textwidth,angle=0,clip]{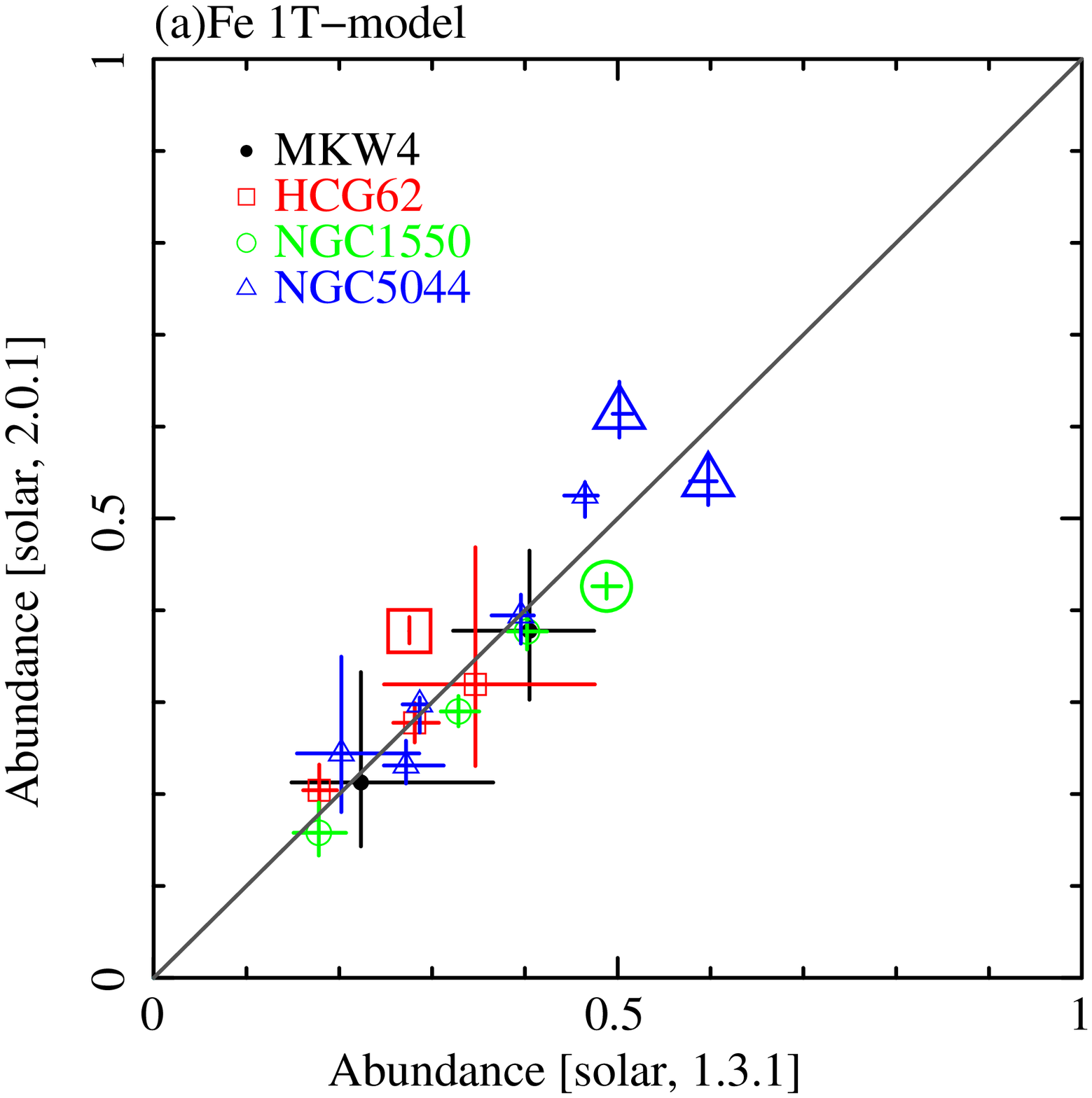} 
    \includegraphics[width=0.4\textwidth,angle=0,clip]{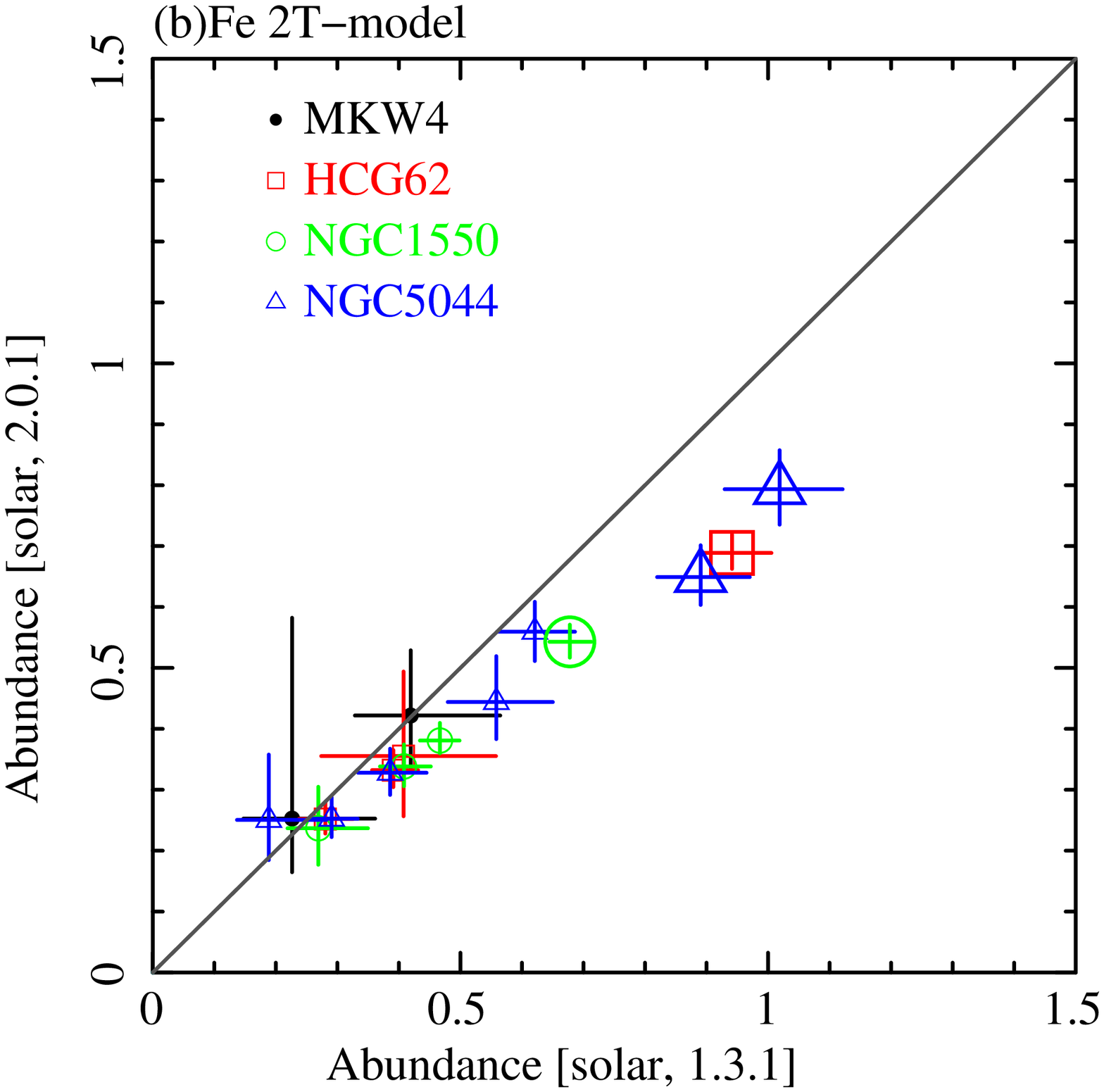} 
     \end{center}
  \caption{
Comparisons of the Fe abundances derived from the 1T and 2T model fits 
with the ATOMDB version 1.3.1 and 2.0.1\@. The colors are the same as in 
 figure \ref{radial_chi}.	
Bigger marks indicate the Fe abundance and the ratios within 
$0.05~r_{180}$.
}
\label{fig:201131_Fe}
\end{figure*}

Figure \ref{fig:201131_Fe} shows the difference of the 
derived Fe abundances with the 1T or 2T model between the 
ATOMDB version 1.3.1 and 2.0.1\@.  The spectral fits with 
the version 2.0.1 gave significantly smaller Fe abundances 
than those with the version 1.3.1\@, particularly within 
$0.05~r_{180}$.  

We also compared the abundance ratios  of O, Mg, Si, and S to Fe 
for the 1T and 2T model fits with the version 1.3.1 and 2.0.1, 
in units of the solar ratios (figure \ref{fig:201131_abund_1T} and \ref{fig:201131_abund_2T} ). 
The spectral fits with either version gave similar Si/Fe and S/Fe 
ratios.  On the other hand, the spectral fits with the version 2.0.1 
gave higher Mg/Fe ratios than those with the version 1.3.1 by 20\%.
This is because the Mg lines suffer from the Fe-L lines overlapping, 
and the differences between the ATOMDB versions would cause such 
systematic differences in the Mg/Fe ratios.  The O/Fe ratios with 
the version 2.0.1 had fairly higher value than those with the 
version 1.3.1\@.

\subsection{Comparisons of the results with previous results}
\label{sec:xmmchandra}

The temperature and abundance profiles
of central regions of our samples, HCG~62, the NGC~1550 group, and the 
NGC~5044 group, were already reported using XMM, Chandra, and Suzaku data.
Therefore, in this subsection, our results using the ATOMDB version 1.3.1
 were compared with these previous results, because the previous 
 results were derived with the ATOMDB version 1.3.1 or earlier versions.
Here, effects of difference in adopted solar abundance tables 
were corrected to \cite{Lodders2003}. 
Temperatures and abundances in this work with the ATOMDB version 1.3.1 
agreed well with the previous 
Suzaku results for the NGC~1550 group out to 0.5~$r_{180}$ \citep{Sato2010}, 
HCG~62 out to 0.2~$r_{180}$ \citep{Tokoi2008}, and 
the NGC~5044 group out to 0.3~$r_{180}$ \citep{Komiyama2009}.

For the NGC~1550 group, temperatures and abundances 
with the 1T and 2T model in this work agreed well 
with the XMM/Chandra results \citep{Sun2003,Kawaharada2009} 
for the whole region observed with the XMM (out to 14$'$).

In 6$'$--9$'$ of HCG~62, temperatures and abundances 
with the 1T and 2T model in this work were
consistent with the Chandra and XMM results \citep{Morita2006}
derived from  {\it vMEKAL} model \citep{Mewe1985, Mewe1986,Liedahl1995,Kaastra1992}  fits of deprojected spectra.
 However, the  Fe abundances within 6$'$ were significantly smaller 
and that at 9$'$--14$'$ were significantly higher than those in
\citet{Morita2006}. 
This difference in the inner regions would be caused by the difference in the PSF,
in the deprojection and projection, and the adopted atomic codes.
The difference in the outer region may be caused by a difference in the treatment of the
 Galactic foreground emission.

Excluding the innermost region of Suzaku, $r < 2'$, 
the temperatures and metal abundances of the NGC~5044 
group agreed well with the XMM 
\citep{Buote2003a, Buote2003b} out to $0.2~r_{180}$.
In the 0.2--0.3~$r_{180}$, the temperature profiles of our 
results agreed well with \citet{Buote2004}.
However, our Fe abundances were higher than the XMM 
abundances by 0.1 solar than that derived with 
 the XMM observations \citep{Buote2004}  ($\sim 0.13$ solar converted 
to the solar abundance table of \citealt{Lodders2003}).
This difference in the Fe abundances may be caused by
the difference in the observed azimuthal directions of the 
Suzaku (north) and XMM (south).
As discussed in \citet{Komiyama2009}, the difference in  the treatment of
the Galactic components may also cause a discrepancy.
At the innermost region, $r < 2'$, of the NGC~5044 group, the temperatures 
derived from the 1T model and cooler component derived from the 
2T model were consistent with 
the XMM results \citep{Buote2003a} within the statistical uncertainties,
although there were small ($\sim 0.3$ keV) discrepancies for the hotter
component of the 2T model fitting.
Although our abundances derived from the 1T and 2T models of this region
were smaller than the XMM results \citep{Buote2003b} by several tens of percent, 
the abundance ratios were consistent each other.
When we used the ATOMDB version 1.10, 
which is used in \citet{Buote2003a, Buote2003b}, 
the discrepancy in the innermost region became smaller.
%

The central region of MKW~4 (out to 0.2~$r_{180}$) was observed 
with XMM \citep{OSullivan2003}.
The radial temperature profiles outside 0.2~$r_{180}$ in this work  
smoothly continued the XMM results within 0.2~$r_{180}$.
The abundance ratios with XMM were also 
consistent with our work beyond 0.2~$r_{180}$.

\subsection{Uncertainties for the spectral fits}
\label{sec:sys}

We estimated the systematic errors in our analysis by changing 
the normalizations of the CXB and the Galactic components by 10\%,
and the NXB levels by 10\% in the spectral fits.  As a result, the 
systematic errors from the background estimations were negligible, 
and the resultant temperatures and abundances did not change within 
the statistical errors by changing the background level.

We examined the influence of the uncertainties from the contaminant 
on the OBF of Suzaku XIS by changing the C/O ratio. We fitted the 
spectra by changing the C/O ratio to be 12.0 and 3.0, which 
corresponded to twice and a half of the default composition number 
ratio, respectively. Consequently, all the parameters except for the 
O abundance did not changed within the statistical errors.  Because 
C and O absorptions were assumed as the contaminant, O abundance 
would suffer from the changing ratio.  Therefore, we note that the
O abundance is not reliably determined because of the contaminant 
uncertainty in our analysis.

The effect of the PSF and stray light of 
Suzaku's X-ray telescope would make photon contaminations from a nearby 
sky (see also \citealt{Sato2007, Urban2013}). In order to estimate 
the effects, we examined simulations of the contamination flux base on 
a ray tracing simulator, "xissim" as shown in \citet{Sato2007}. For 
example, in the case of the NGC~5044, we assumed 
$\sim 2^{\circ} \times 2^{\circ}$ size of the $\beta$--model 
surface brightness profile derived from the XMM observations by 
\citet{Nagino2009} with a 300 ksec exposure time.  As a result, the 
photon fractions originated from each extracted annulus in the outside 
of 6$'$ $\sim 0.1~r_{180}$ were 80\% and over.  On the other hand, although the 
spectra within 6$'$ suffer from the contaminations by $\sim$ 30\%, the resultant 
parameters in the central regions agreed with the previous XMM results 
within the statistical errors, and they came mostly from the adjacent 
regions.  As for the stray light, we examined the spectral fit in the 
outermost region including the estimated contamination flux from the 
bright core in the model. The resultant fit parameters did not change 
within the statistical errors.  Consequently, we proceeded to the 
spectral analysis without making photon corrections due to the 
contaminations for the PSF and stray light.

\subsection{$K$-band luminosity of galaxies}
\label{sec:K-band}

\begin{deluxetable}{llllll}
\tabletypesize{\scriptsize}
\tablewidth{0pt}
\tablecaption{
Summary of basic properties for calculating the total $K$-band luminosity within $r_{180}$ for each group.
\label{tb:lk}
}
\tablehead{
\colhead{group} & \colhead{$D_{L}^{a}$} & \colhead{$A_{K}^{b}$} & \colhead{$m_{K}^{c}$} & \colhead{$L_{K,BGD}^{d}$} & \colhead{$L_{K, r_{180}}$} \\
\colhead{} & \colhead{Mpc}&\colhead{} & \colhead{}& \colhead{$\times 10^{7}~L_{\odot,K}/{\rm arcmin}^{2}$} & \colhead{ $\times 10^{12}~L_{\odot,K}$ } 
}
\startdata
MKW~4 & 87.0 & 0.008 & 7.1 & 1.72 & 1.91 \\

HCG~62 & 64.3 & 0.016 & 4.7 & 2.90 & 2.13\\
NGC~1550 & 53.6 & 0.050 & 2.8 & 2.78 & 0.73\\
NGC~5044 & 40.0 & 0.026 & 2.2 & 0.93 & 0.82\\
\enddata
\tablenotetext{a}{The luminosity distance for each galaxy group from NED.}
\tablenotetext{b}{The Galactic extinction value from NED\citep{Schlegel1998}.}
\tablenotetext{c}{The apparent magnitude of the cD galaxy for MKW~4,
the NGC~1550 group, and the NGC~5044 group, and the most luminous galaxy for HCG~62.}
\tablenotetext{d}{The luminosity in the background region between 1 and
 2 $r_{180}$. For details, see in text.}
\end{deluxetable}

\begin{figure*}[!th]
 \begin{center}
     \includegraphics[width=0.34\textwidth,angle=0,clip]{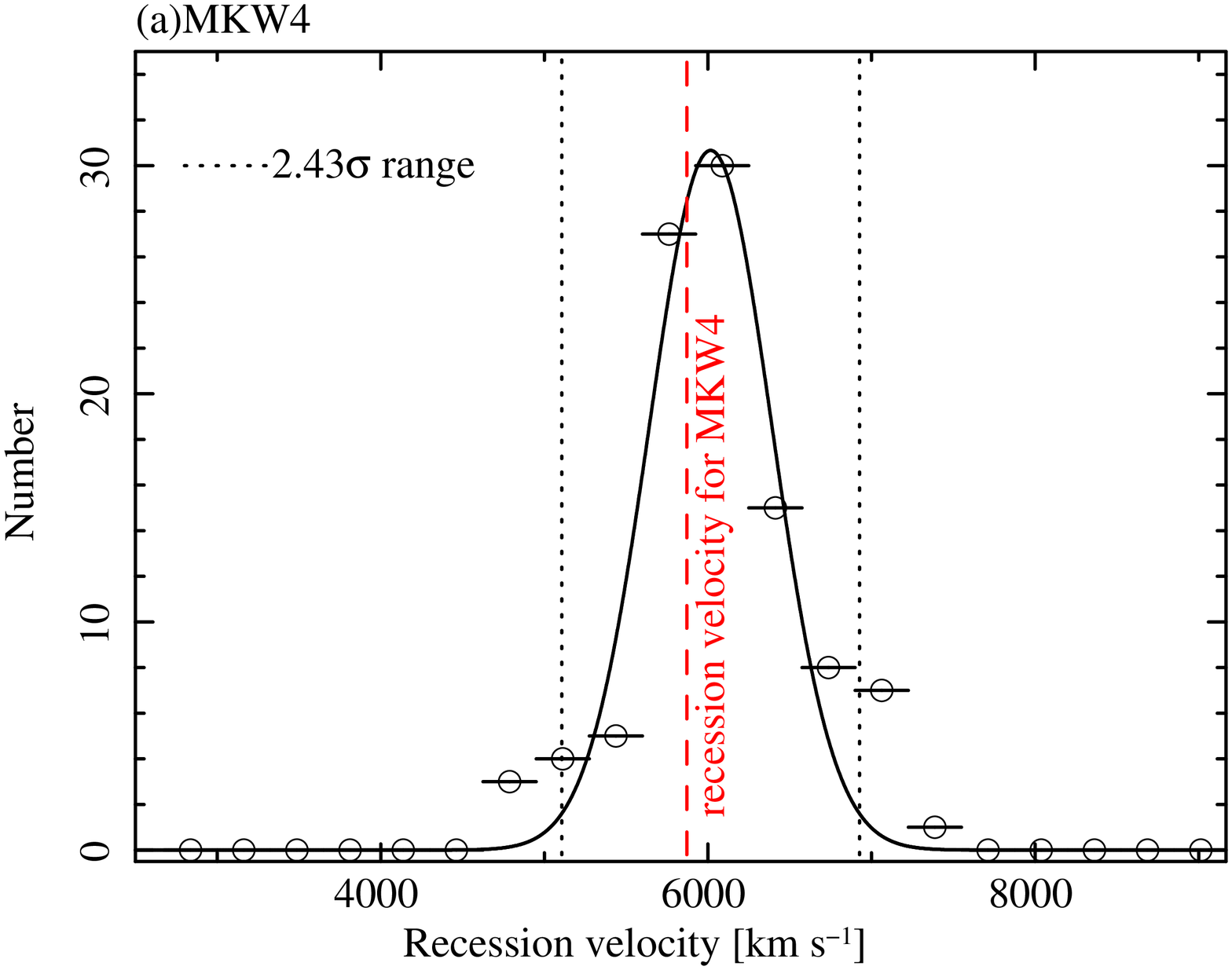}
     \includegraphics[width=0.34\textwidth,angle=0,clip]{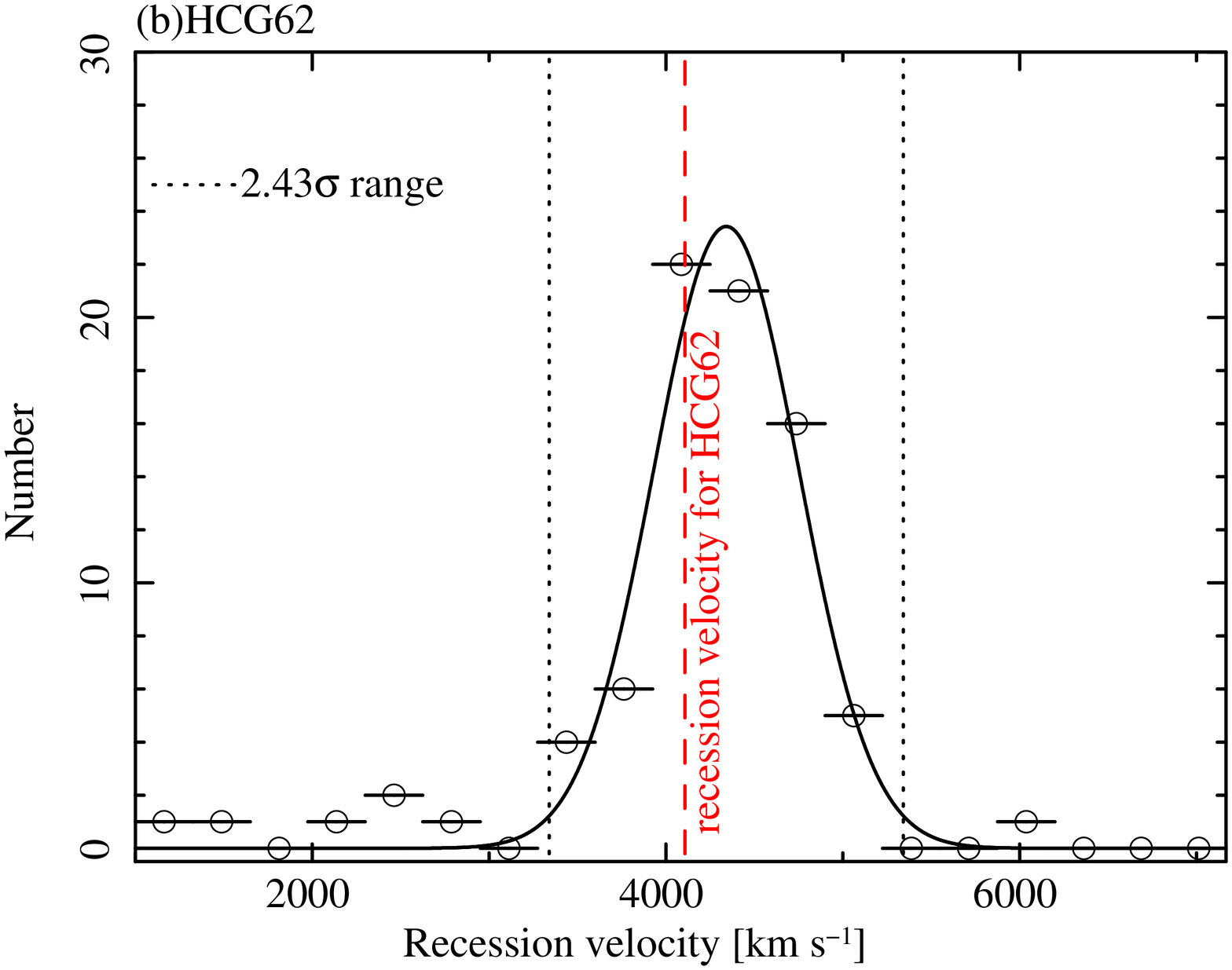}
     \includegraphics[width=0.34\textwidth,angle=0,clip]{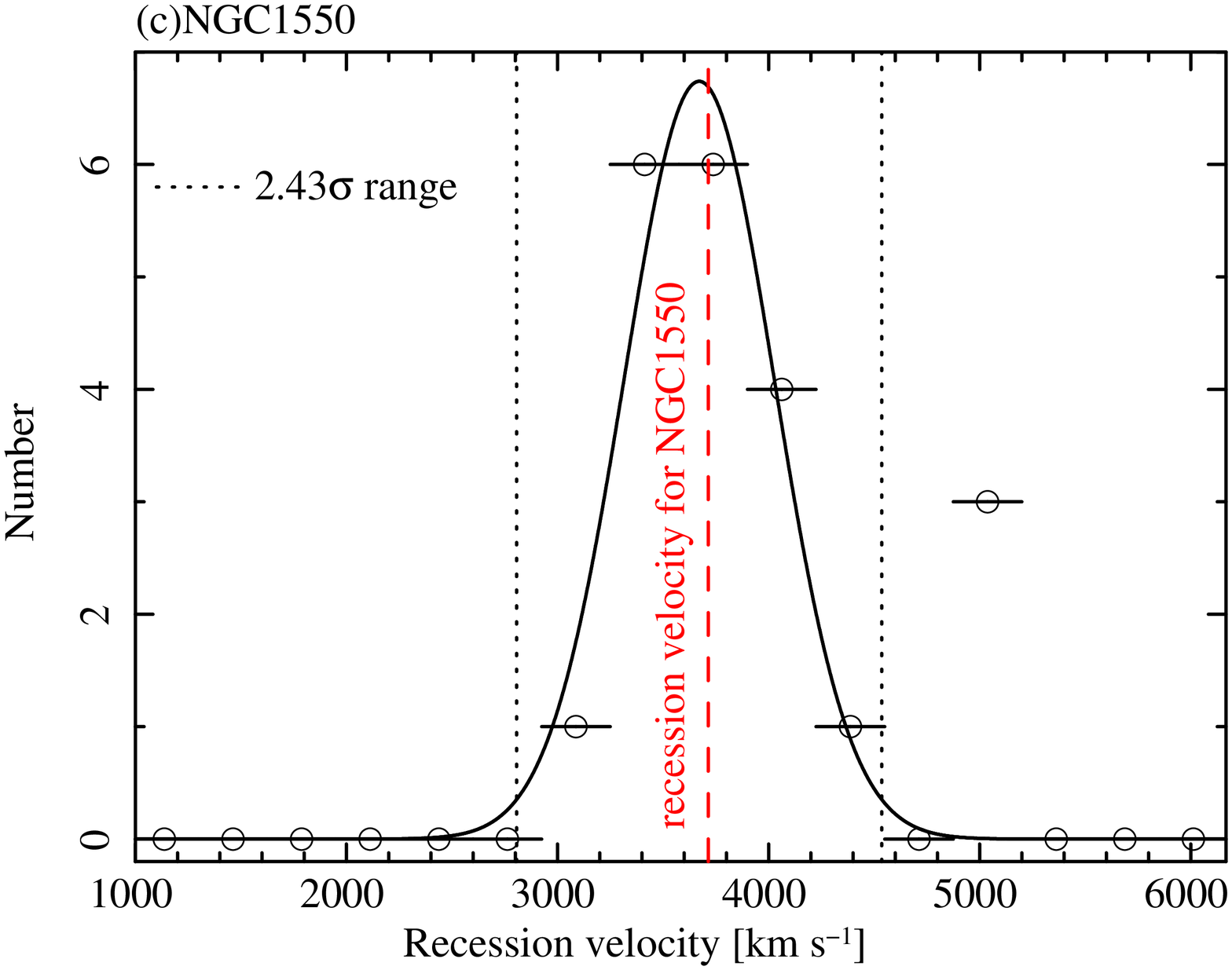}
     \includegraphics[width=0.34\textwidth,angle=0,clip]{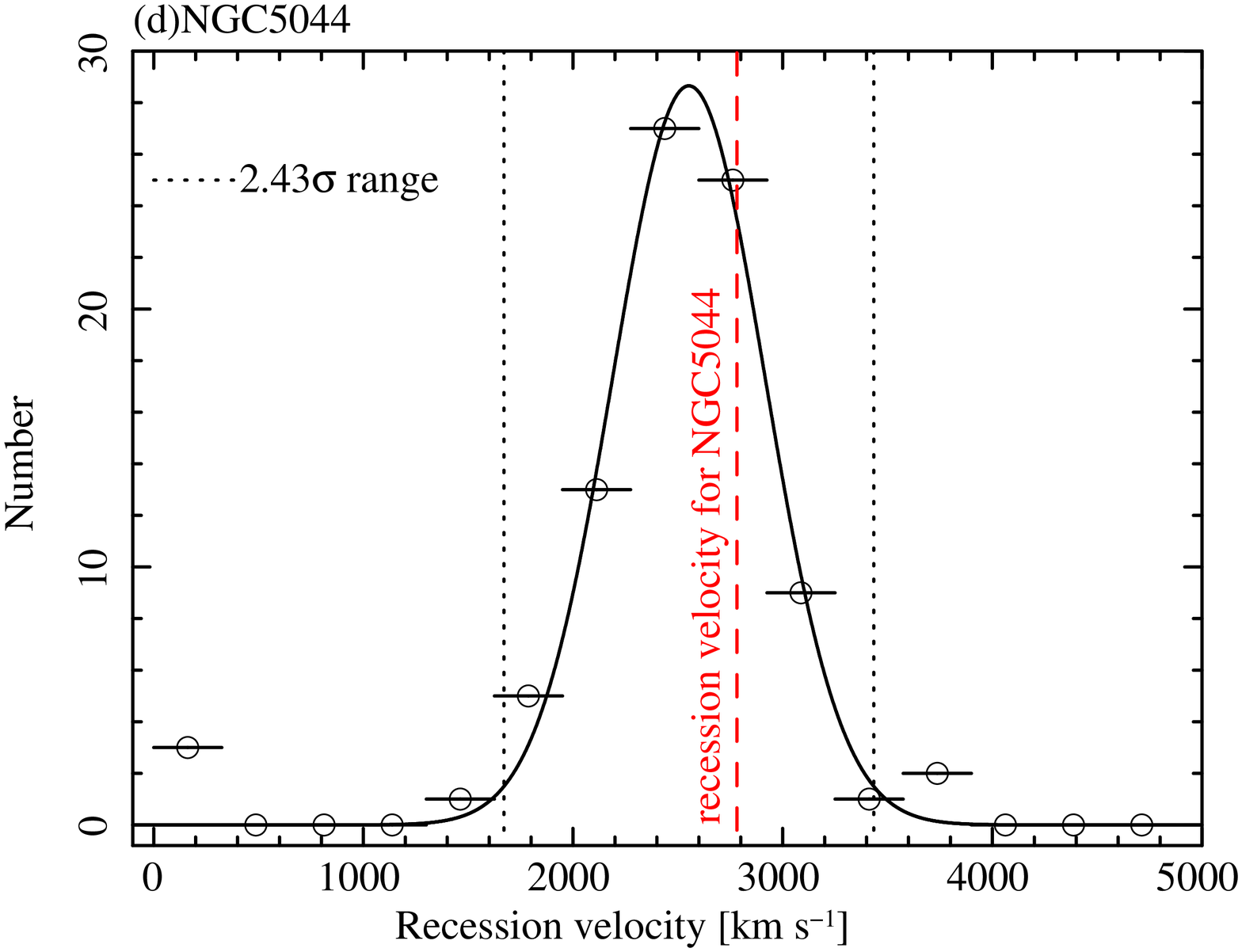}
     \end{center}
\caption{
Recession velocity distributions of the galaxies which were 
identified by the redshift in the catalogues for each group.
Black dotted lines indicate $2.43~\sigma$ around the means which 
are close to the recession velocity for each group from NED as 
shown by red dashed lines.
}
\label{fig:dispersion}
\end{figure*}

\begin{figure*}[!th]
 \begin{center}
     \includegraphics[width=0.34\textwidth,angle=0,clip]{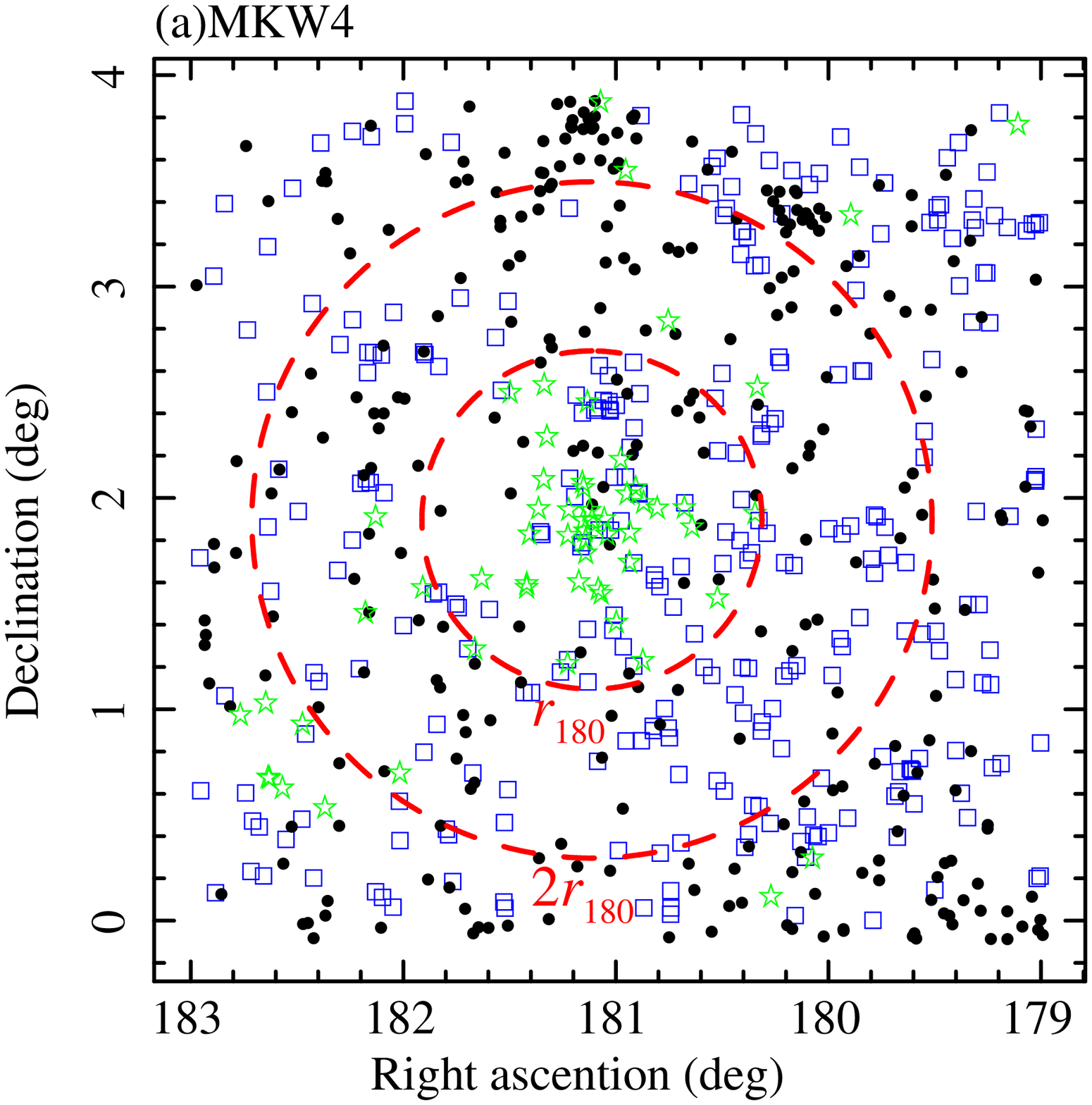}
     \includegraphics[width=0.34\textwidth,angle=0,clip]{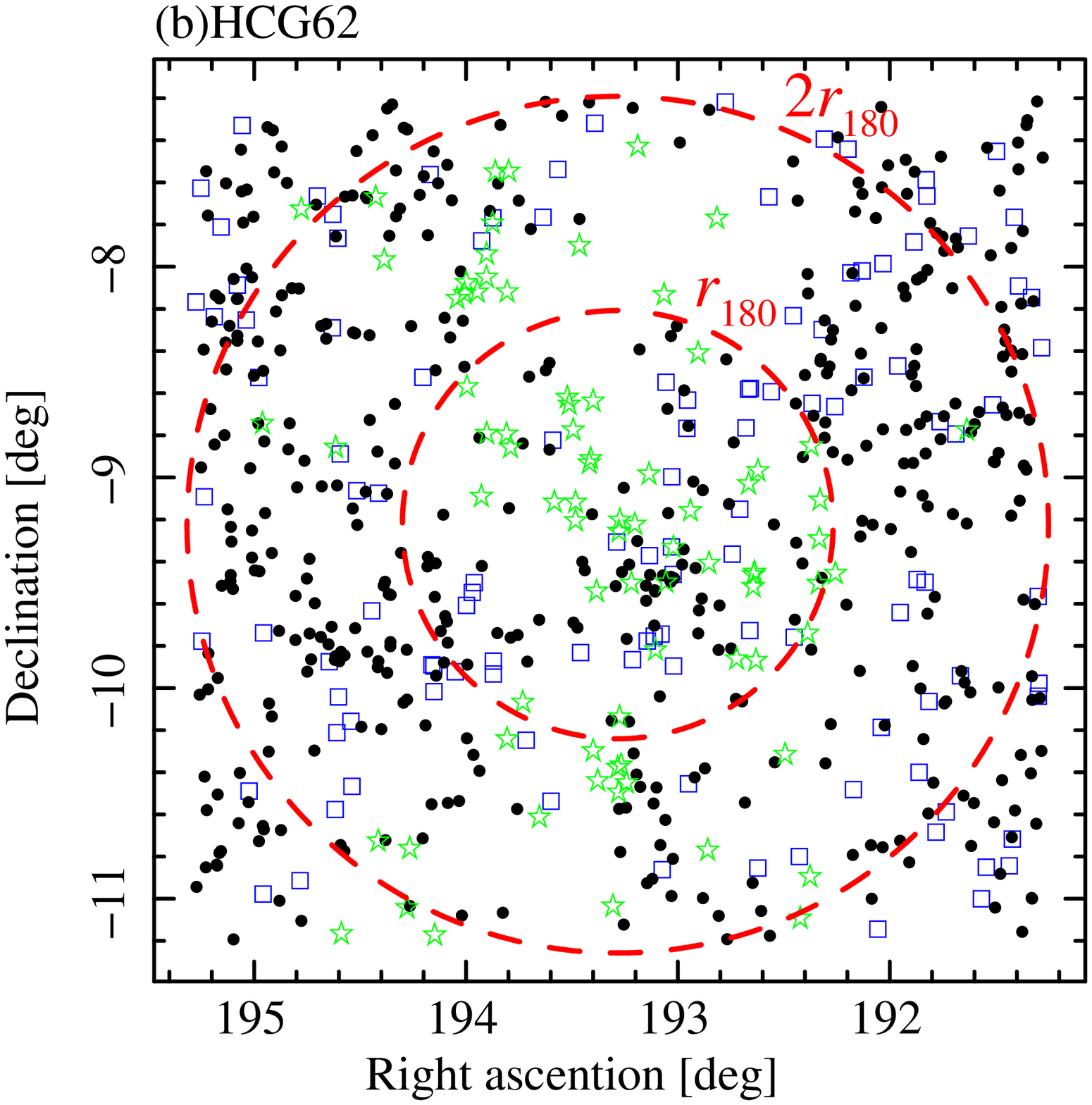}
     \includegraphics[width=0.34\textwidth,angle=0,clip]{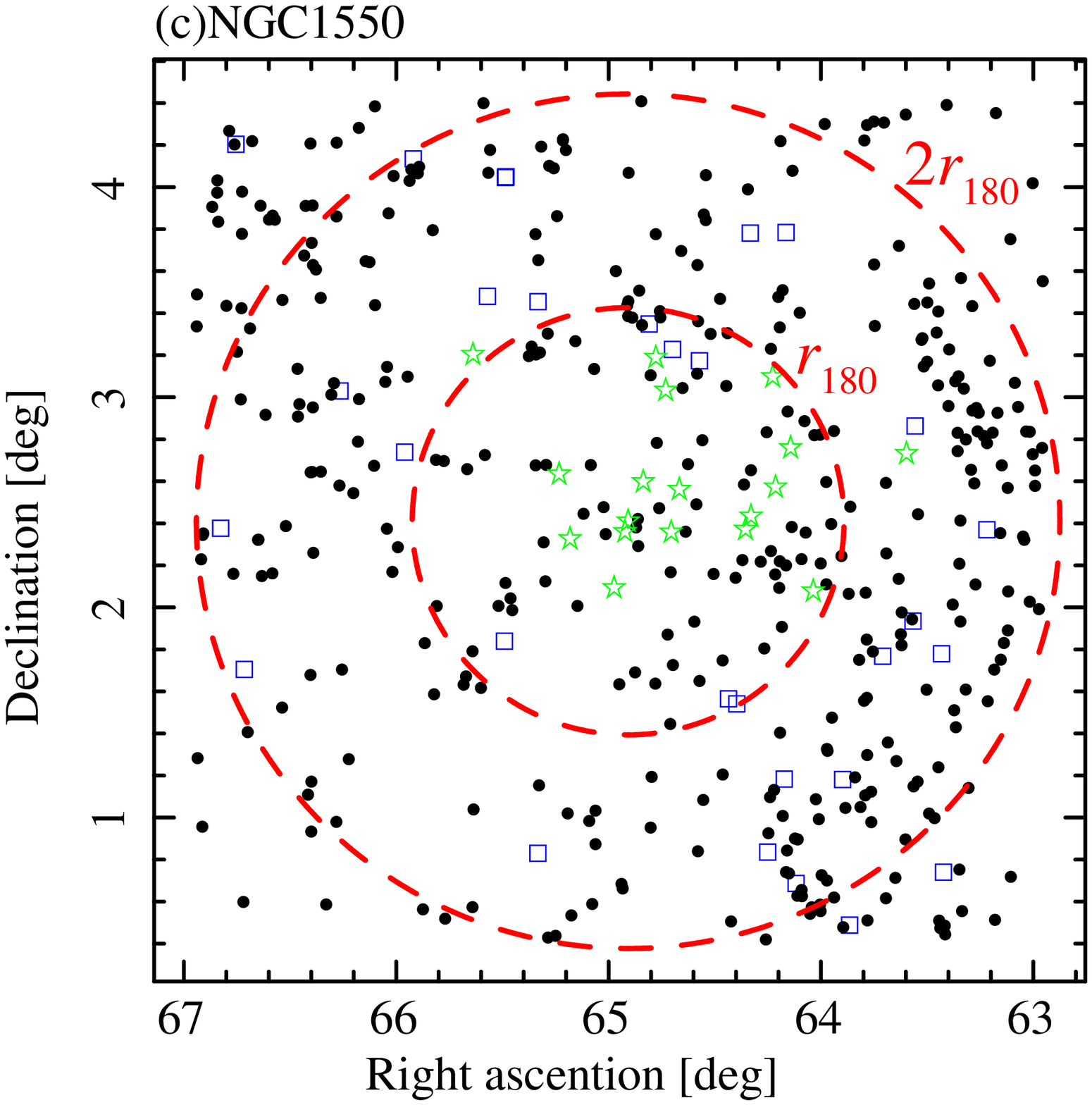}
     \includegraphics[width=0.34\textwidth,angle=0,clip]{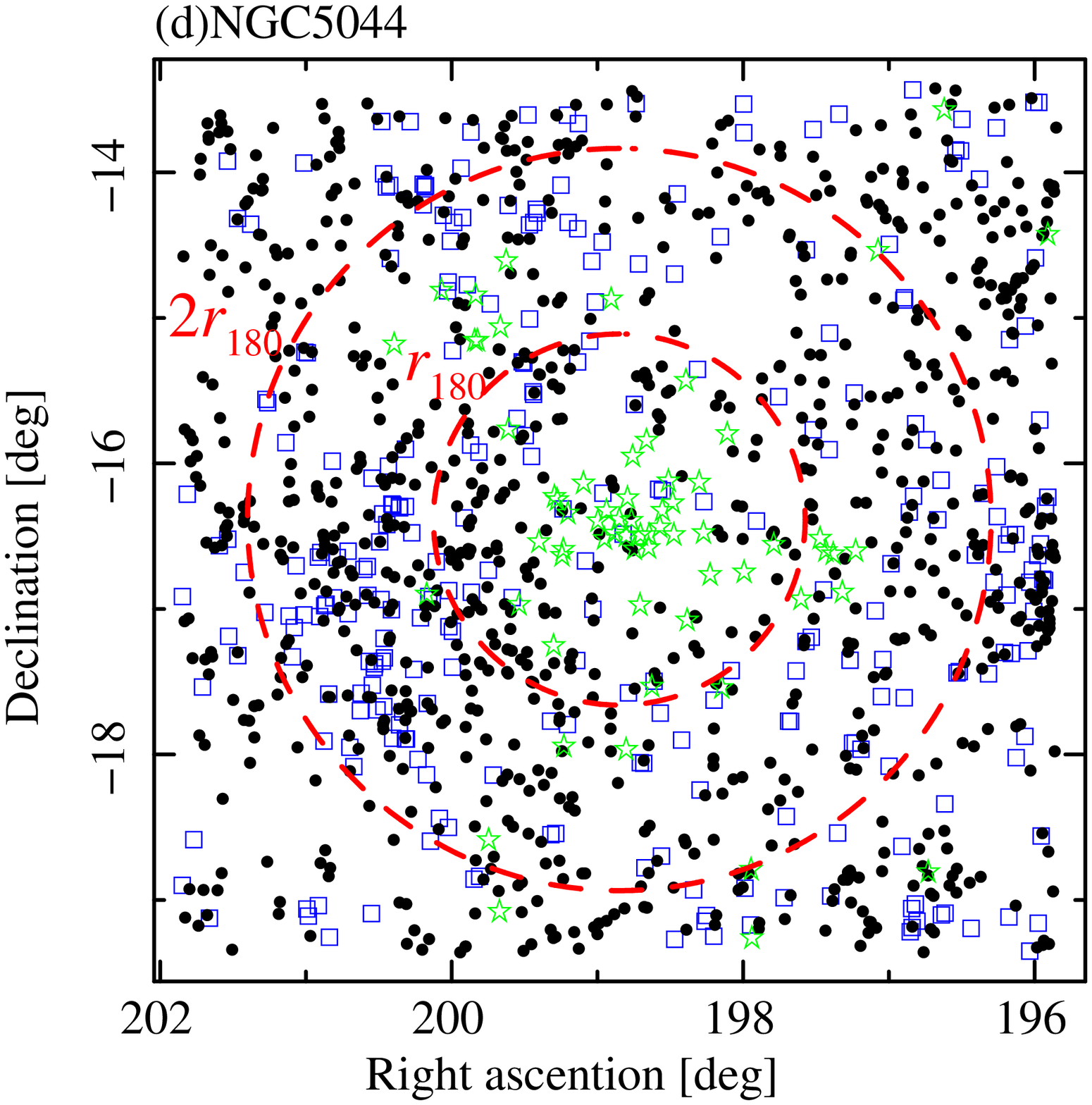}
     \end{center}
\caption
{
Galaxy distributions for MKW~4, HCG~62, the NGC~1550 group, and the 
NGC~5044 group in $K$-band from the 2MASS catalogue.  Red dashed lines 
indicate the projected radii of the $r_{180}$ and $2~r_{180}$. 
Green stars and blue boxes indicate member and non-member galaxies, 
respectively, which are identified by the redshift and recession 
velocity. Black filled circles indicate the unidentified galaxies 
by the redshift. 
(Color version of these figures are available in the online journal.)
} 
\label{fig:distribution}
\end{figure*}

\begin{figure*}[!th]
 \begin{center}
     \includegraphics[width=0.37\textwidth,angle=0,clip]{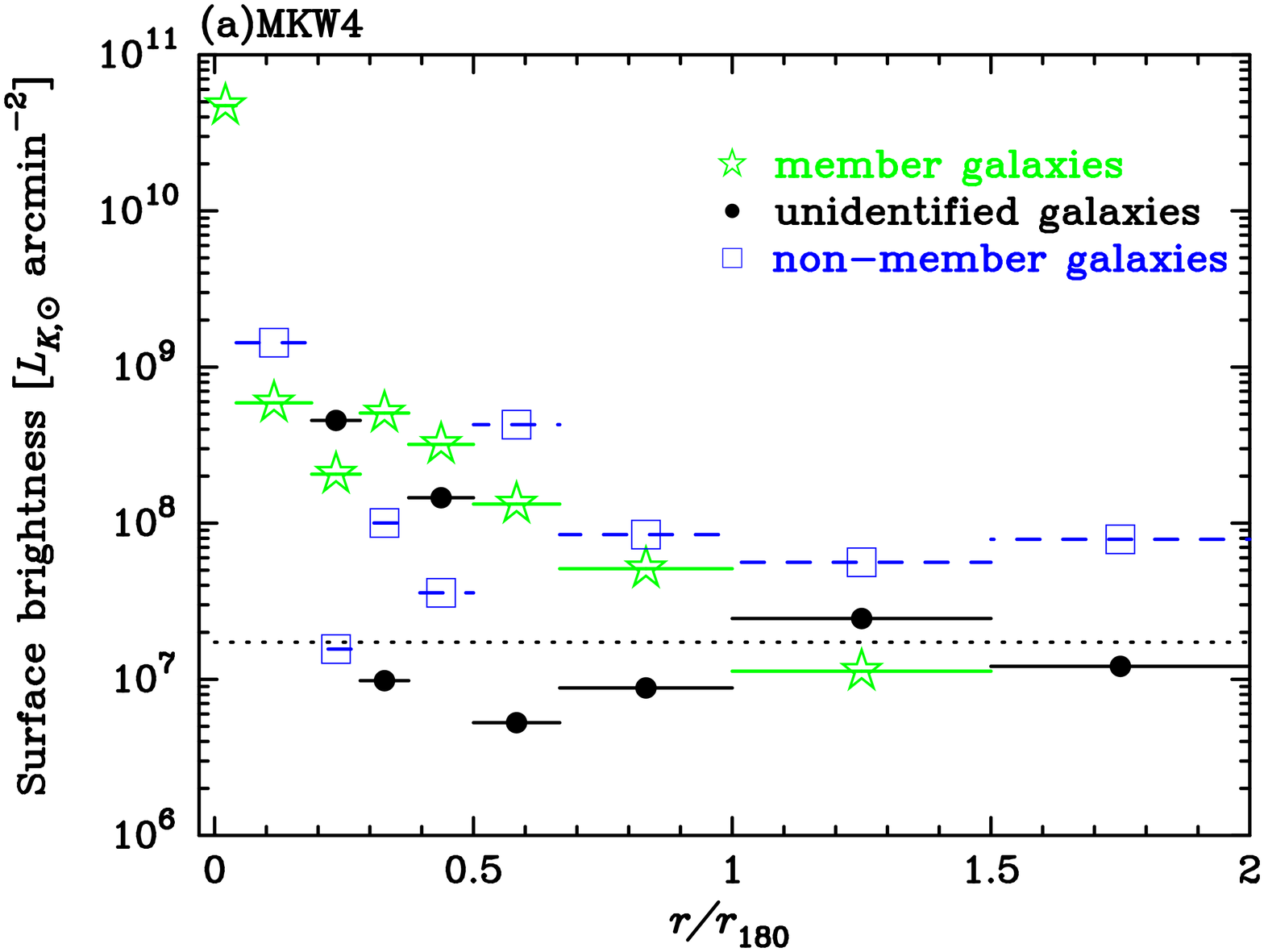}
     \includegraphics[width=0.37\textwidth,angle=0,clip]{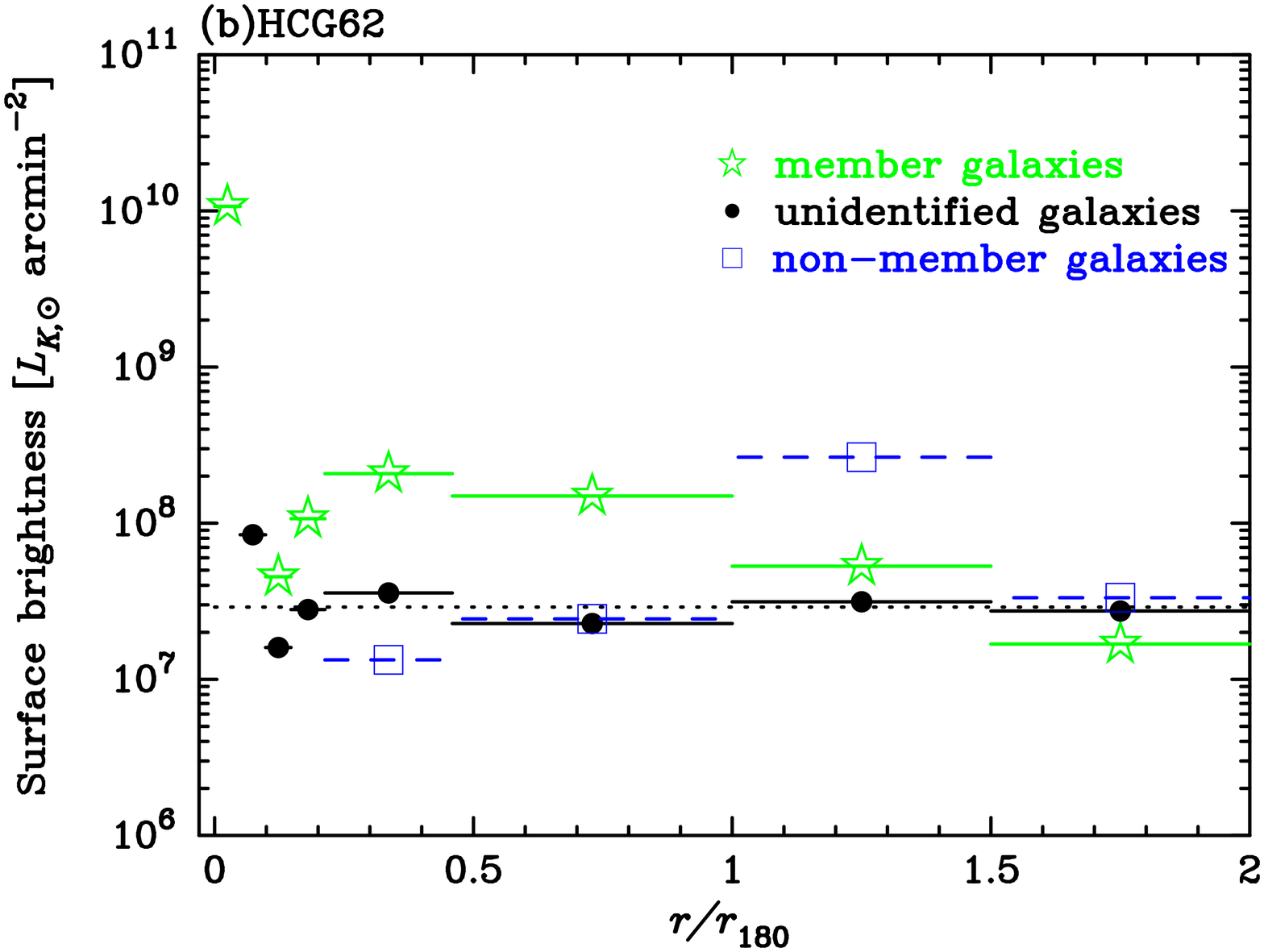}
     \includegraphics[width=0.37\textwidth,angle=0,clip]{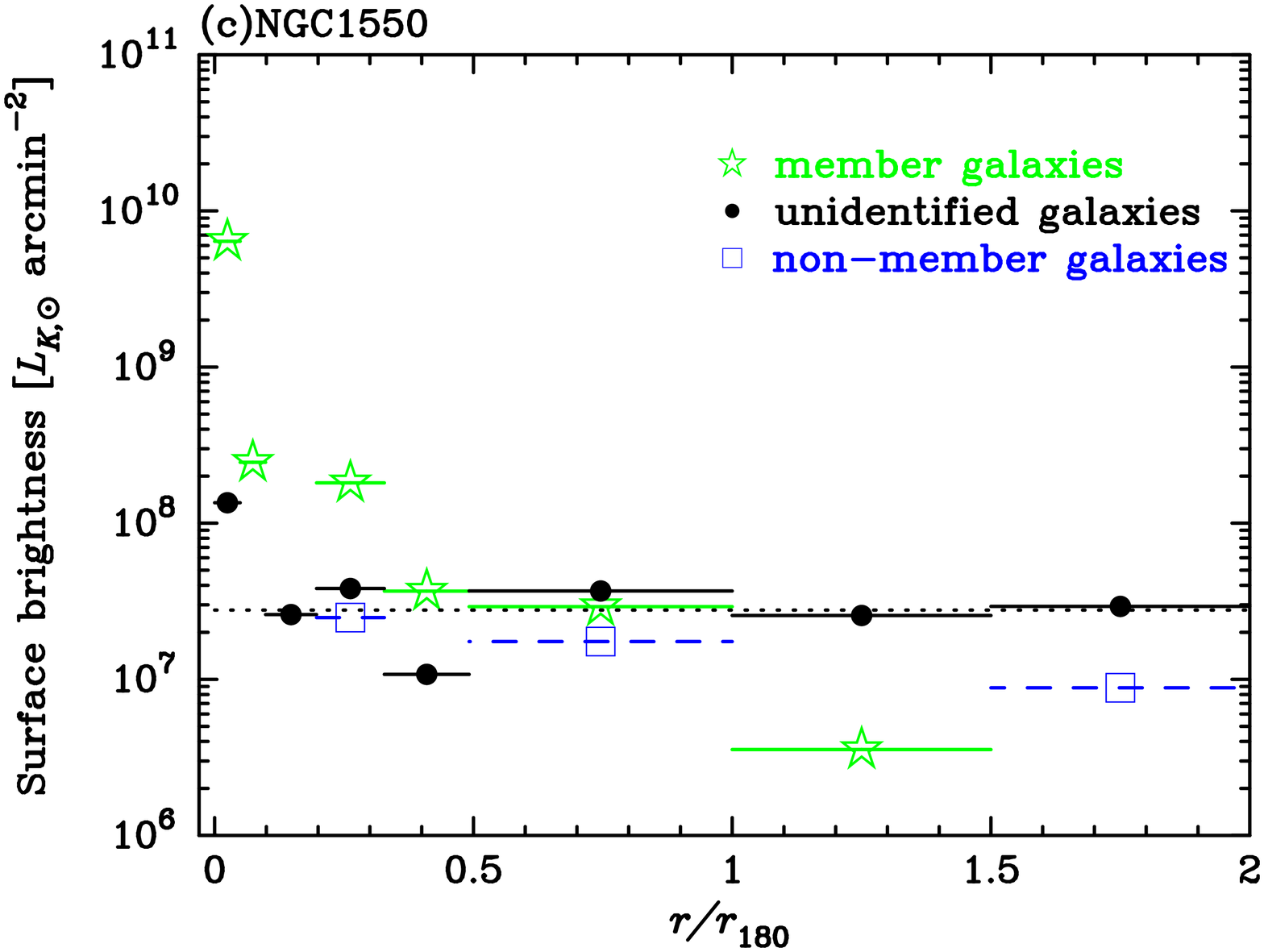}
     \includegraphics[width=0.37\textwidth,angle=0,clip]{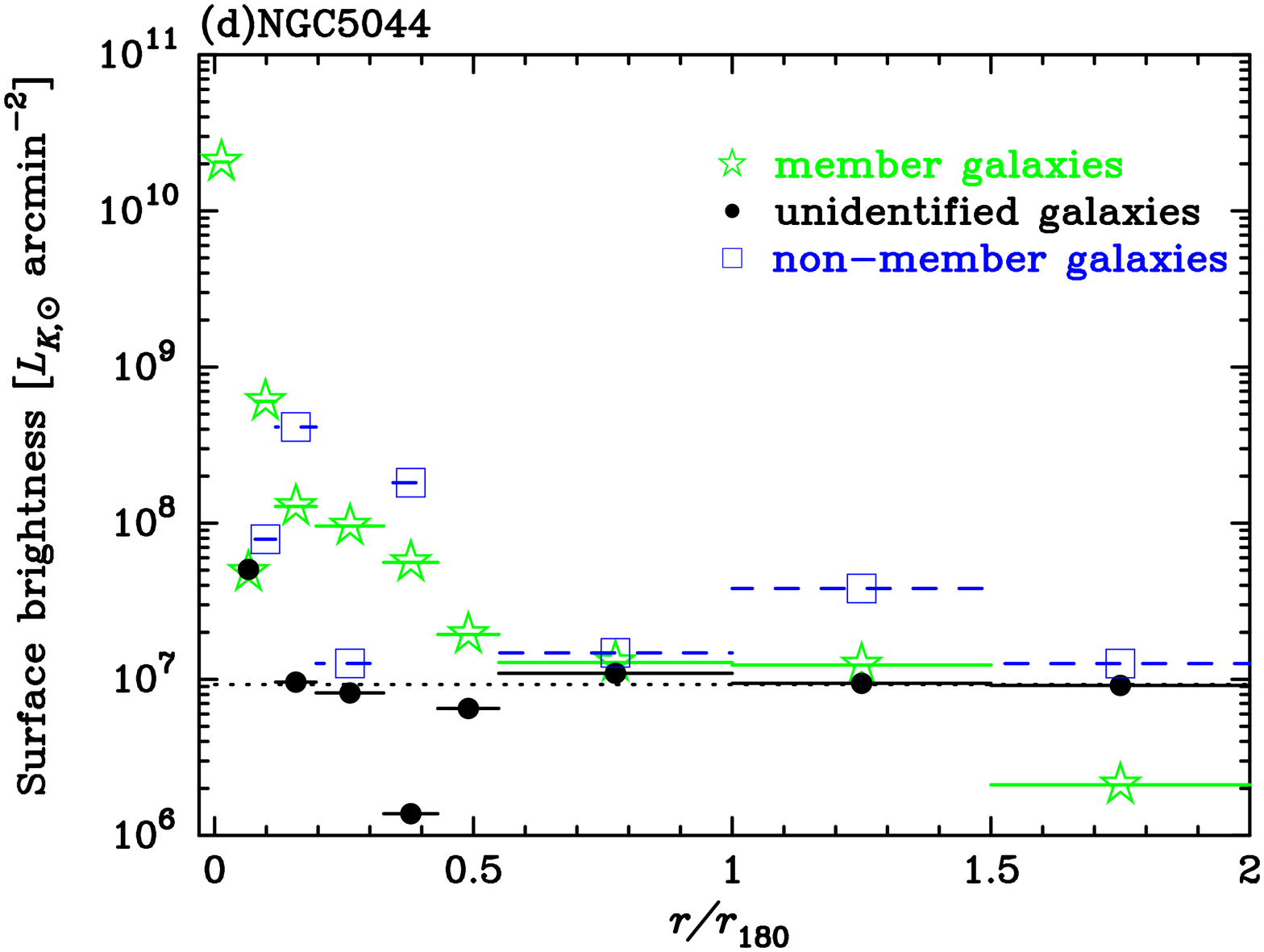}          
 \end{center}
\caption{
Radial profiles of the surface brightness derived from the $K$-band
luminosity of the member galaxies for each group. Green stars, black 
filled circles, and blue squares correspond to the surface brightness 
of the member galaxies identified by the redshift in the catalogues, 
the member galaxies unidentified by the redshift, and non-member 
galaxies, respectively. Dashed lines indicate the background level 
derived from the luminosity between the 1 and 2 $r_{180}$ region.
For details, see in text.
}
\label{fig:surface}
\end{figure*}

We investigated radial profiles of the IMLR, which is a useful 
parameter for a comparison of the ICM metal profiles with the galaxy 
mass or luminosity distribution, because most metals in the ICM 
were synthesized in the member galaxies. In order to calculate the 
IMLR profiles, we derived $K$-band luminosity profiles which 
well-trace the stellar (galaxy) mass distribution \citep{Nagino2009}, 
from the Two Micron All Sky Survey (2MASS) catalogue\footnote{http://www.ipac.caltech.edu/2mass/}.
In the catalogue, we used all of the data in 
$4^{\circ}\times4^{\circ}$ region for MKW~4, HCG~62, and the NGC~1550 
group, and $6^{\circ}\times6^{\circ}$ region for the NGC~5044 group, 
centered on the cD galaxy for each group.  We summarize the parameters 
such as each luminosity distance, the foreground Galactic extinction 
to calculate the luminosity for each member galaxy, and the appnt 
magnitude of central galaxy in $K$-band as shown in table \ref{tb:lk}.

As for an identification of the redshift for each galaxy in the 2MASS 
catalogue, we assumed an available redshift of each member galaxy as 
the one by which the galaxy was identified in the Sloan Digital Sky 
Survey catalogue\footnote{http://www.sdss.org/} for MKW~4, 
in the 6dF Galaxy Survey 
database\footnote{http://www.aao.gov.au/6dFGS/} for HCG~62 and 
the NGC~5044 group, and in the 2MASS redshift survey 
\citep{Crook2007, Crook2008} for the NGC~1550 group.  As for the 
galaxies which were identified by the redshift or recession velocity 
within the projected radius of 2~$r_{180}$, we calculated the 
recession velocity distribution with 325 km sec$^{-1}$ intervals 
as shown in figure \ref{fig:dispersion}, and then obtained the 
resultant member galaxies which were within $\pm2.43~\sigma$ around 
the mean. The spacial distributions of the galaxies inside or 
outside $2.43~\sigma$ are shown in figure \ref{fig:distribution}, 
and the radial profile of the surface brightness are shown in 
figure \ref{fig:surface}.

For the galaxies in the 2MASS catalogue which were not identified 
by the redshift, such as dwarf galaxies, we examined the background 
subtraction as follows.  We derived the surface brightness profile 
of the galaxies between the projected radius of $r_{180}$ and 
2~$r_{180}$\@.  And then, we subtracted the surface brightness level 
between the $r_{180}$ and 2~$r_{180}$ region as the background from 
the surface brightness within $r_{180}$ region.  The radial profile 
of the surface brightness up to $r_{180}$ and the background level 
derived from the 1--2 $r_{180}$ region are shown in figure 
\ref{fig:surface}.

\begin{figure*}[!ht]
 \begin{center}
     \includegraphics[width=0.47\textwidth,angle=0,clip]{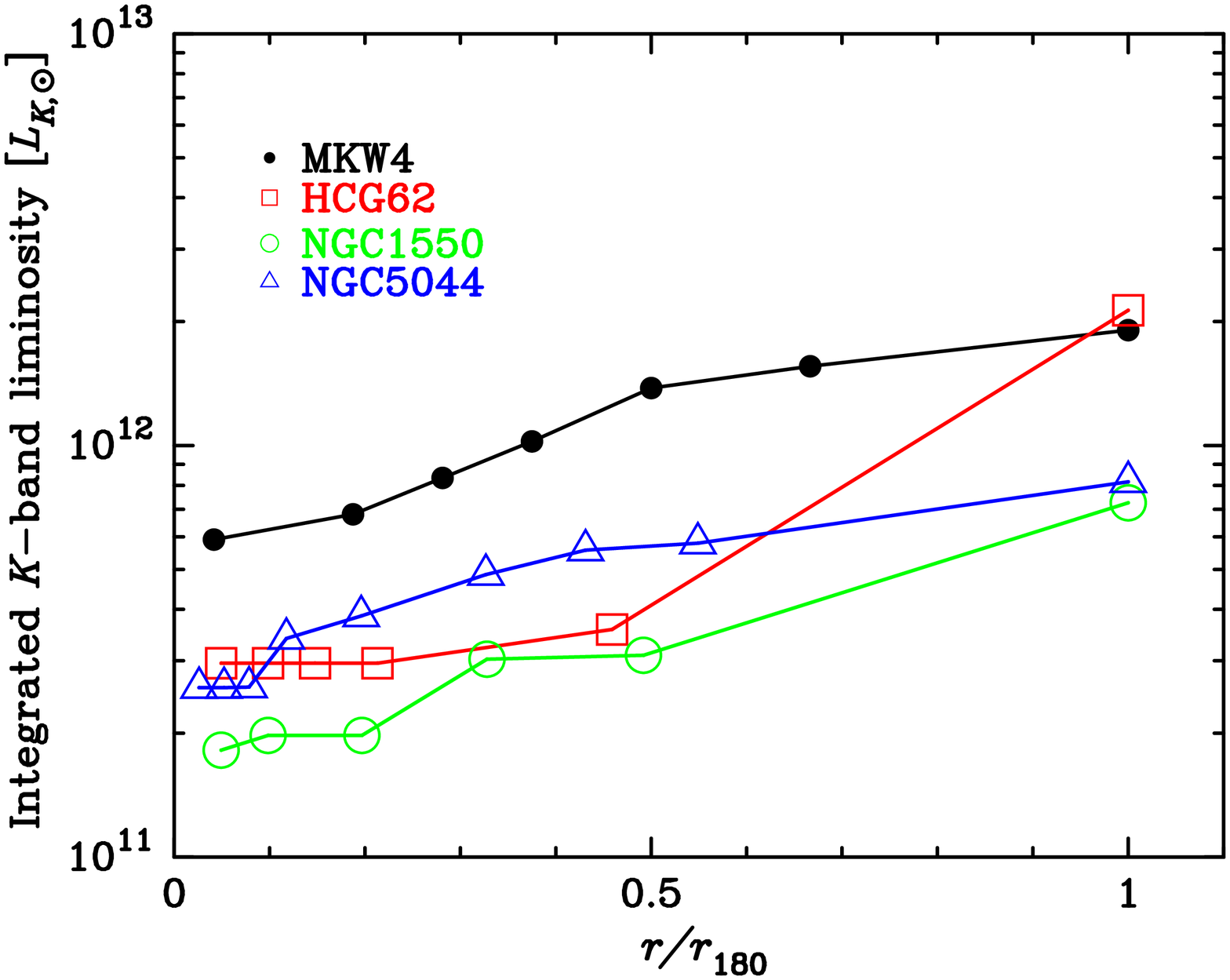}
 \end{center}
\caption{
Radial profiles of the integrated $K$-band luminosity (solid lines) 
for MKW~4, HCG~62, the NGC~1550 group, and the NGC~5044 group.  
The notations of colors and marks are the same as in  
figure \ref{radial_chi}.
}
\label{fig:lk}
\end{figure*}

As a result, we regarded the sum of the luminosity of the member 
galaxies which were selected by the redshift and recession velocity, 
and the galaxies which were not identified by the redshift but for 
which we performed the background subtraction, as the total luminosity 
for each group.  And then, we calculated radial profiles of the 
integrated $K$-band luminosity.  We next deprojected the luminosity 
profiles as a function of the radius assuming spherical symmetry, and 
then derived three-dimensional radial profiles of the integrated 
$K$-band luminosity up to $r_{180}$ as shown in figure \ref{fig:lk} 
and table \ref{tb:lk}.  Here, even if we took into account the 
luminosity of dwarf galaxies which were not detected in the 2MASS 
catalogue, assuming the luminosity function integrated over 
the magnitude from the cD galaxy to the fully faint-end galaxy 
for each group, the resultant total $K$-band luminosity within 
$r_{180}$ would not change by a factor of 2.


\subsection{Gas-mass and Iron-mass to light ratio}
\label{sec:imlr}

\begin{figure*}[htbp]
\begin{center}
\includegraphics[width=0.37\textwidth,angle=0,clip]{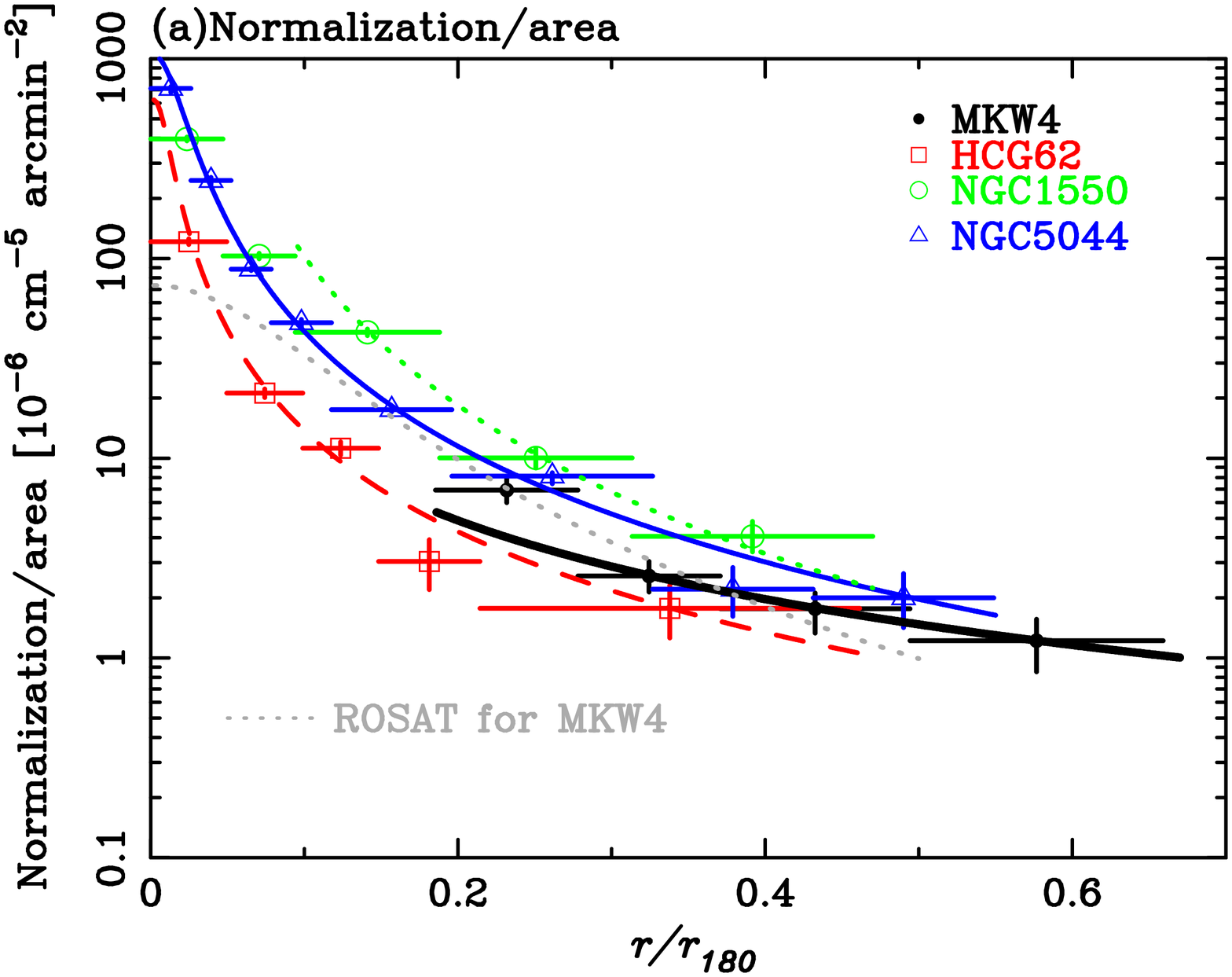}
\includegraphics[width=0.4\textwidth,angle=0,clip]{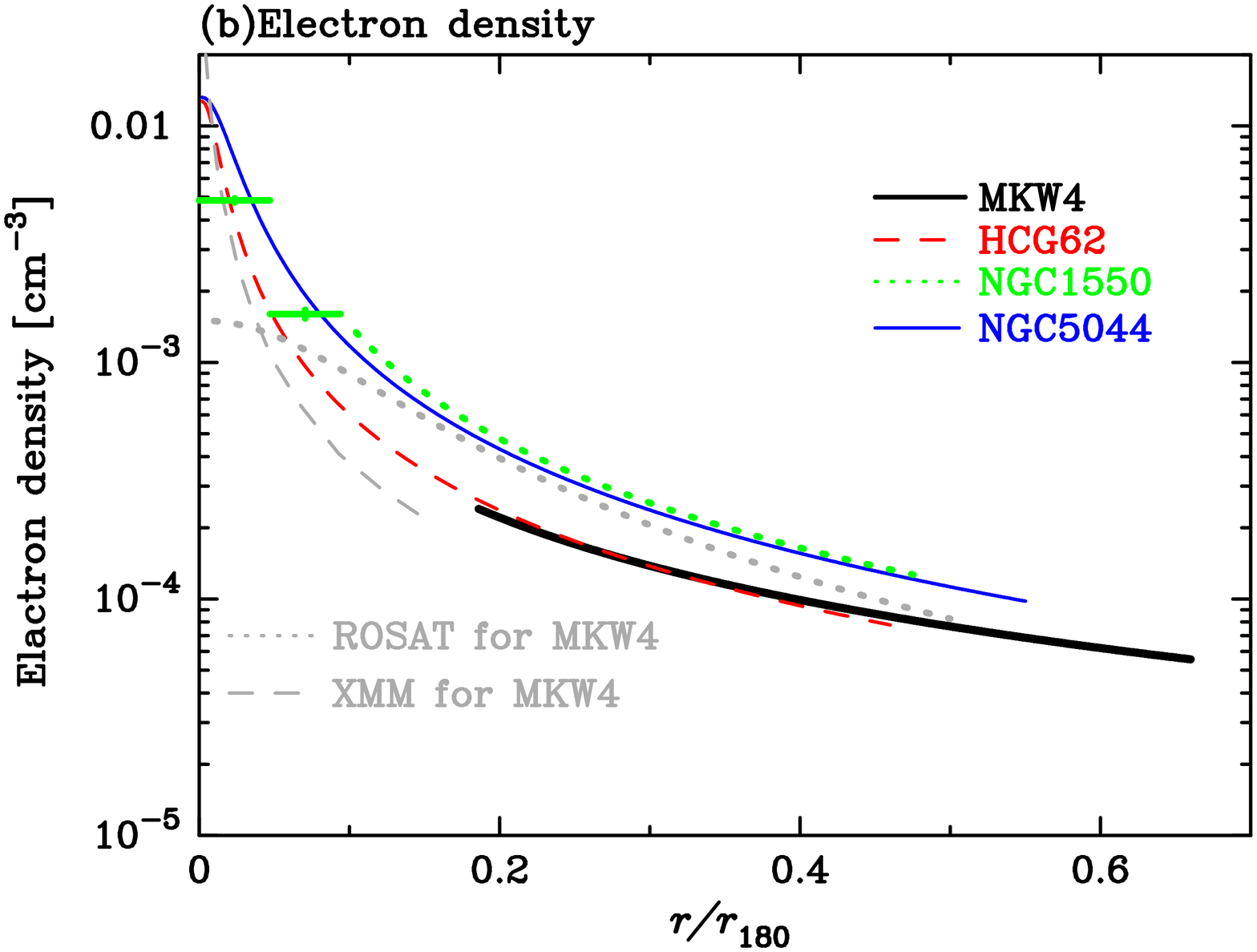}
\caption{
(a) Radial profiles of the normalizations divided by the area 
from the spectral fits with the 1T model.  
Solid black and blue, and dashed red lines correspond 
to the best-fit $\beta$-model for MKW~4, the NGC~5044 group, and 
HCG~62, respectively.
A dotted line is the best fit power-law model for the NGC~1550 group 
from the normalizations per area.
A dotted light gray line shows the surface brightness profile of MKW~4 
derived from the previous ROSAT result \citep{Sanderson2003}, which is
normalized to the brightness level observed with Suzaku by fitting with 
the fixed $\beta$ and $r_c$ to be 0.64 and 5.45 arcmin, respectively.  
The notations of colors and marks are the same as in figure 
\ref{radial_chi}.
(b) Radial profiles of the electron density, $n_{e}$ (cm$^{-3}$), 
derived from the normalizations.  
Crosses indicate the deprojected electron density derived from 
the normalizations per area for the NGC~1550 group.
Gray dotted and dashed lines indicate 
the electron density profiles of MKW~4 derived from ROSAT 
\citep{Sanderson2003} and XMM \citep{OSullivan2003}, respectively. 
For details, see in text.
}
\label{fig:dens}
\end{center}
\end{figure*}

\begin{figure*}[!th]
\begin{center}
\includegraphics[width=0.40\textwidth,angle=0,clip]{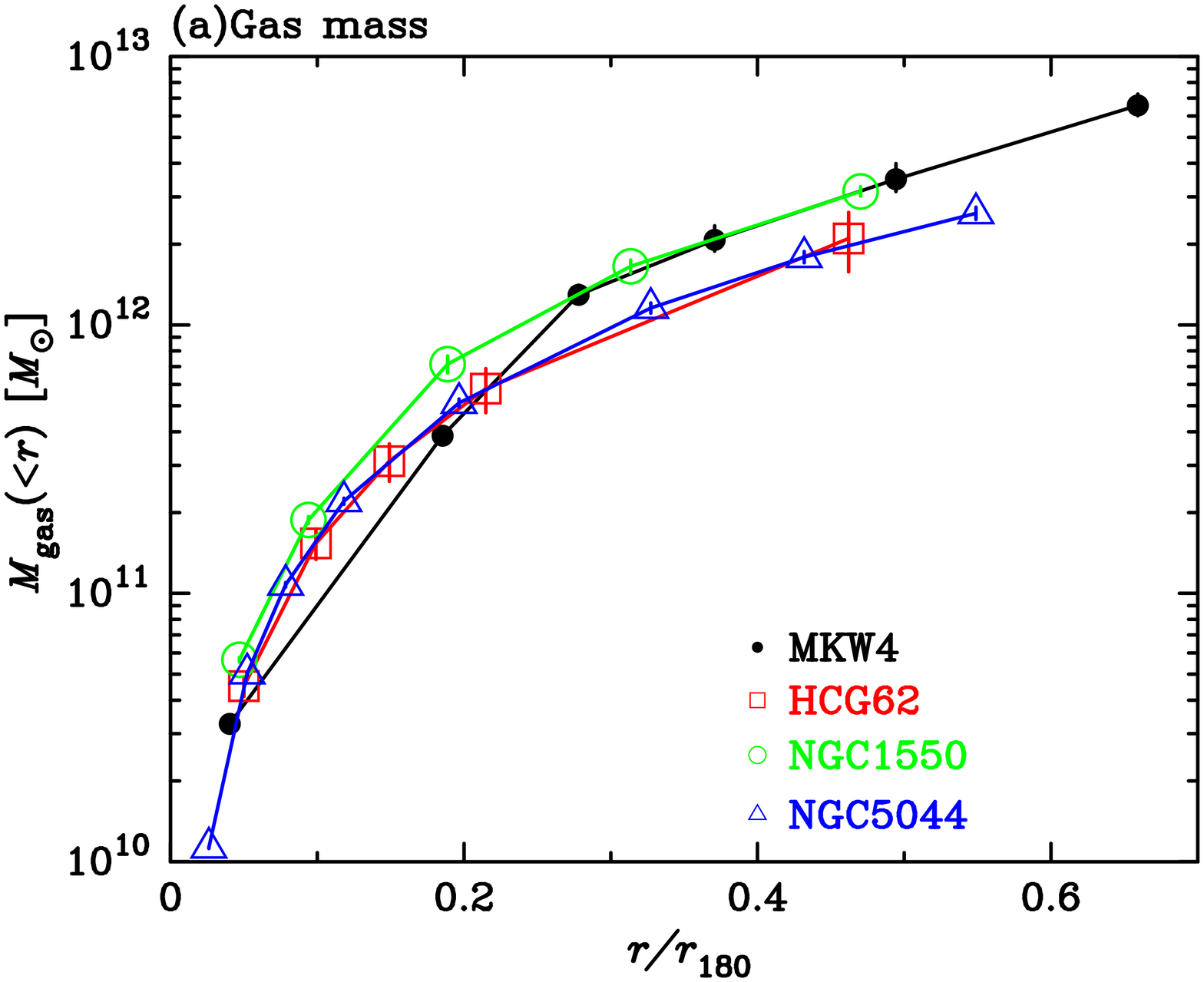}
\includegraphics[width=0.40\textwidth,angle=0,clip]{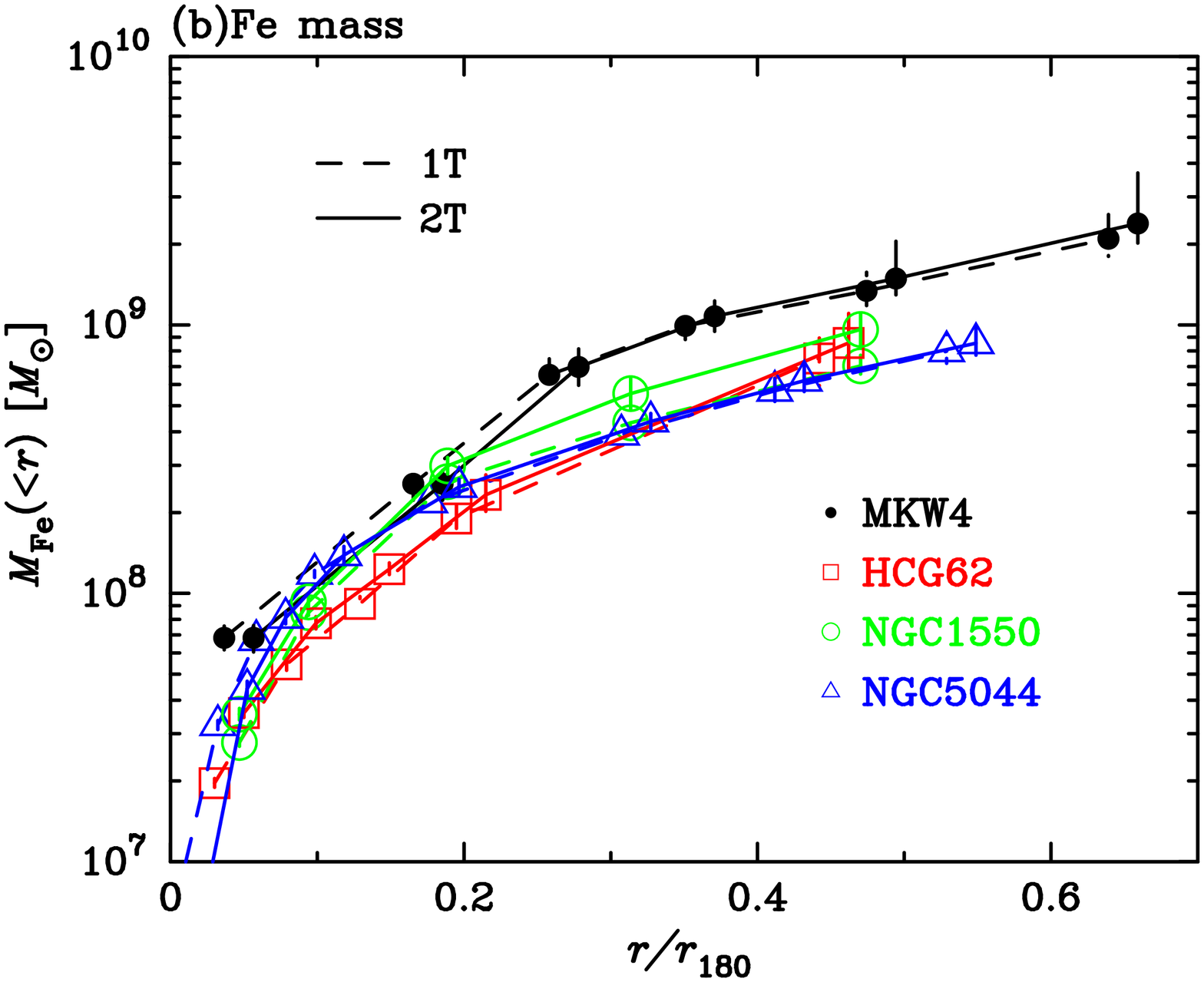}
\end{center}
\caption{
(a) Radial profiles of the integrated gas mass for MKW~4, HCG~62, 
the NGC~1550 group, and the NGC~5044 group.
(b) Radial profiles of the integrated iron mass for MKW~4, HCG~62, 
the NGC~1550 group, and the NGC~5044 group.  Dashed and solid lines 
correspond to the iron mass derived from the Fe abundances with the 
1T and 2T model, respectively.  The notations of colors and marks are 
the same as in figure \ref{radial_chi}.
}
\label{fig:mass}
\end{figure*}

\begin{figure*}[!th]
 \begin{center} 
     \includegraphics[width=0.4\textwidth,angle=0,clip]{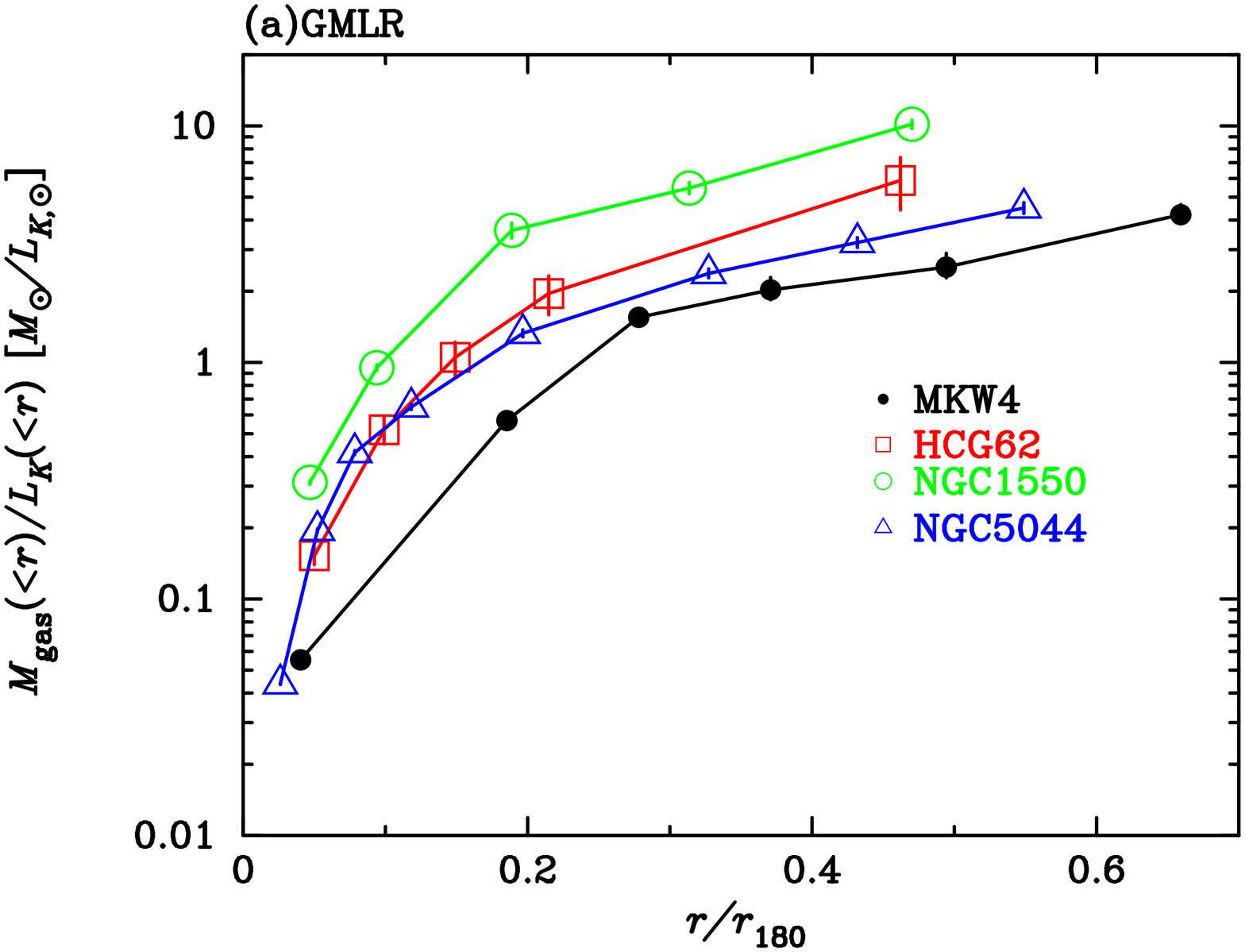}    
     \includegraphics[width=0.4\textwidth,angle=0,clip]{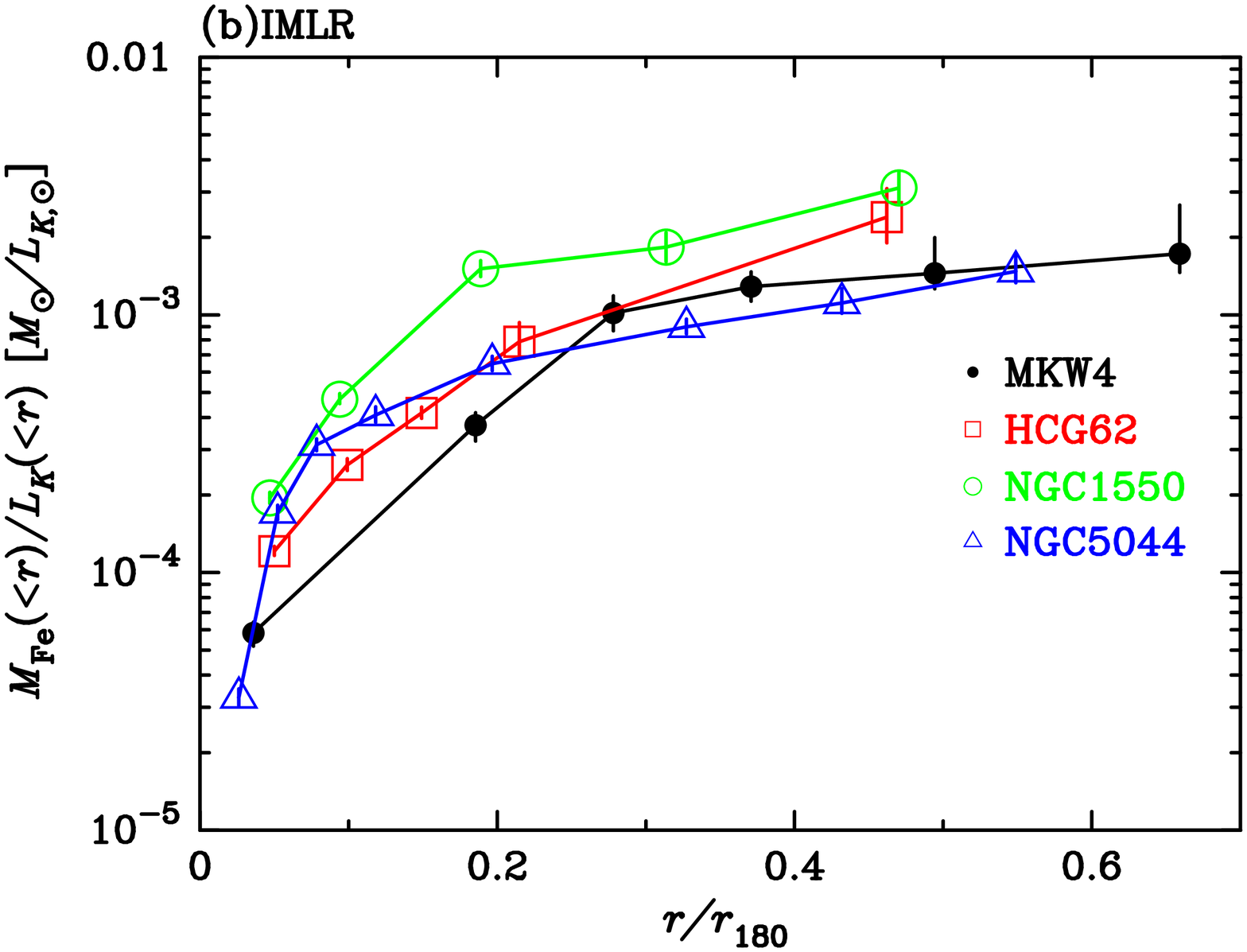}    
       \end{center}
\caption{
(a) Radial profiles of the integrated GMLRs with the $K$-band luminosity 
in units of $M_{\odot}/L_{\odot}$.  The color and mark notations are 
the same as in figure \ref{radial_chi}.  (b) Radial profiles of the 
integrated IMLRs with the $K$-band luminosity in units of 
$M_{\odot}/L_{\odot}$.  The notations of color and mark are the same as 
in figure \ref{radial_chi}.
}
\label{fig:imlr_5044}
\end{figure*}

In order to derive gas mass profile, we calculated the electron 
density from the resultant normalizations with the 1T model because 
the spectra in the outer regions were well-represented with the 1T 
model.  Figure \ref{fig:dens} (a) shows radial profiles of the 
normalizations with the 1T model divided by the area, which corresponded
to the surface brightness, for each group.  We fitted the surface 
brightness profiles 
 with a single $\beta$--model formula,
$S(r)=S_{0} \left[ 1+\left(r/r_{\rm c} \right)^{2}\right]^{-3\beta+0.5}$.
Here, $S_{0}$, $r_{c}$, and $\beta$ are the normalization, 
core radius, and index, respectively.
Beyond $0.1~r_{180}$ of the NGC~1550 group, 
however, there is an discrepancy between the data and the best-fit
single $\beta$--model. An power-law model well represented the
data between 0.1~$r_{180}$ and 0.5~$r_{180}$.
We, therefore, adopted the results of the power-law model fitting for
the region outside 0.1~$r_{180}$.
In figure \ref{fig:dens} (a), for MKW~4, we also plotted 
the surface brightness profile derived from the previous ROSAT result 
\citep{Sanderson2003}, which were normalized to the brightness level 
observed with Suzaku by fitting with the fixed $\beta$ and $r_c$ to be 
0.64 and 5.45$'$, respectively.  And then, we derived the electron 
density from the parameters by fitting the surface brightness profile.
The derived electron density profiles for each group were shown 
in figure \ref{fig:dens} (b).
We also plotted electron density profiles of MKW~4 derived from
the previous ROSAT \citep{Sanderson2003} and XMM \citep{OSullivan2003} 
results in figure \ref{fig:dens} (b).  By integrating the electron 
density profiles, we derived the integrated gas mass profiles for 
each group.  Figure \ref{fig:mass} (a) shows the derived gas mass profile.
For MKW~4, we used the gas mass derived from XMM \citep{OSullivan2003} 
up to $0.2~r_{180}$, and, beyond $0.2~r_{180}$, we integrated the 
electron density derived from Suzaku observations.  Note that, because 
MKW~4 has asymmetrical structure as mentioned in \citet{OSullivan2003}, 
the gas mass profile for MKW~4 would have an uncertainty of a factor of 2\@.
For the NGC~1550 group, we derived the deprojected electron density from 
the fitting result of normalizations up to $0.1~r_{180}$, and multiplied the 
volume of each rings to derive the integrated gas mass.
Beyond $0.1~r_{180}$, we calculated the electron density profiles derived from 
power-law fitting from the normalizations per area with deprojection analysis.
By integrating the electron density, we derived the integrated gas mass 
beyond $0.1~r_{180}$

In figure \ref{fig:mass} (b), 
we derived the integrated iron mass profiles from the gas mass from 
the resultant normalizations with the 1T model and the Fe abundances 
derived from the 1T and 2T model fits with the ATOMDB version 2.0.1\@.  
The radial iron mass profiles with the 2T model for each group were
higher by $\sim 20\%$ at the most than those with the 1T model.
The difference of Fe mass between the 1T and 2T model reflects 
the multiphase ICM in the center of groups and the consequent Fe-bias
(e.g. \citealt{Buote2000, Buote2003a, Buote2003b, Johnson2011}).
The spectral fits with the version 1.3.1 gave higher Fe abundance by 
several tens of \% than those with the version 2.0.1\@.  The total 
iron mass, therefore, would have systematic uncertainties of several 
tens of \% due to uncertainties of the temperature structure and the 
ATOMDB versions.  We also calculated the integrated radial profile 
of GMLRs and IMLRs with the $K$-band luminosity as described in 
subsection \ref{sec:K-band}, up to $\sim 0.5~r_{180}$ as shown in 
figure \ref{fig:imlr_5044}.

\section{Discussion}
\label{sec:discussion}

We derived radial profiles of metal abundances in the ICM of 
the four groups of galaxies, MKW~4, HCG~62, the NGC~1550 group, 
and the NGC~5044 group, observed with Suzaku out to $\sim$ 0.5~$r_{180}$.
In this section, we compare our results with those of clusters 
of galaxies observed with Suzaku and XMM, 
the Coma cluster ($kT \sim 7.8~{\rm keV}$,  $z=0.023$, 
\citealt{Matsushita2013Coma}),
the Perseus cluster ($kT \sim 6.1~{\rm keV}$, $z=0.018$,  
\citealt{Matsushita2013Perseus}), 
the AWM~7 cluster ($kT \sim 3.5~{\rm keV}$, $z=0.017$,  
\citealt{Sato2008}),
the Hydra~A cluster ($kT \sim 3.0~{\rm keV}$, $z=0.054$, 
\citealt{SatoT2012}), 
and the Abell~262 cluster ($kT \sim 2.0~{\rm keV}$, $z=0.016$, 
\citealt{Sato2009b}).

\begin{figure*}[!ht]
 \begin{center}
   \includegraphics[width=0.4\textwidth,angle=0,clip]{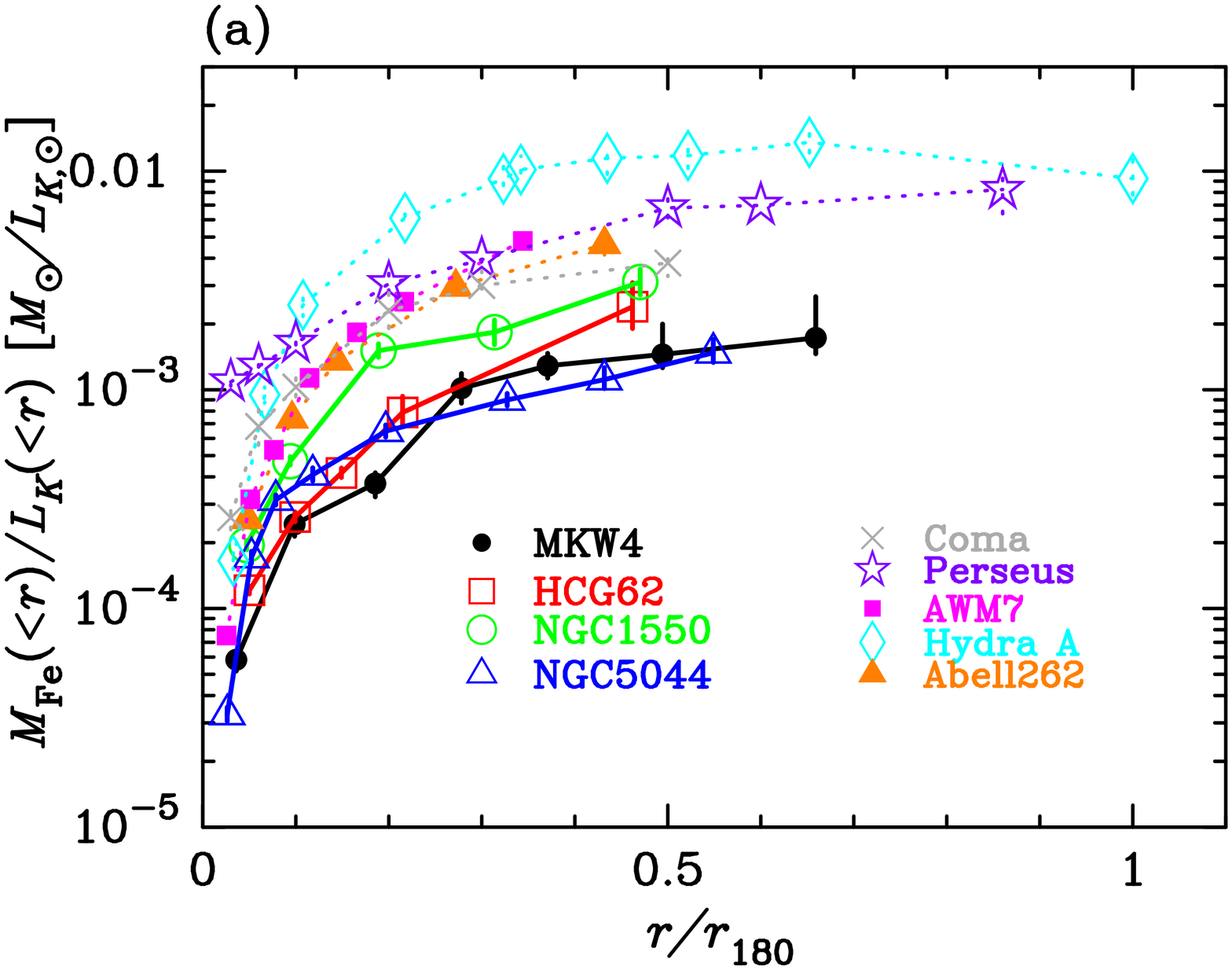}
   \includegraphics[width=0.4\textwidth,angle=0,clip]{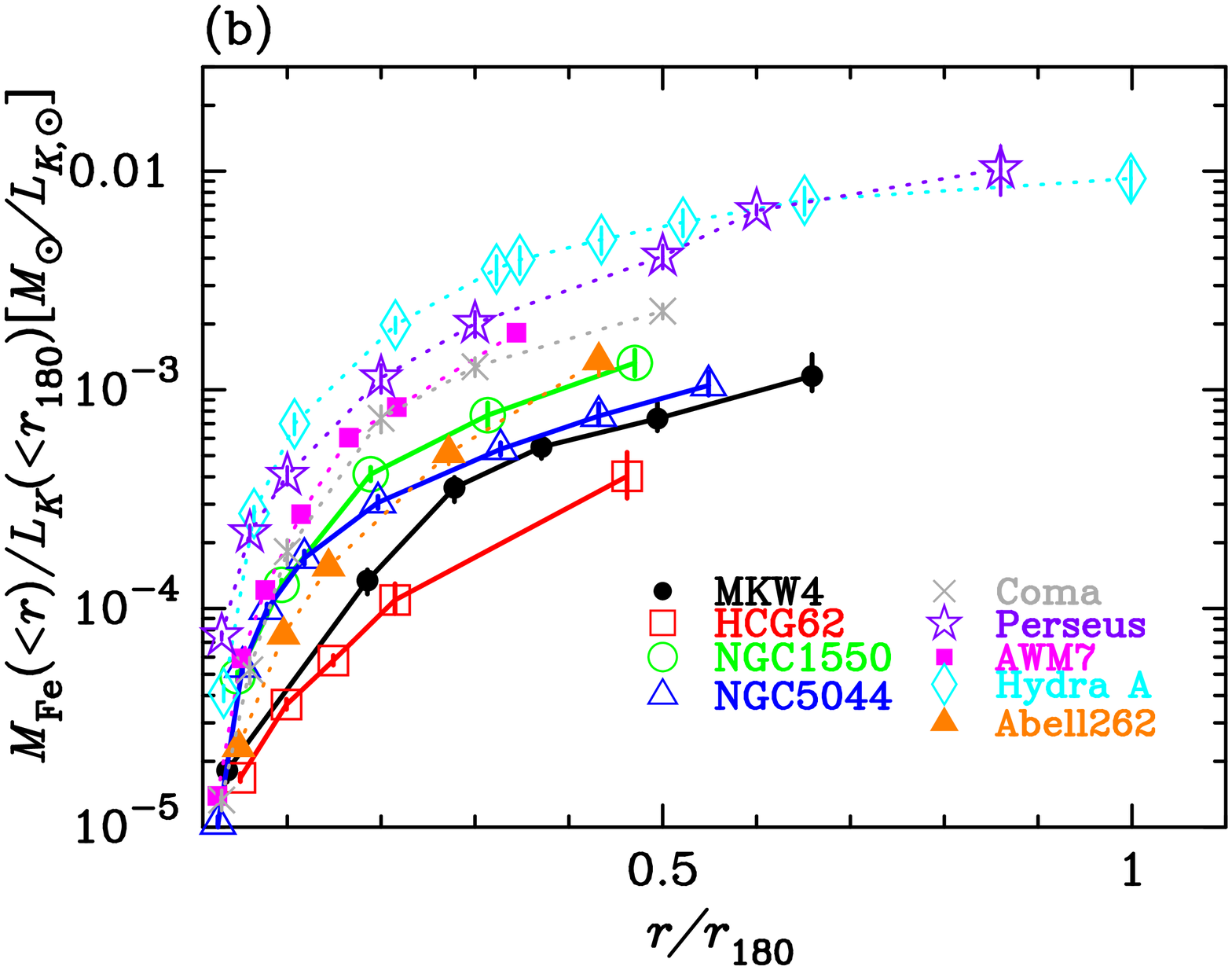}  \end{center}
\caption{
(a) Radial profiles of the integrated IMLRs, 
$M_{\rm Fe}(<r)/L_K(<r)$, in units of $M_{\odot}/L_{\odot}$ 
with the $K$-band luminosity for the Coma cluster \citep{Matsushita2013Coma},
the Perseus cluster \citep{Matsushita2013Perseus},
the Centaurus cluster \citep{Sakuma2011}, the AWM~7 \citep{Sato2008}, 
Hydra~A cluster \citep{SatoT2012}, and the Abell~262 \citep{Sato2009b},.
(b) Radial profiles of the integrated IMLRs with total $K$-band luminosity 
within~$r_{180}$, $M_{\rm Fe}(<r)/L_K(<r_{180})$.
}
\label{fig:imlr}
\end{figure*}

If all galaxies synthesized a similar amount of metals per unit 
stellar mass, and groups and clusters of galaxies contain all the 
metals synthesized in the past, the IMLR values should be similar.
Then, systems with lower gas mass would be expected to have higher 
ICM abundances.  In subsection \ref{subsec:comp_imlr} and 
\ref{subsec:Fe}, we study the dependence of the IMLR and the Fe 
abundance on the system mass.  The relative timing of metal enrichment 
and heating should affect on the present metal distributions in the 
ICM, and therefore, in subsection \ref{subsec:comp_ento},
we study the correlation of IMLR and entropy excess
and discuss the early-metal enrichment in groups and clusters.
The abundance ratios of Si/Fe and Mg/Fe in the ICM constrain the 
relative contributions from SNe Ia and SNecc, and in subsection 
\ref{subsec:comp_pattern} and \ref{subsec:cont_Ia_cc}, 
we compare these abundance ratios of the groups and clusters and 
nucleosynthesis models.

\subsection{The dependence of the iron-mass-to-light ratios 
on the system scale}
\label{subsec:comp_imlr}

Figure \ref{fig:imlr} (a) compares the radial profiles of the 
integrated IMLR, $M_{\rm Fe}(<r)/L_K(<r)$, of the galaxy groups 
with those of several other clusters observed with Suzaku and XMM.
Here, $M_{\rm Fe}(< r)$ and $L_K(< r)$ are integrated Fe mass and 
$K$-band luminosity within a radius $r$, respectively.  The 
integrated IMLR of each cluster or group increases with radius out 
to 0.5~$r_{180}$, and beyond the radius, those of the Hydra A cluster 
and the Perseus cluster become flatter.
In other words,  the distribution of Fe in the ICM are much
more extended than the stellar distribution at least out to 0.5~$r_{180}$.  
Outside the core of galaxy groups, 
if metal enrichment occurs after the formation of clusters, 
the metal distribution would be expected to follow 
the stellar distribution.  
Because of the limited spatial resolution of Suzaku, 
the core of galaxy groups, where are filled the newly formed 
metals spread out by the AGN action, did not discuss here.
Therefore, the increase of IMLR with 
radius indicates that a significant fraction of Fe synthesized in 
an early phase of cluster evolution.


As shown in figure \ref{fig:imlr} (a), the IMLR of the groups 
were significantly smaller than those of the other clusters at 
a given radius in units of $r_{180}$.  Since the IMLR profiles 
increase with radius, comparison of the total IMLR values requires 
observations of IMLR profiles out to the virial radii.
However, if we extrapolate the observed IMLR profiles of groups 
out to the virial radii, it may be difficult to reach the level 
of clusters.

\begin{figure*}[!ht]
 \begin{center}
   \includegraphics[width=0.45\textwidth,angle=0,clip]{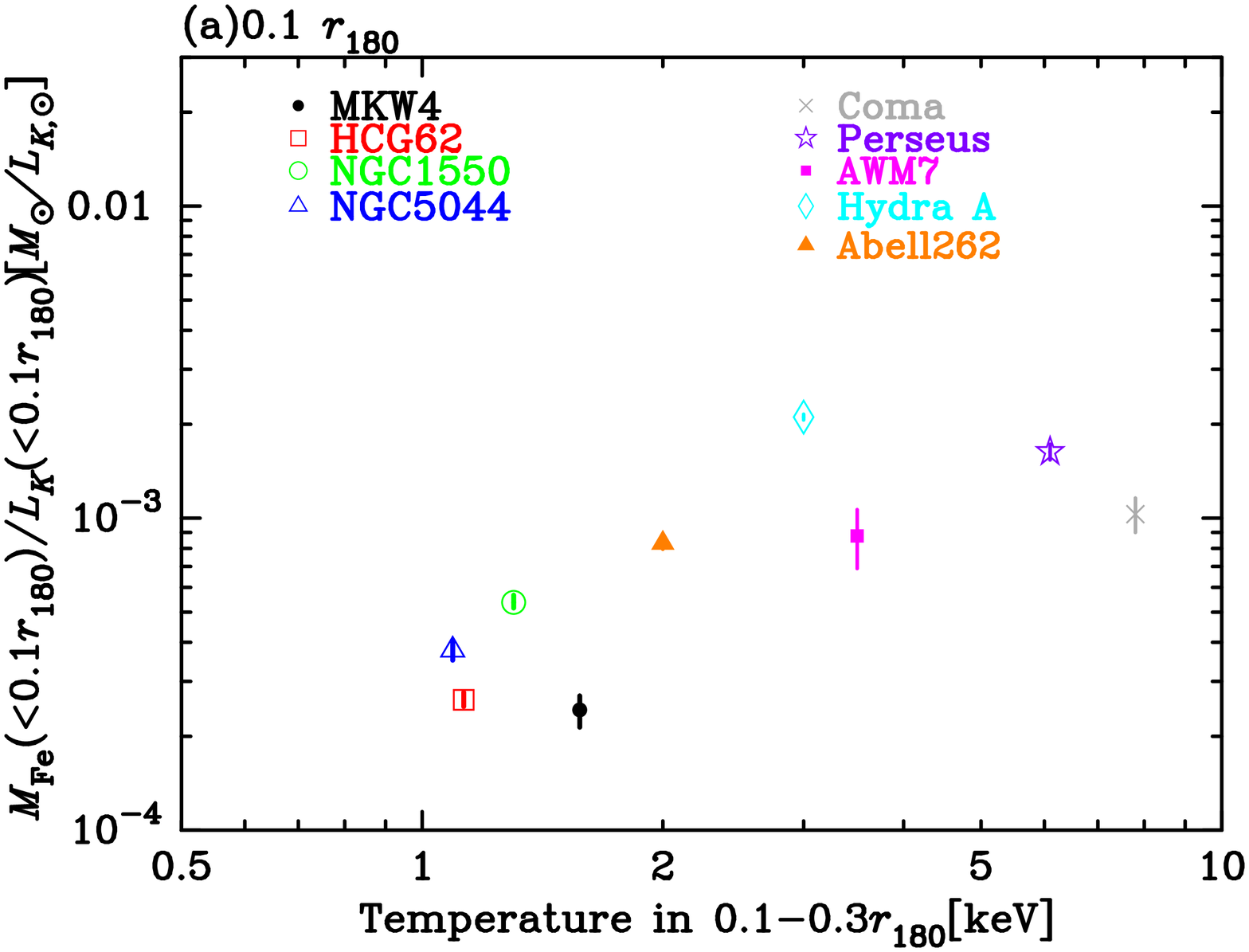}
   \includegraphics[width=0.45\textwidth,angle=0,clip]{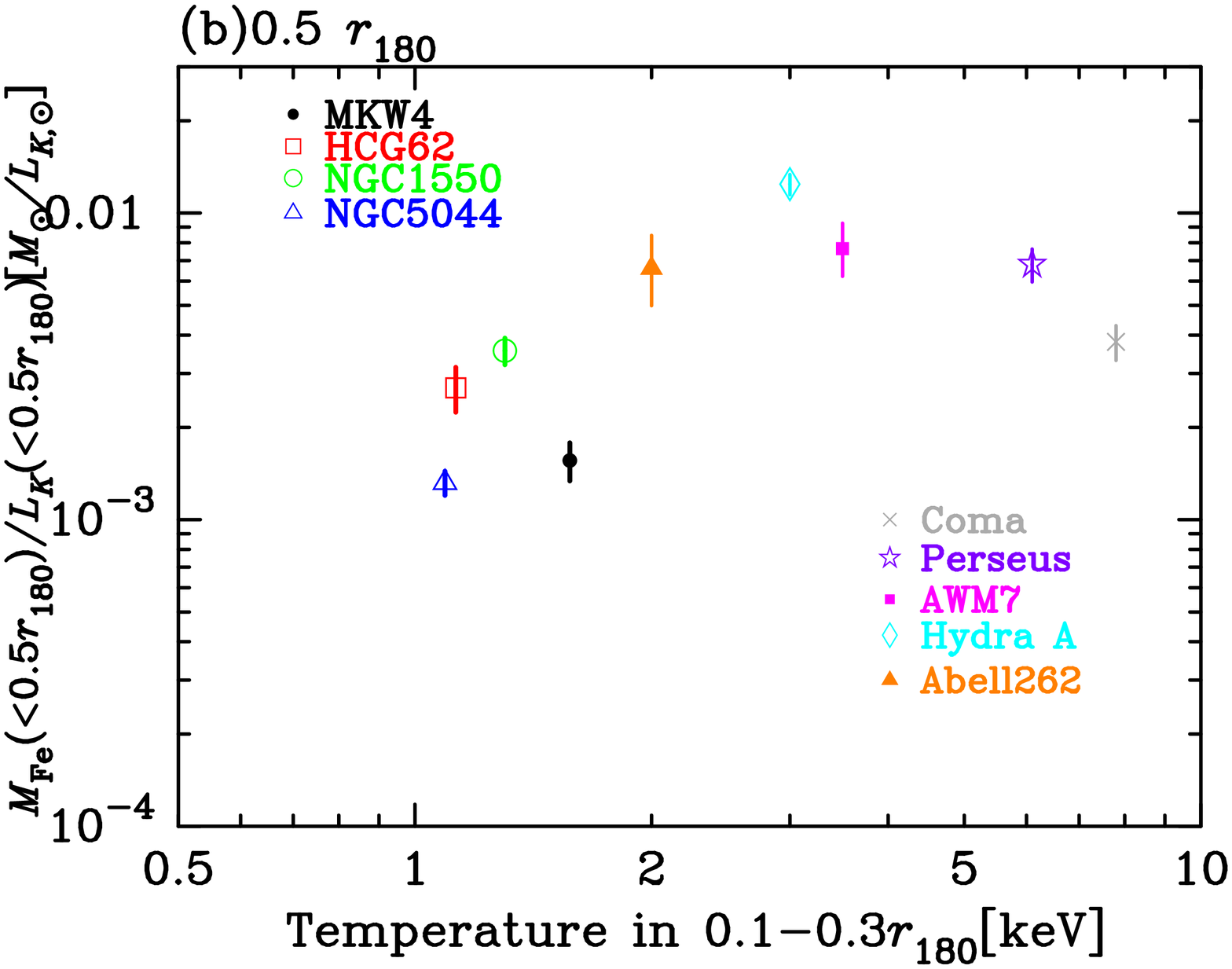}
     \end{center}
\caption{
Integrated IMLRs within $\sim 0.1~r_{180}$ (a) and 
$\sim 0.5~r_{180}$ (b) as a function of the ICM temperature 
at 0.1--0.3 $r_{180}$ of galaxy groups compared with the Coma 
cluster \citep{Matsushita2013Coma}, the Perseus cluster 
\citep{Matsushita2013Perseus}, the Centaurus cluster 
\citep{Sakuma2011}, the AWM~7 \citep{Sato2008}, 
Hydra~A cluster \citep{SatoT2012}, and the Abell~262 \citep{Sato2009b}.
}
\label{fig:imlr_k}
\end{figure*}

To derive the three dimensional $K$-band luminosity profiles,
we deprojected the radial luminosity profiles in the $K$-band
assuming a spherical symmetry.  However, the luminosity 
distribution of galaxies in each cluster was far from the 
spherical symmetry, especially for luminous galaxies. As a result, 
several luminous galaxies could cause artificial breaks in the 
luminosity profiles, especially for poor systems with a smaller 
number of galaxies. Therefore, in figure \ref{fig:imlr} (b),
we also show the ratio of integrated iron mass, $M_{\rm Fe}(<r)$, 
to the total $K$-band luminosity within $r_{180}$, 
$M_{\rm Fe}(<r)/L_K(<r_{180})$\@, to study the difference in the 
IMLR between groups and clusters.  Here, the total $K$-band
luminosities of the other groups and clusters were calculated in 
the same way with our sample groups from the 2MASS catalogue. 
Then, $M_{\rm Fe}(<r)/L_K(<r_{180})$ of groups are also 
systematically lower than those in clusters.  The derived profiles 
of $M_{\rm Fe}(<r)/L_K(<r_{180})$ became smoother than those of 
$M_{\rm Fe}(<r)/L_K(<r)$\@, 
that are reflecting smooth iron density profiles.
The radial profiles of 
$M_{\rm Fe}(<0.5~r_{180})/L_K(<r_{180})$ of the Hydra~A cluster 
and the Perseus cluster agree very well, although those of 
$M_{\rm Fe}(<r)/L_K(<r)$ have a discrepancy of a factor of 2--3
at a given radius in units of $r_{180}$.  Because of the higher 
redshift of the Hydra~A cluster and the lower ICM temperature,
the number of galaxies detected in Hydra~A with 2MASS might not 
be sufficient to deprojected the luminosity profiles in the $K$-band.
However, at $\sim r_{180}$, the systematic uncertainties due to 
the limited number of galaxies are relatively small.

In this paper, we use the scaling radius, $r_{180}$, using the 
average ICM temperature, $\left<kT\right>$, expected from numerical 
simulations.  However, non-gravitational energy inputs in the past 
may change the $r_{180}$--$\left<kT\right>$ relation.  
The $r_{500}$--$\left<kT\right>$ relation estimated for groups by 
\citet{Finoguenov2006, Finoguenov2007} and that used for clusters
by \citet{Pratt2009} differ by $\sim$ 20\% at average ICM temperature 
of 1~keV\@.  This uncertainty is even smaller than the radial range 
in a annular region used in our analysis.  If there is a systematic 
uncertainty in the scaling radius of 20\%, Figure \ref{fig:imlr} 
shows that the systematic uncertainties in the IMLR values
at 0.5 $r_{180}$ are about 10--30\%, which are much smaller than 
the variation in the IMLR.




To study the dependence of the IMLR on the average ICM temperature,
in figure \ref{fig:imlr_k}, we plotted the integrated IMLRs of the 
galaxy groups and clusters at $\sim 0.1~r_{180}$ 
and $\sim 0.5~r_{180}$ with $K$-band, 
$M_{\rm Fe}(<0.1~r_{180})/L_K(<0.1~r_{180})$ 
and  $M_{\rm Fe}(<0.5~r_{180})/L_K(<0.5~r_{180})$,
against the  ICM temperature at 0.1--0.3 $r_{180}$.
At $\sim 0.1~r_{180}$ and $\sim 0.5~r_{180}$,
systems with smaller IMLR values have relatively lower ICM 
temperatures, although there is a significant scatter in the 
IMLR  below 2 keV\@.  The NGC~5044 group and MKW~4 show the smallest 
$M_{\rm Fe}(<0.5~r_{180})/L_K(<0.5~r_{180})$ values, which are an 
order of magnitude smaller than those of rich clusters.

\begin{figure*}[!ht]
 \begin{center}
   \includegraphics[width=0.4\textwidth,angle=0,clip]{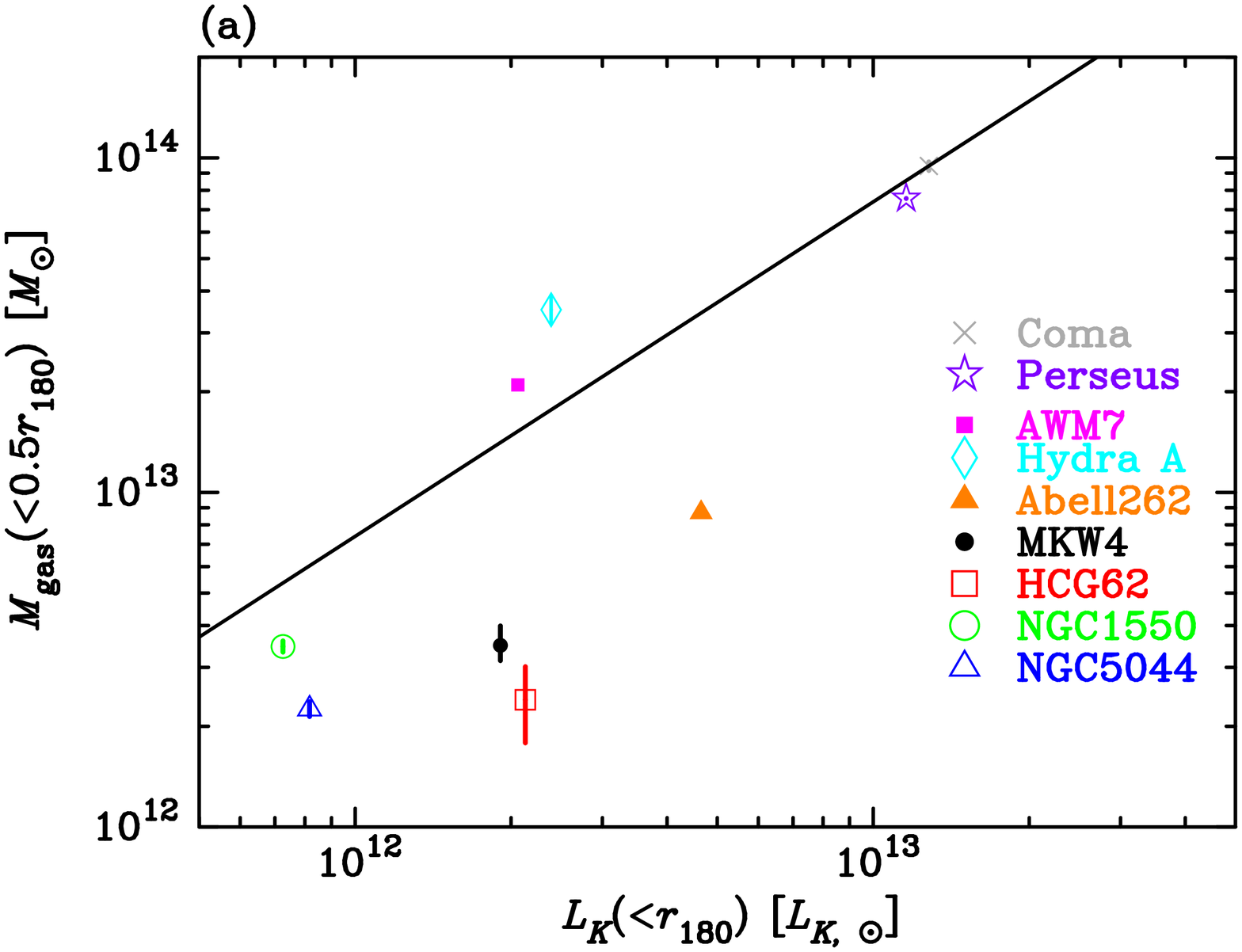}
   \includegraphics[width=0.4\textwidth,angle=0,clip]{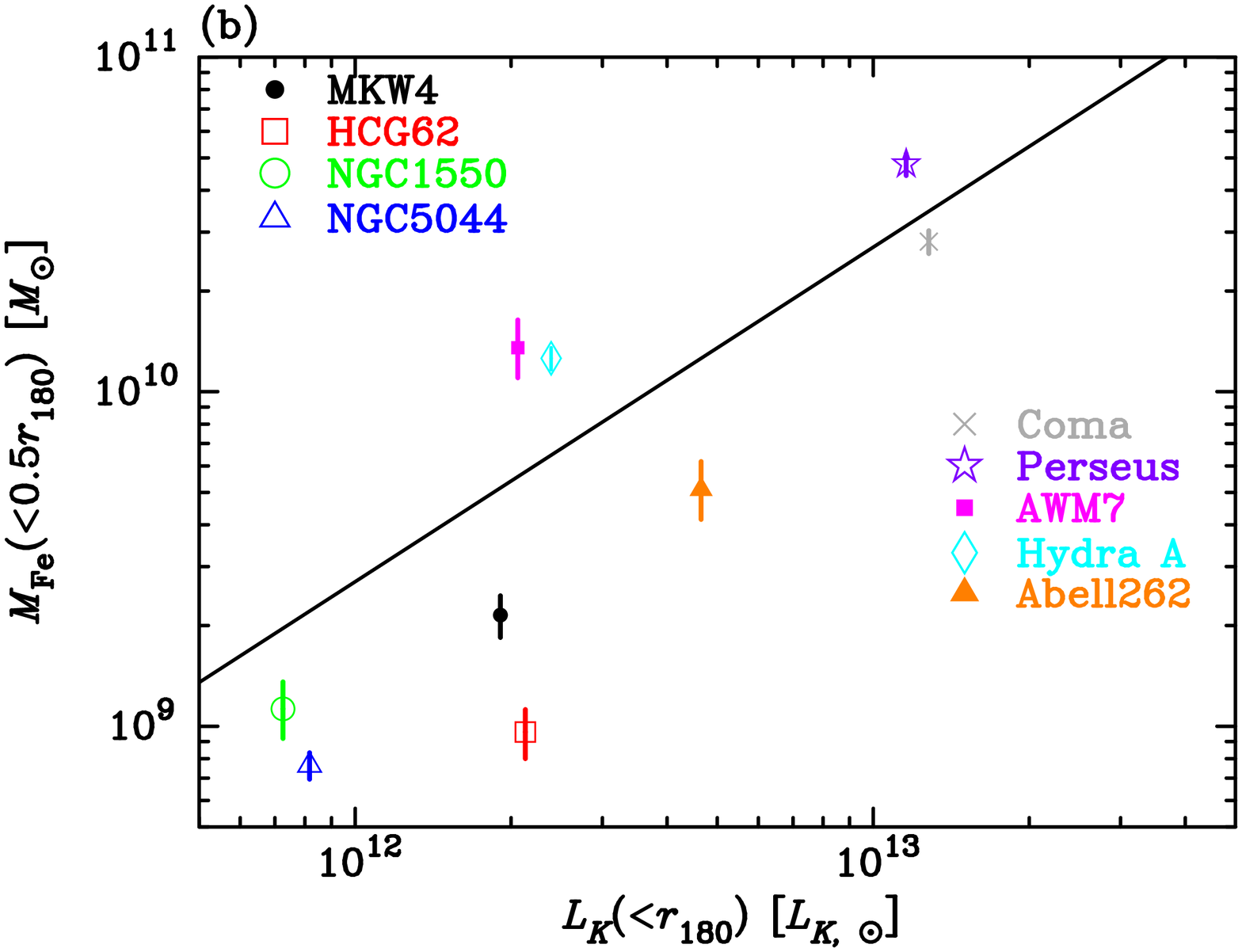}
     \end{center}
\caption{
(a) The integrated gas mass within 
$0.5~r_{180}$, $M_{\rm gas}(<0.5~r_{180})$, is plotted 
against the integrated $K$-band luminosity within 
$r_{180}$, $L_{K}(<r_{180})$.
The color and mark notations are the same as figure \ref{fig:imlr_k}.
The solid line corresponds to a constant 
$M_{\rm gas}(<0.5~r_{180})$ to $L_{K}(<r_{180})$ ratio of 7.39 
$M_\odot/L_\odot$
, which is averaged for the Coma cluster, the Perseus cluster, and the AWM~7.
(b) The integrated iron mass, 
$0.5~r_{180}$, $M_{\rm Fe}(<0.5~r_{180})$, is plotted
against  $L_{K}(<r_{180})$.
The solid line corresponds to a constant $M_{\rm Fe}(<0.5~r_{180})$ 
to $L_{K}(<r_{180})$ ratio of $2.7 \times 10^{-3} M_\odot/L_\odot$
, which is averaged for the Coma cluster, the Perseus cluster, and the AWM~7.
}
\label{fig:MFe_LK}
\end{figure*}

In figure \ref{fig:MFe_LK}(a), we plotted integrated gas mass out to
$0.5~r_{180}$, $M_{\rm gas}(<0.5~r_{180})$, of the four groups of 
galaxies in our sample with those of the other clusters as a function 
of $L_{K}(<r_{180})$.
$M_{\rm gas}(<0.5~r_{180})$ of these clusters of galaxies and 
the NGC~1550 group are mostly proportional to $L_{K}(<r_{180})$.
In contrast, the other groups and Abell~262 cluster have smaller 
$M_{\rm gas}(<0.5~r_{180})$ to $L_{K}(<r_{180})$ ratio by a factor 
of $\sim$ 5.  The right panel of figure \ref{fig:MFe_LK} shows the 
integrated Fe mass out to $0.5~r_{180}$, $M_{\rm Fe}(<0.5~r_{180})$, 
against  $L_{K}(<r_{180})$.  The groups with lower 
$M_{\rm gas}(< 0.5~r_{180})$ for a given $L_{K}(<r_{180})$ have 
lower $M_{\rm Fe}(<0.5~r_{180})$ values.

In summary, some groups of galaxies have lower gas-mass-to-light 
ratios with lower IMLR values.  If all galaxies synthesized a similar 
amount of metals per unit stellar mass, the observed lower IMLR values 
indicate that a significant fraction of Fe synthesized in the
past is not located within 0.5~$r_{180}$ of these groups of galaxies.



\subsection{The dependence on the Fe abundance on the system scale}
\label{subsec:Fe}
\label{subsec:comp_fe_abundance}




\begin{figure*}[!ht]
 \begin{center}
     \includegraphics[width=0.47\textwidth,angle=0,clip]{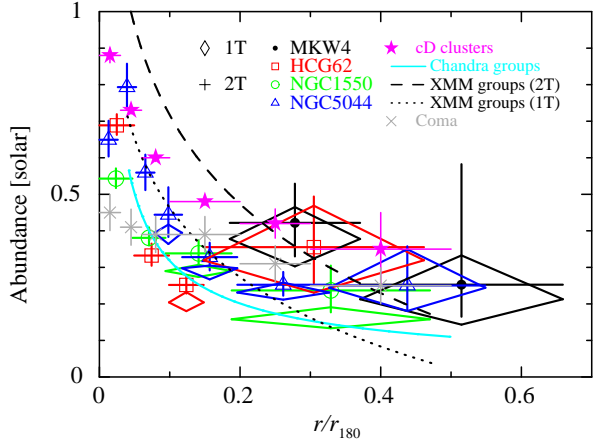}
  \end{center}
\caption{
Radial profiles of the Fe abundance derived from the 2T model (crosses) 
with the ATOMDB version 2.0.1\@.  We also plotted the Fe abundance 
profiles with the 1T model (diamonds) beyond  $0.1~r_{180}$ with the 
ATOMDB version 2.0.1.  The notations of colors and symbols are the same 
as in figure \ref{fig:kTprofile}\@.
The weighted average Fe abundance of relaxed clusters with a cD galaxy
at their center observed with XMM \citep{Matsushita2011}
are shown by filled stars (magenta).  The best-fit regression relation 
for groups observed with Chandra \citep{Rasmussen2007} is indicated by 
the solid (cyan) line, and those for cool core groups with XMM 
(2T: dashed line, 1T: dotted line, \citealt{Johnson2011}) are 
also plotted.  Here, we corrected the XMM and Chandra results for the 
differences of the definition on the solar abundance table and the 
virial radius. 
}
\label{fig:fe_radial_clusters}
\end{figure*}

The radial profiles of the Fe abundance of the galaxy groups with 
previous measurements for clusters of galaxies are shown 
in figure \ref{fig:fe_radial_clusters}. 
Absolute values of the Fe abundance derived from the Fe-L spectral 
fitting could have larger systematic uncertainties than the IMLR, 
considering that a higher Fe abundance gives a lower normalization.
The 2T model fits on the Fe-L spectra sometimes give significantly 
higher Fe abundances than the 1T model fits by several tens of percents
\citep{Buote2000, Johnson2011, Murakami2011}.
As shown in subsection \ref{sec:atomdb}, using the 2T model fits, 
the new version of ATOMDB yielded lower Fe abundances
than the old one, especially at the group center.

\begin{figure*}[!t]
\begin{center}
   \includegraphics[width=0.42\textwidth,angle=0,clip]{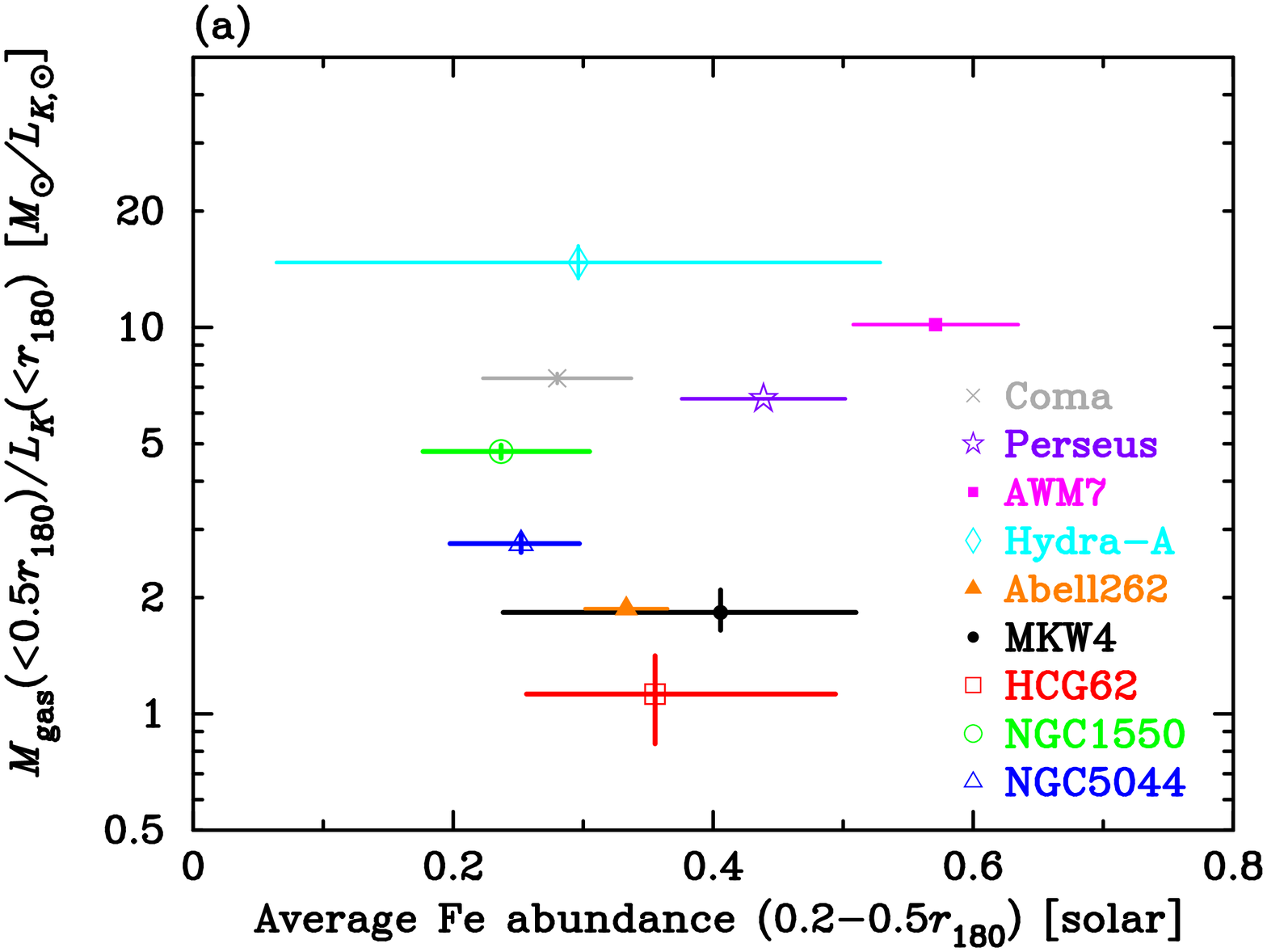}
   \includegraphics[width=0.45\textwidth,angle=0,clip]{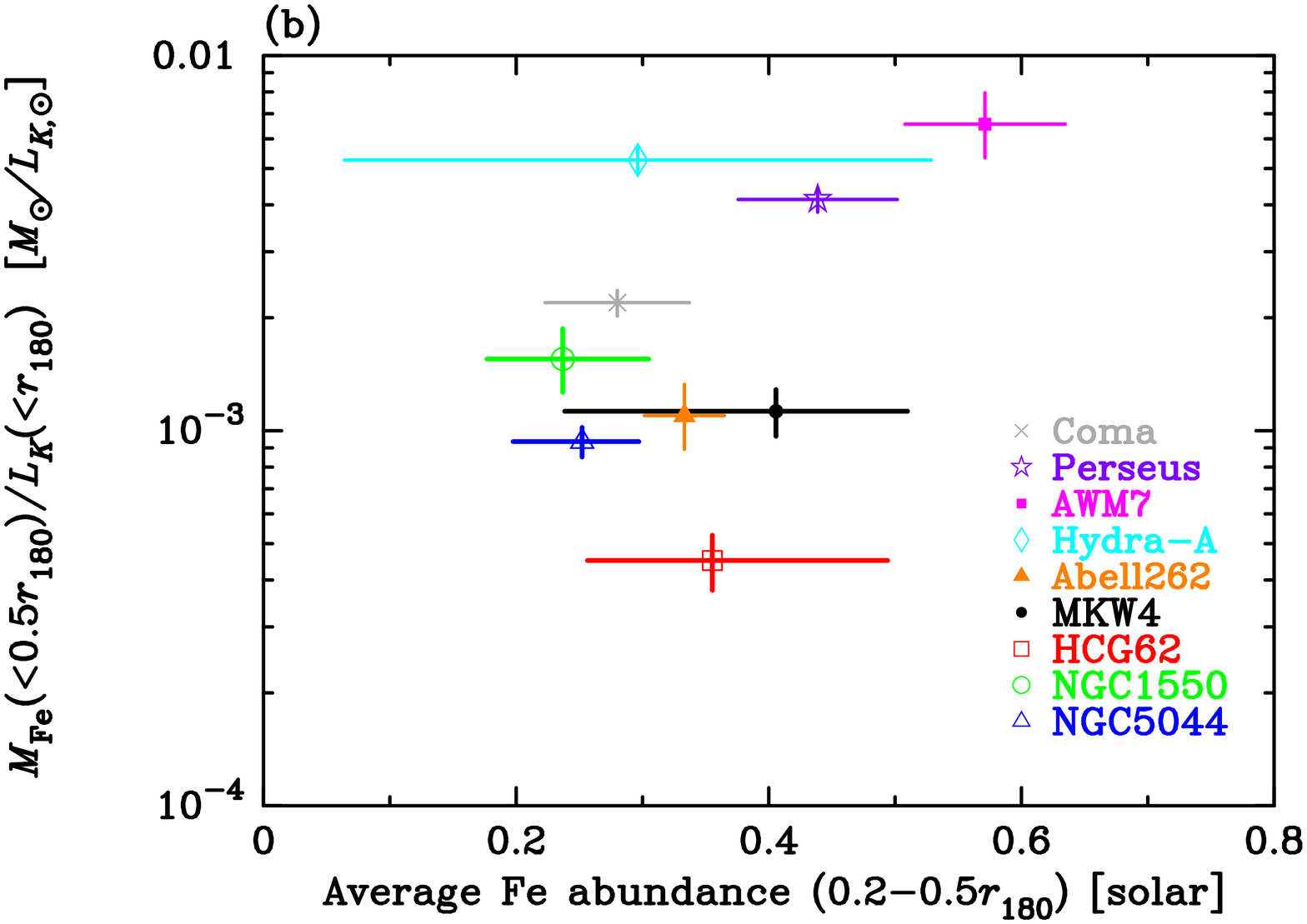}
 \end{center}
\caption{
(a)The $M_{gas}(< 0.5~r_{180})$ to $L_{K}(<r_{180})$ ratio as a function of 
the average Fe abundances in 0.2--0.5 $r_{180}$ for galaxy groups 
compared with the Coma cluster \citep{Matsushita2013Coma},
the Perseus cluster \citep{Matsushita2013Perseus},
the AWM~7 \citep{Sato2008}, 
Hydra~A cluster \citep{SatoT2012},
and the Abell~262 \citep{Sato2009b}. 
Here, the result of the AWM 7 cluster is an extrapolation from the
results within $\sim 0.4~r_{180}$.
(b)The same as (a), but Fe mass, $M_{Fe}(<0.5~r_{180})$.
}
\label{fig:Fe_IMLR}
\end{figure*}

Beyond $0.1~r_{180}$, the Fe abundances of our sample groups 
derived from the 1T and 2T model fits are about 0.2--0.4 solar, 
except for the 1T model result for the NGC~1550 group.
Beyond $0.1~r_{180}$, the Fe abundance profiles of the galaxy 
groups tend to be smaller than the weighted average of the nearby 
relaxed clusters with a cD galaxy at their center by several tens 
of percents, while agree well with that of the Coma cluster, 
which is one of the largest cluster in nearby Universe, observed 
with XMM \citep{Matsushita2011}.
The Fe abundances of clusters were derived from the K$\alpha$ lines 
of Fe and as a result,  systematic uncertainties should be smaller 
than those from the Fe-L lines.  Even considering the systematic 
uncertainties in the derived Fe abundances from the Fe-L spectral 
fitting, the Fe abundances in the groups are not much  higher
than those of clusters.

\begin{figure*}[!t]
 \begin{center}
   \includegraphics[width=0.3\textwidth,angle=0,clip]{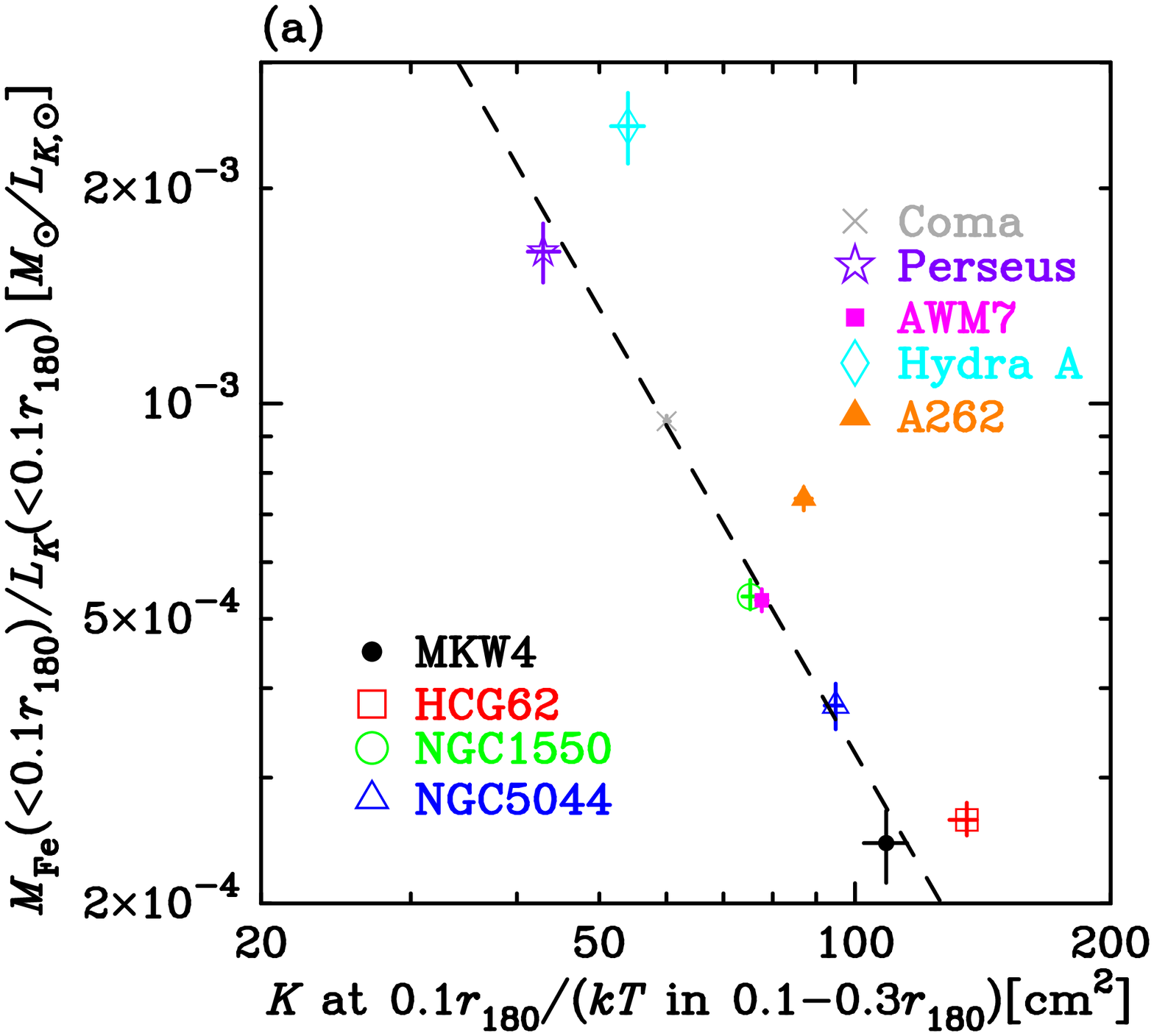}
   \includegraphics[width=0.3\textwidth,angle=0,clip]{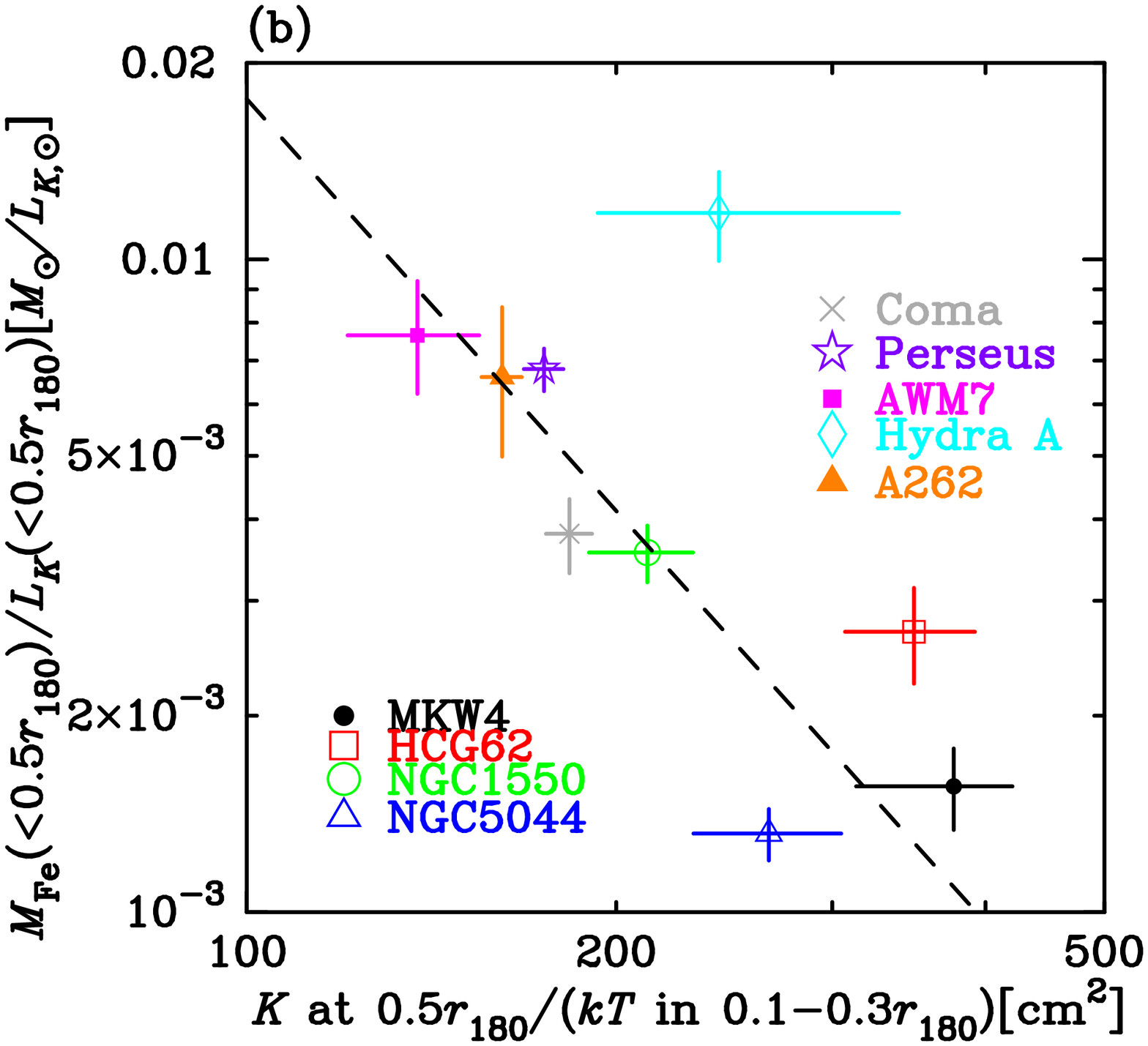}
   \includegraphics[width=0.3\textwidth,angle=0,clip]{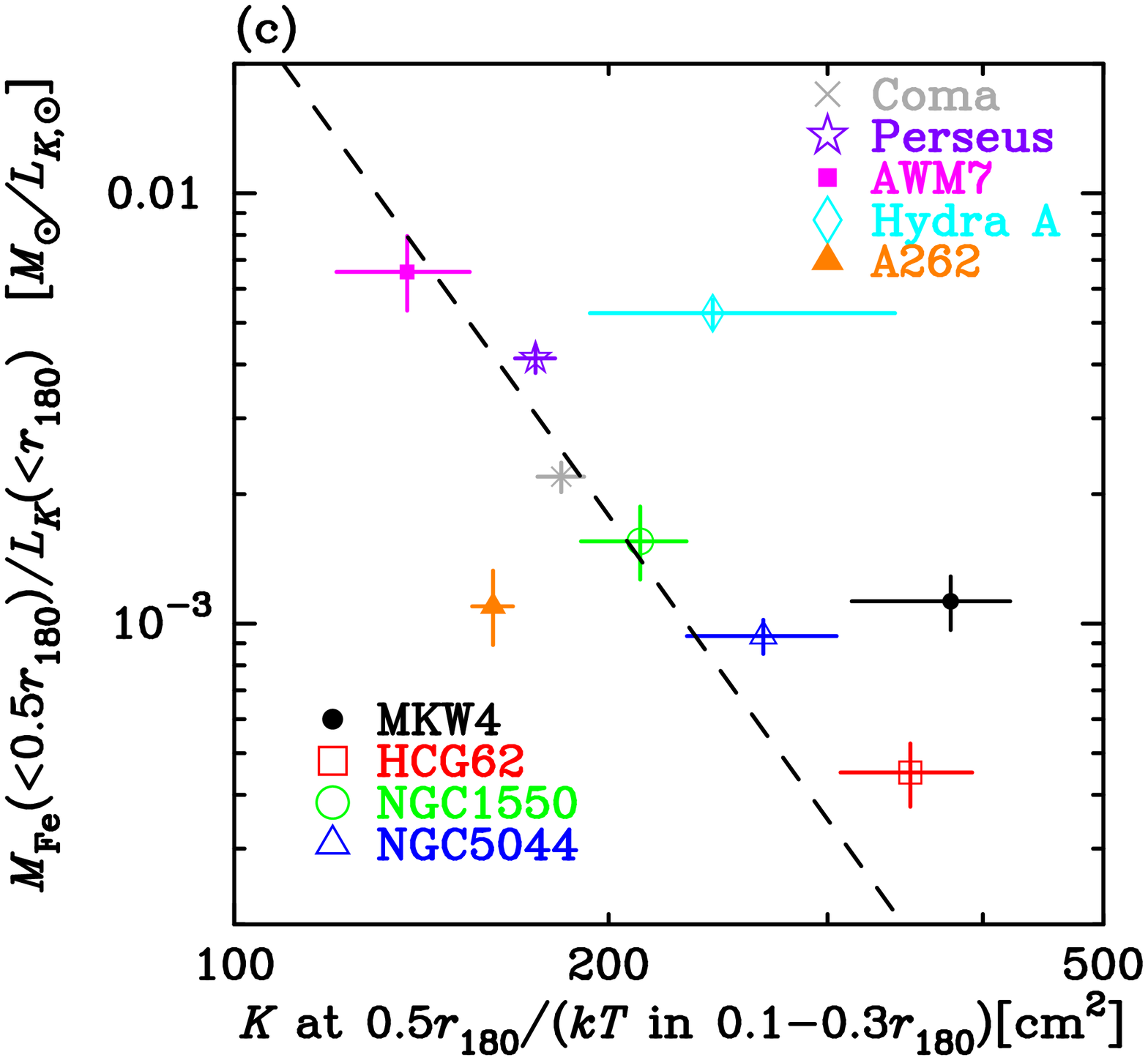}
  \end{center}
\caption{
(a) The integrated IMLRs within $0.1~r_{180}$ as a function of the 
scaled entropy profiles at $\sim 0.1~r_{180}$.
The dashed lines indicate the best--fit power-law model.
(b)The same as (a), but at 0.4--0.5~$r_{180}$.
(c) The $M_{\rm Fe}(< 0.5~r_{180})$ to $L_{K}(< r_{180})$ ratios against 
scaled entropy at 0.4--0.5~$r_{180}$.
The color and mark notations are the same as figure \ref{fig:imlr}.
}
\label{fig:S_IMLR}
\end{figure*}

Beyond $0.1~r_{180}$, our Fe abundances derived from the 2T model
tend to be lower and flatter than the best-fit regression relation 
from the 2T model fits on the results of the groups
observed with XMM \citep{Johnson2011} and higher than the relation 
of the 1T model fits on Chandra data \citep{Rasmussen2007}.
Some part of this difference may be caused by an difference in the 
adopted atomic data, since the previous results were derived using 
the old version of ATOMDB.  The differences in the sample may also 
cause the discrepancy, since our sample is limited within relatively 
X-ray luminous groups.
%


The observed dependency of stellar to gas mass ratios within 
$r_{500}$ on the total system mass has been sometimes interpreted 
that star formation efficiency from baryons also depends on the 
system mass.  Then, the Fe abundance in the ICM of systems with 
lower gas mass-to-light ratios is expected to be higher than those 
with higher values.
In figure \ref{fig:Fe_IMLR}, 
$M_{\rm gas}(<0.5~r_{180})$/$L_{K}(<r_{180})$  and
$M_{\rm Fe}(<0.5~r_{180})$/$L_{K}(<r_{180})$ 
are plotted against the Fe abundances at 0.2--0.5 $r_{180}$.
There is no significant correlation between the Fe abundance and 
$M_{\rm gas}(<0.5~r_{180})$/$L_{K}(<r_{180})$ or 
$M_{\rm Fe}(<0.5~r_{180})$/$L_{K}(<r_{180})$.
Among clusters of galaxies, the Fe abundance in 0.1--0.3 $r_{180}$ 
in the ICM does not depend on the ICM temperature 
\citep{Matsushita2011}, although the dependence of the stellar to 
gas mass ratio have been found among clusters.





\subsection{Comparison of the entropy profiles with other systems}
\label{subsec:comp_ento}

The difference in the ratio of gas-mass-to-stellar-mass would 
reflect differences in distributions of gas and stars, which in turn 
reflects the history of energy injection from galaxies to the ICM.  
If metal enrichment occurred before energy injection, the poor 
systems would carry relatively smaller metal mass with a smaller gas 
mass than rich clusters, whereas, the metal abundance would be quite 
similar to those in rich clusters.

Numerical simulations indicate that when clusters are radially scaled 
to the virial radius, or $r_{180}$, distribution of dark matter and 
baryons become self-similar(e.g. \citealt{Navarro1995}).  Entropy carry important information 
about the thermal history of the ICM. Usually, the entropy parameter, 
$K$ is defined as,
\begin{equation}
K(r)= \frac{kT(r)}{n_{e}(r)^{2/3}},
\end{equation}	
where $kT(r)$ and $n_{e}(r)$ are temperature and deprojected 
electron density, respectively, at a radius $r$ from the cluster center.
Especially, feedback from galaxies, such as galactic winds, changes 
ICM entropy rather than temperature.  If there was no feedback, entropy 
profile is determined by pure gravitational heating.
Then, at a given radius in units of $r_{180}$, the entropy, $K$,  of 
the ICM is expected to be proportional to the average ICM temperature, 
$\left<kT\right>$ \citep{Ponman1999, Ponman2003}.
In other words, considering pure gravitational heating only, the scaled 
entropy, $K/\left<kT\right>$ is expected to be a constant.
However, the relative entropy level in groups of galaxies is 
systematically higher than that for clusters of galaxies 
\citep{Ponman1999, Ponman2003}, and the gas density profiles in the 
central regions of groups and poor clusters were observed to be 
shallower than those in the self-similar model
(e.g. \citealt{Cavagnolo2009}).


In order to investigate the correction between entropy and IMLR profiles, 
we  plotted  IMLR profiles as a function of entropy profiles  
scaled by the ICM temperature in the 0.1--0.3~$r_{180}$.
In figure \ref{fig:S_IMLR} (a) , we plotted the IMLR within 
$0.1~r_{180}$ against the scaled entropy at $0.1~r_{180}$ .
Figure \ref{fig:S_IMLR} (b) is the same as (a), but 
the radius is changed to 0.5~$r_{180}$.
Figure \ref{fig:S_IMLR} (c) is plotted 
$M_{\rm Fe} (< 0.5~r_{180})/L_K (< r_{180})$ as a function of 
the scaled entropy profiles at 0.4--0.5~$r_{180}$.
We note that the ICM temperatures at this radial range of groups of 
galaxies are close to the average ICM temperature
$\left<kT\right>$ \citep{Rasmussen2007}.
Excluding the Hydra A data, the IMLR correlate inversely well with 
the scaled entropy at each radius.  When we plotted 
$M_{\rm Fe}(< 0.5~r_{180})/L_K(< r_{180})$ against the scaled entropy 
at 0.5~$r_{180}$, the Hydra~A data became closer to the relation of 
the other groups and clusters.

The increase of integrated IMLR with radius, the Fe 
abundance profiles, and the relationship between the scaled entropy 
and IMLR indicate early metal enrichments in clusters and groups of 
galaxies as already suggested by \citet{Matsushita2011}.
If these systems synthesized Fe in an early phase of cluster 
evolution, the ICM is polluted in the same way.
Then, the relative importance of the non-gravitational energy inputs 
in poor systems causes the difference in the gas distribution and 
as a result, the excess entropy inversely correlates with the 
integrated gas-mass-to-light ratio and IMLR.  In systems with higher 
excess entropy,  the gas is much more extended than stars, and as a 
result, the metals are also extended than stars.


\subsection{Comparison of the Si/Fe and Mg/Fe ratios 
with other groups and clusters of galaxies }
\label{subsec:comp_pattern}

\begin{figure}[!ht]
 \begin{center}
        \includegraphics[width=0.47\textwidth,angle=0,clip]{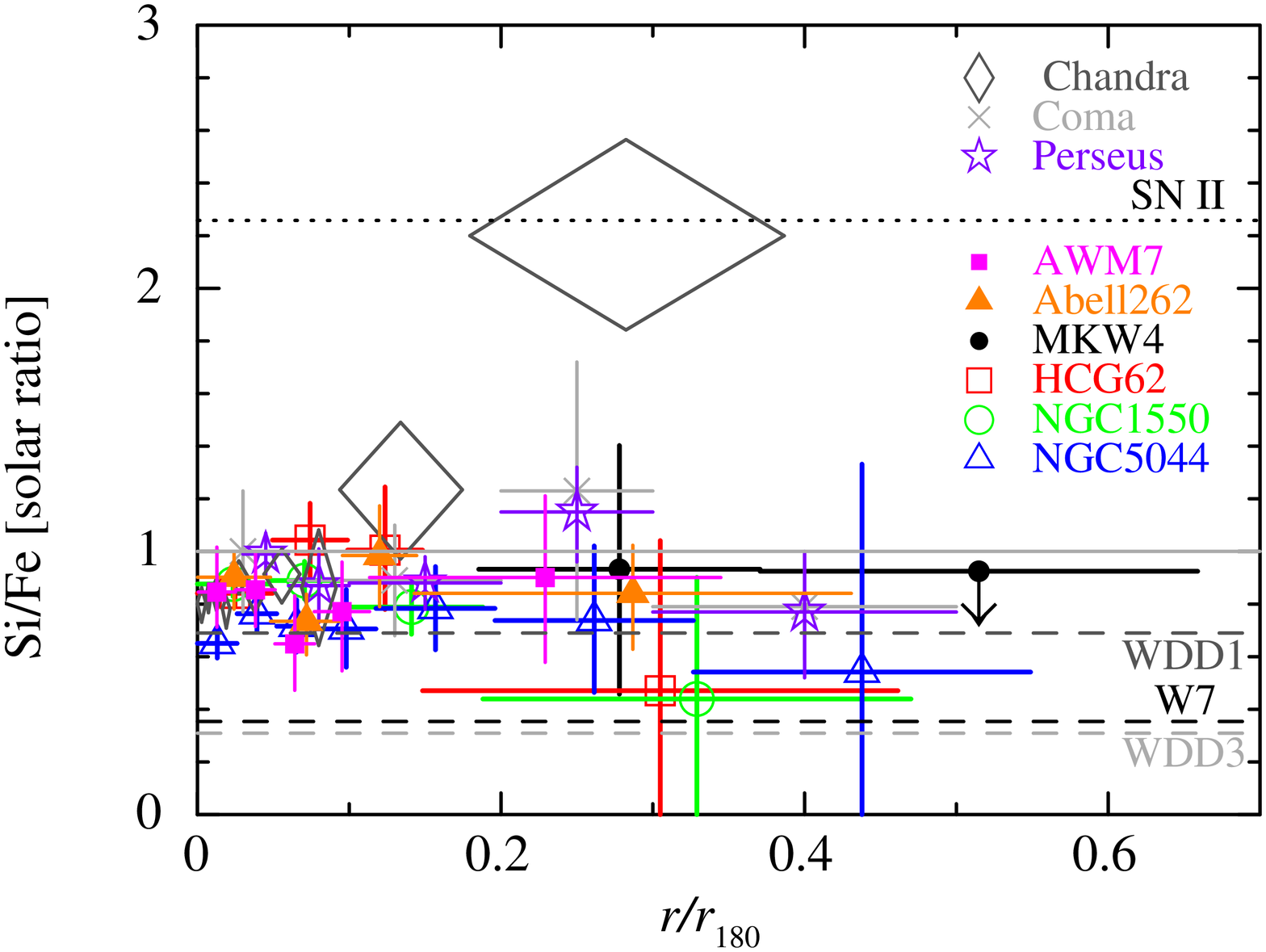}
  \end{center}
\caption{
Radial profiles of the Si/Fe ratios derived from the 2T model fits 
with the ATOMDB version 2.0.1.
The color and symbol notations are the same as in figure
\ref{fig:kTprofile}.
The crosses, stars, filled squares, and filled triangles correspond
to the Coma cluster \citep{Matsushita2013Coma}
  the Perseus cluster \citep{Matsushita2013Perseus},
AWM~7 \citep{Sato2008}, and Abell~262 cluster \citep{Sato2009b},
 respectively.
Diamonds show the radial profile of the 
weighted average of the Si/Fe ratio of groups observed with Chandra
\citep{Rasmussen2007}. Here, effects of differences in adopted solar 
abundance tables were corrected.
Black, light gray, and dark gray dashed lines show the Si/Fe ratios of 
SN Ia yields of W7, WDD1, WDD3 model \citep{Iwamoto1999}, respectively.
Black dotted line also shows the ratios of SNcc yields 
\citep{Nomoto2006}\@. 
}
\label{fig:si_radial_clusters}
\end{figure}

The ratios of $\alpha$-elements to Fe abundance give a strong 
constraint for the nucleosynthesis contribution from SNe Ia and SNecc.
As shown in figure \ref{fig:si_radial_clusters},
the Si/Fe ratios in the ICM of the four galaxy groups in our sample 
are almost constant at $\sim 1$ solar ratio out to 0.2--0.3 $r_{180}$. 
Beyond 0.2--0.3 $r_{180}$ out to 0.5 $r_{180}$, the upper limit of 
the Si/Fe ratios are about unity in solar units.
Here, we only plotted the Si/Fe ratios derived from the 2T model fits, 
since the  1T and 2T models gave almost the same values as shown in 
subsection \ref{sec:atomdb}.
Figure \ref{fig:si_radial_clusters} also shows the radial profiles of
the Si/Fe ratios in the ICM of several clusters observed with Suzaku 
and XMM.  The Si/Fe ratios in these clusters, the flat radial profiles
at the solar ratio, agree well with those of our sample groups of galaxies.
Although these previous measurements on abundance ratios were derived 
using the old ATOMDB except for the Perseus cluster, the systematic 
uncertainties in the Si/Fe ratios caused by the different version of 
ATOMDB may be relatively small as shown in subsection \ref{sec:atomdb}.


\begin{figure}[!tpd]
\begin{center}
        \includegraphics[width=0.470\textwidth,angle=0,clip]{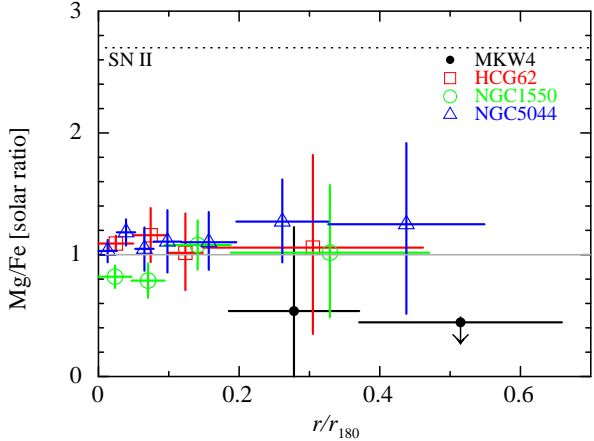}
    \end{center}
\caption{
Radial profiles of the Mg/Fe ratios in the four groups derived 
from the 2T model with the ATOMDB version 2.0.1\@.
The color and symbol notations are the same as in figure \ref{fig:kTprofile}\@.
Black dotted line shows the ratios of SNe II yields 
\citep{Nomoto2006}\@.
}
\label{fig:mg_radial_clusters}
\end{figure}

In figure \ref{fig:si_radial_clusters}, we also plotted the average 
Si/Fe ratios in the ICM of galaxy groups observed with 
Chandra \citep{Rasmussen2007}.
Within 0.1~$r_{180}$, the Si/Fe ratios of the groups and clusters 
observed with Suzaku agree well with the weighted average of groups 
observed with Chandra.  However, beyond 0.2~$r_{180}$, the average 
Si/Fe ratio from the Chandra data is significantly higher than the 
Suzaku results.  Some part of the difference in the Si/Fe ratio can 
be caused by the differences in the sample, since the Si/Fe ratios 
of the some X-ray luminous galaxy groups observed with 
Chandra are consistent with a flat radial profile.
Especially, the Si/Fe ratios of three groups, the NGC~5044 group, 
HCG~62, and MKW~4, observed with Suzaku are consistent within error 
bars with those with Chandra, except for the outermost region 
($\sim 0.2~r_{180}$) of the NGC~5044 group observed with Chandra.


Figure \ref{fig:mg_radial_clusters} shows the radial profiles of 
the Mg/Fe ratios of the four groups.
As in the Si/Fe ratios, the Mg/Fe ratios are mostly consistent with 
the solar ratio and exhibit no significant radial dependence out to 
0.5~$r_{180}$\@.  The Mg/Fe ratios of  other groups and poor clusters 
of galaxies at 0.1--0.3 $r_{180}$ with Suzaku using the old version 
of ATOMDB are mostly consistent with the solar ratio 
\citep{Sato2008, Sato2009a, Sato2009b, Komiyama2009, Murakami2011, 
Sakuma2011}.
The systematic differences in the derived Mg/Fe ratios between the 
1T and 2T model fits and between the two versions of ATOMDBs
are smaller than a few tens of \%. 
(subsection \ref{sec:fittingresults} and \ref{sec:atomdb}).


\subsection{Contribution of SNe Ia and SNecc}
\label{subsec:cont_Ia_cc}

The Si/Fe ratios of  SNe Ia and SNecc yields from nucleosynthesis 
models are shown in figure \ref{fig:si_radial_clusters}.
Here, SNecc yields by \citet{Nomoto2006} refer to an average over 
the Salpeter IMF of stellar masses from 10 to 50 $M_\odot$,
with a progenitor metallicity of $Z=0.02$. The SNe Ia yields of
the classical deflagration model, W7, and a delayed detonation (DD) 
models, WDD1, and WDD3, were taken from \citet{Iwamoto1999}.
The derived Si/Fe ratios were located between those of SNe Ia and 
SNecc yields.  However, the systematic uncertainties in the Si/Fe 
ratio in SNe Ia nucleosynthesis models give an uncertainty in the 
relative contribution of both SN types as discussed in 
\citet{Degrandi2009}.
Because O and Mg are predominantly synthesized in SNecc, 
the O/Fe and Mg/Fe ratios give more unambiguous information on the 
relative contribution from SNecc and SNe Ia.
The Mg/Fe ratio of SNecc yields assuming the Salpeter IMF
with a progenitor metallicity of 0.02 is shown in figure 
\ref{fig:mg_radial_clusters}.  the Mg/Fe ratios of those groups are
a factor of 2--3 smaller than that of the SNecc yields.

The number ratio of SNecc and SNe Ia to synthesize metals in the ICM 
was estimated with Suzaku and XMM data 
(e.g. \citealt{dePlaa2007, Sato2007, Sato2008, Sato2010, 
Matsushita2013Perseus, Matsushita2013Coma}).
The observed abundance pattern of O, Mg, Si and Fe  in the ICM 
observed with Suzaku is more consistent with a mixture of yields 
of SNecc with W7 or WDD3 models rather than that with the WDD1 model 
\citep{Sato2007}.  The similarity of Mg/Si/Fe pattern of our sample 
with the previous Suzaku measurements also indicates a contribution 
of the W7 or WDD3 yields rather than WDD1 ones.  The  number ratio 
of SNecc and SNe Ia to synthesis the observed Mg, Si and Fe in the 
ICM of this work should close to the previous estimations, which are 
about 3--4 using SNe Ia yield of W7.  Then, most of the Fe in the ICM 
should have been synthesized by SNe Ia.

The solar ratio of the Mg/Fe and Si/Fe ratios of our sample groups 
and those of clusters observed with Suzaku indicates that the 
contributions from two SN types to the metals in the ICM are universal.
When systems are old enough, most SNe Ia and SNecc would have already 
exploded. To explain the abundance pattern of the stars in the solar 
neighborhood, the lifetimes of SNe Ia are confined within 0.5--3 Gyr, 
with a typical lifetime of 1.5 Gyr \citep{Yoshii1996}. 
\citet{Stolger2010} estimated a  delay-time distribution for SNe Ia is 
about 3--4 Gyr and the observed SNe Ia rate in clusters of galaxies 
per unit stellar mass increases with redshift
(e.g. \citealt{Sand2012}).
In clusters of galaxies, to account for the Fe mass in the ICM, the 
past average rate of SNe Ia was much larger than the present rate 
of elliptical galaxies 
(e.g. \citealt{Renzini1993, Matsushita2013Perseus}); 
accumulating the present SNe Ia rate over the Hubble time, the 
expected total Fe mass synthesized from SNe Ia in the past is an 
order of magnitude smaller than the observed total Fe mass out to 
the virial radius \citep{SatoT2012, Matsushita2013Perseus}.
These results indicate that the lifetimes of most of SNe Ia are much 
shorter than the Hubble time.  If stars in clusters have a similar 
initial mass function (IMF) with those of our Galaxy, the abundance 
pattern should naturally be similar to the solar abundance pattern.

\section{Summary and Conclusions}

We analyzed Suzaku data of the four galaxy group, MKW~4, HCG~62, 
the NGC~1550 group, and the NGC~5044 group out to 0.5~$r_{180}$. 
The temperature and metal abundance distributions were derived from 
the 1T and 2T model fits for the ICM with the ATOMDB version 1.3.1 
and 2.0.1\@.  The dependence on the temperature modeling and the 
versions of ATOMDB of these abundance ratios were relatively small.
Beyond 0.1 $r_{180}$, the derived Fe abundance in the ICM was 
0.2--0.4 solar, and consistent or slightly smaller than those 
of clusters of galaxies.  The abundance ratios of Mg/Fe and Si/Fe 
of these groups and clusters are close to the solar ratio and 
exhibited no significant radial dependence.  However, at 
0.5~$r_{180}$, the integrated IMLR of some groups are systematically 
smaller than those of clusters of galaxies.  The systems with 
smaller gas mass to light ratios have smaller IMLR values
and the entropy excess is inversely correlated with the IMLR.
These results indicate early metal enrichments in groups and 
clusters of galaxies.

\acknowledgments




{\it Facilities:} \facility{Suzaku}.



\appendix
\twocolumn
\section{Estimations of the Galactic Foregrounds and Cosmic X-ray Background}
\label{sec:bgd}
\renewcommand{\thefigure}{A.\arabic{figure}}
\setcounter{figure}{0}

\begin{figure}
  \begin{center}
    \includegraphics[width=0.33\textwidth,angle=0,clip]{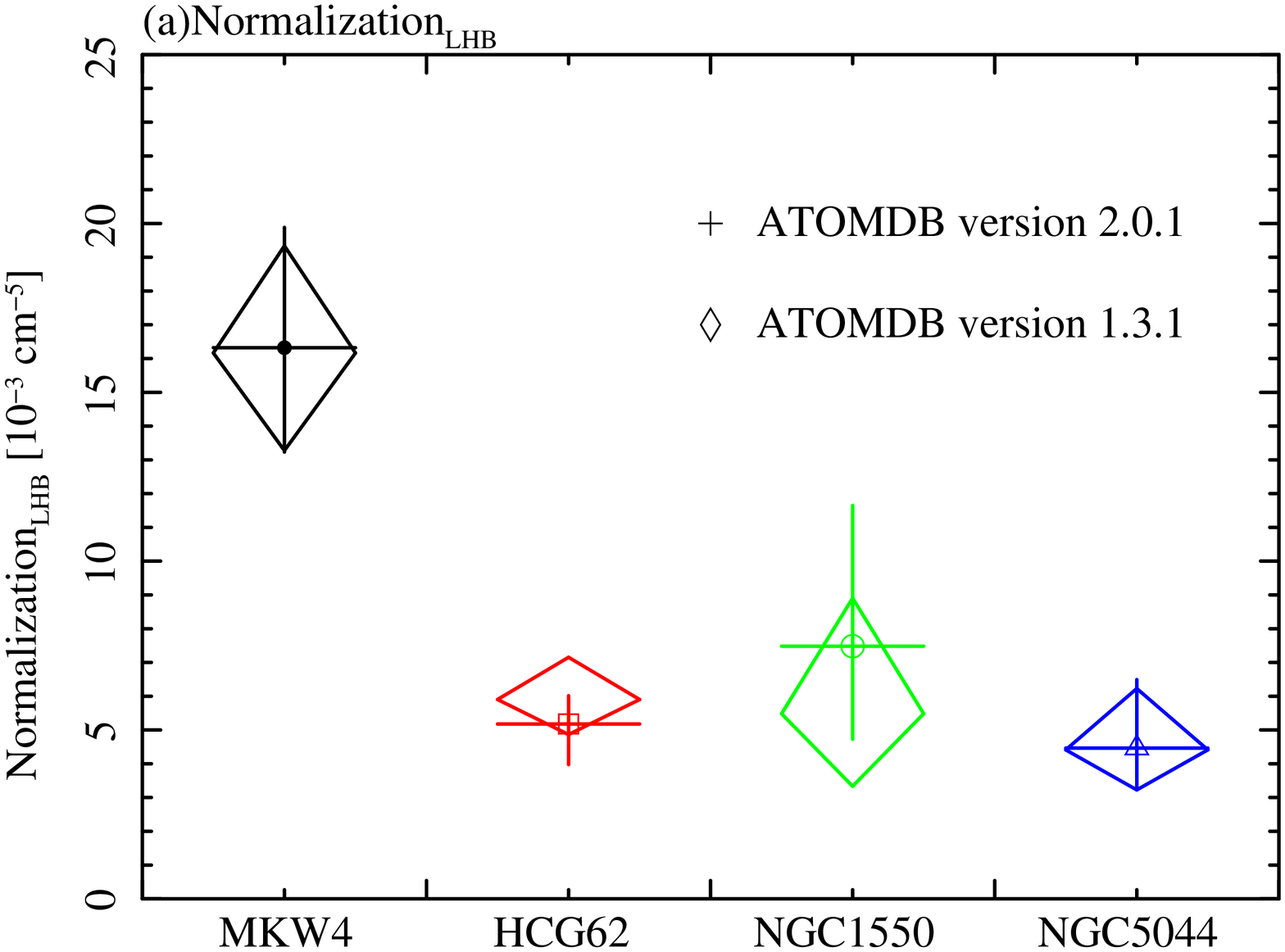}
    \includegraphics[width=0.33\textwidth,angle=0,clip]{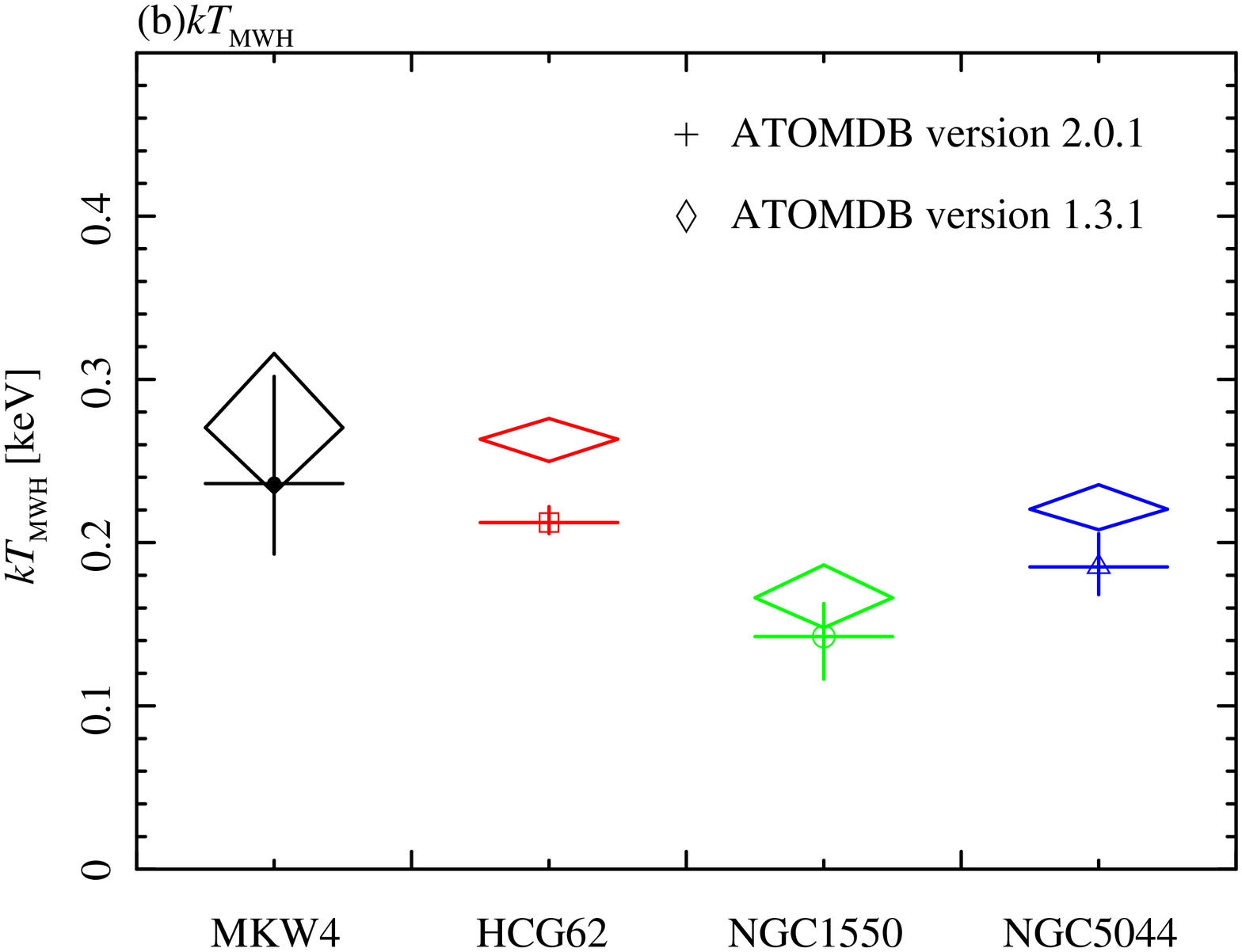}
    \includegraphics[width=0.33\textwidth,angle=0,clip]{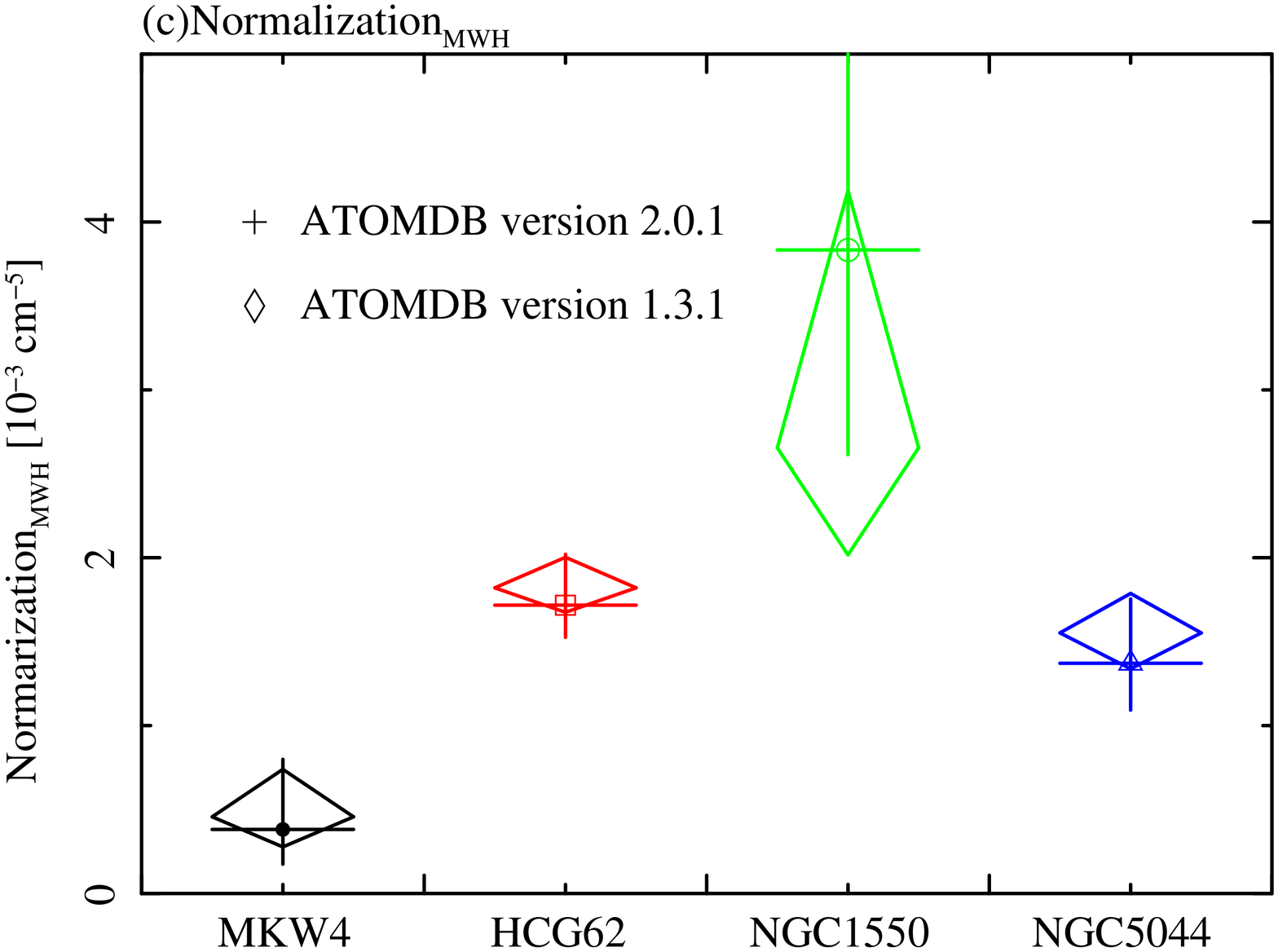}
    \includegraphics[width=0.33\textwidth,angle=0,clip]{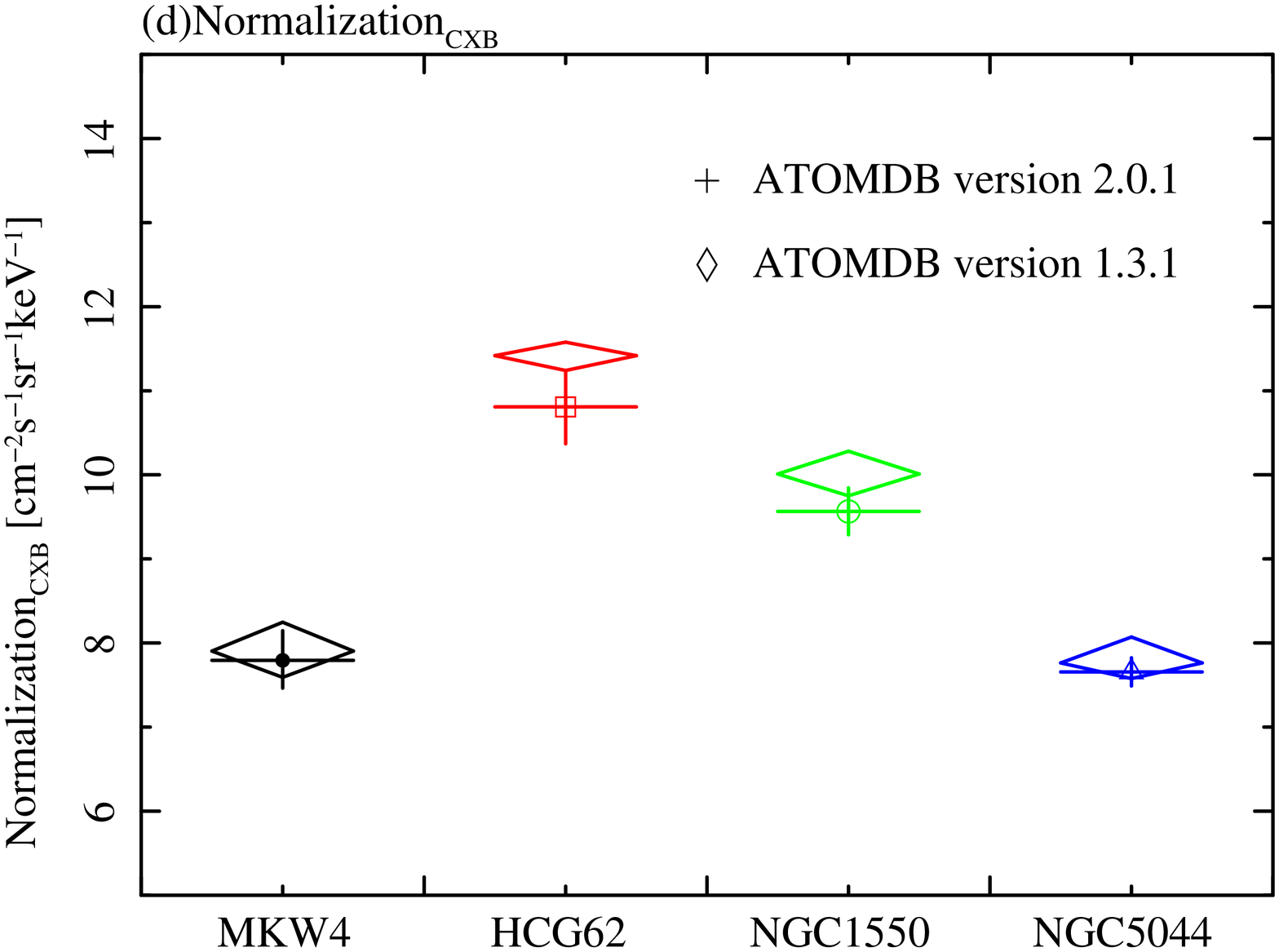}
 \end{center}
\caption{
Comparisons of the background component parameters, (a) LHB
 normalizations, (b) MWH temperature, 
(c) MWH normalizations, and (d) CXB normalizations. 
Normalization of the LHB and MWH ({\it apec} model) components 
divided by the solid angle, 
$\Omega^{U}$, assumed in the uniform-sky ARF calculation (20$'$ radius), 
$Norm = \int n_{\rm e} n_{\rm H} dV \,/~\,[4\pi\,(1+z)^2 D_{\rm A}^{~2}] \,/\, \Omega^{U}$ $\times 10^{-14}$ cm$^{-5}$~400$\pi$~arcmin$^{-2}$, 
where $D_{\rm A}$ is the angular distance to the source. On the other hand, normalization of 
the CXB ({\it power-law}) is  units of photons cm$^{2}$ s$^{-1}$ sr$^{-1}$ keV$^{-1}$ at 1 keV.
}
\label{fig:bgd} 
\end{figure}

It is important to estimate the Galactic and CXB emissions accurately,
because the spectra, particularly in outer region of groups, suffer 
from the Galactic and CXB emissions strongly.  
We assumed the two Galactic emissions, the local hot bubble (LHB) and 
the Milky-Way halo (MWH), with a thermal plasma models ({\it apec} 
model; \cite{Smith2001}) as shown in \citet{Yoshino2009}. The
temperature of the LHB was fixed at 0.10 keV\@, while the normalization 
was a free parameter.  We also allowed to vary the normalization and 
temperature of the MWH.  The redshift and abundance were fixed at 0 and 
1 solar, respectively, for the LHB and MWH components.  We assumed the 
CXB component with a power-law model of a photon index, $\Gamma= 1.4$.  
We simultaneously fitted the spectra for the outermost and other annuli 
regions with the following model formula;
$constant~\times~(apec_{\rm LHB}~+~phabs~\times~(apec_{\rm MWH}~+$ 
$power{\rm-}law~+~vapec_{\rm ICM}~+~vapec_{\rm ICM}))$.
Here, {\it phabs} model indicates the Galactic absorption. 
We assumed the 1T or 2T components for the ICM of the annular regions.
Assumed two-temperature components for the ICM of the outermost region, 
these results did not change within systematic error compared with the 
results with the 1T model.
The result of each simultaneous fits was consistent each other, and 
the weighted average results are shown in figure \ref{fig:bgd}.

With the ATOMDB version 2.0.1, the temperature of the MWH was 0.05 keV 
lower than that with the ATOMDB version 1.3.1, while the normalizations 
were consistent between the ATOMDB versions within statistical error ranges.
The estimated background level with the version 1.3.1 were consistent 
with the typical values for the Galactic emissions \citep{Yoshino2009}.

\newpage

\newpage
\section{Figures of comparisons of the results with the ATOMDB versions}
\label{sec:apend_atomdb}
\renewcommand{\thefigure}{B.\arabic{figure}}
\setcounter{figure}{0}
\renewcommand{\thetable}{B.\arabic{table}}
\setcounter{table}{0}

\begin{figure}[!th]
  \begin{center}
     \includegraphics[width=0.33\textwidth,angle=0,clip]{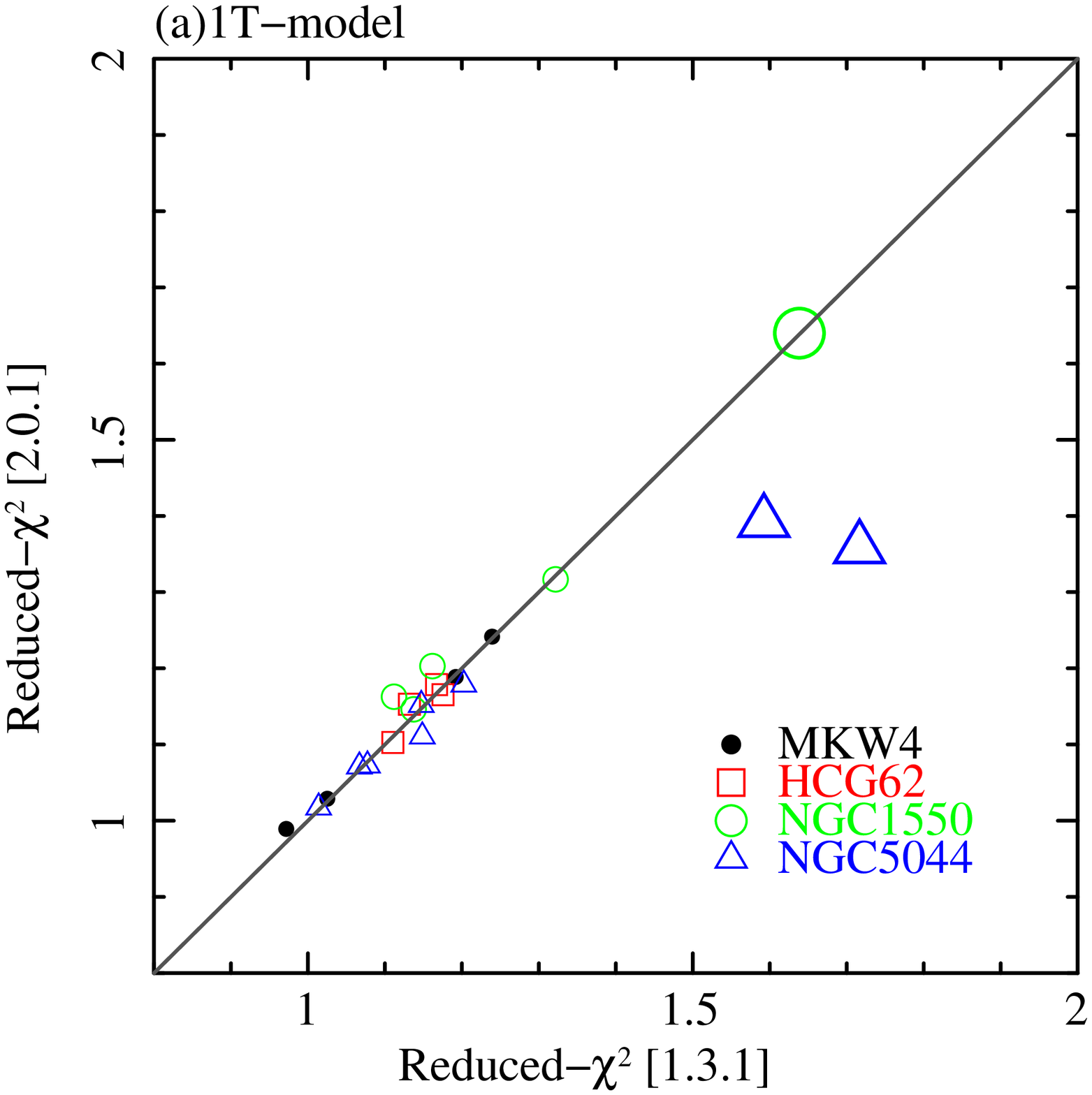}
     \includegraphics[width=0.33\textwidth,angle=0,clip]{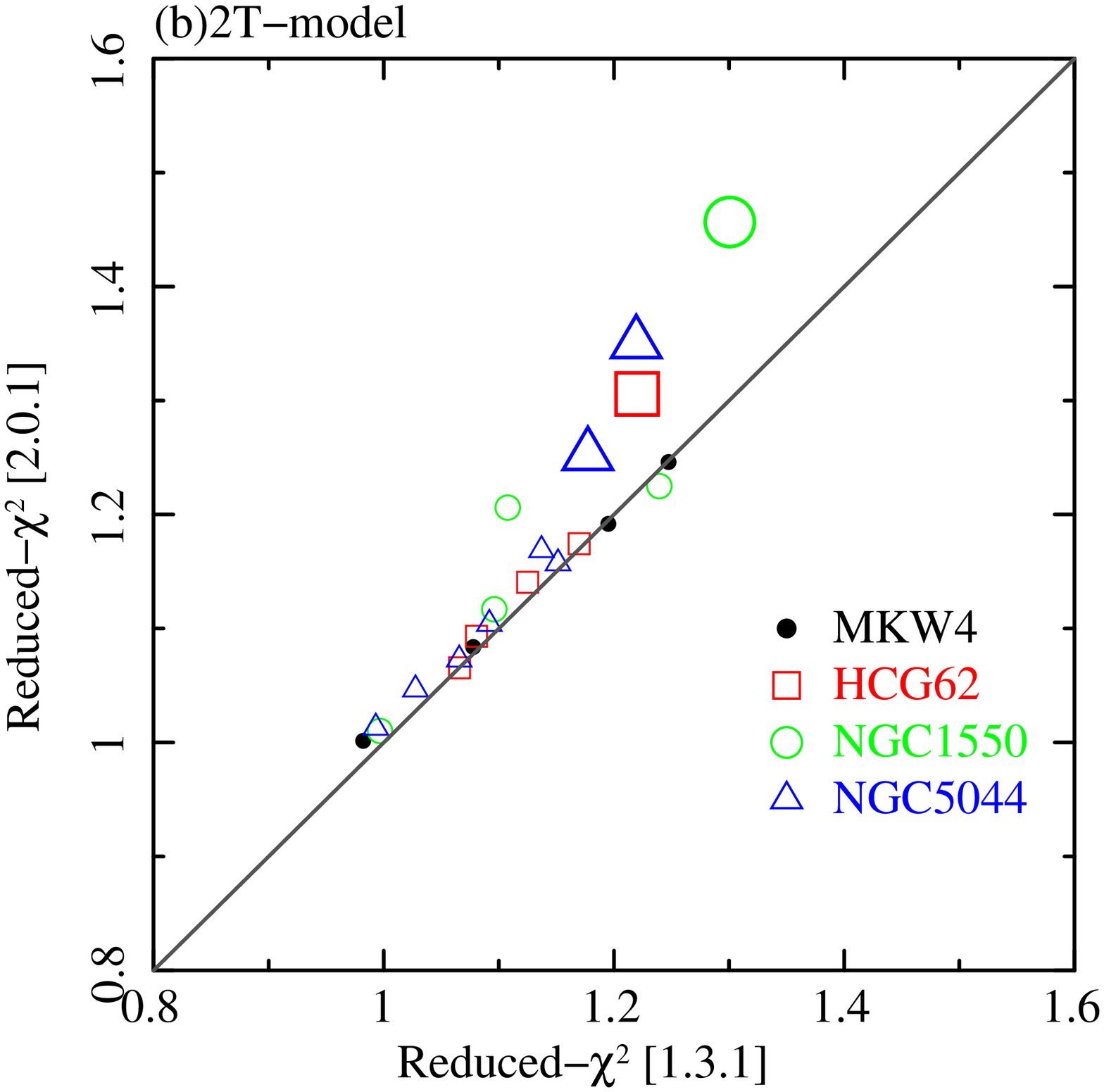}
  \end{center}
  \caption{
Comparisons of the reduced $\chi^{2}$ for the spectral fits with 
the 1T (left) and 2T (right) models for both the ATOMDB versions.
Note that the reduced $\chi^{2}$ in the innermost region for HCG~62 
with the 1T model for the ATOMDB version 1.3.1 was higher than 3,
thus, we did not plot in this figure.  The notations are the same 
as in figure \ref{radial_chi}.  Bigger marks indicate the $\chi^2$ 
within $0.05~r_{180}$.
}
\label{fig:chi_201131}
\end{figure}

\begin{figure*}
  \begin{center}
    \includegraphics[width=0.30\textwidth,angle=0,clip]{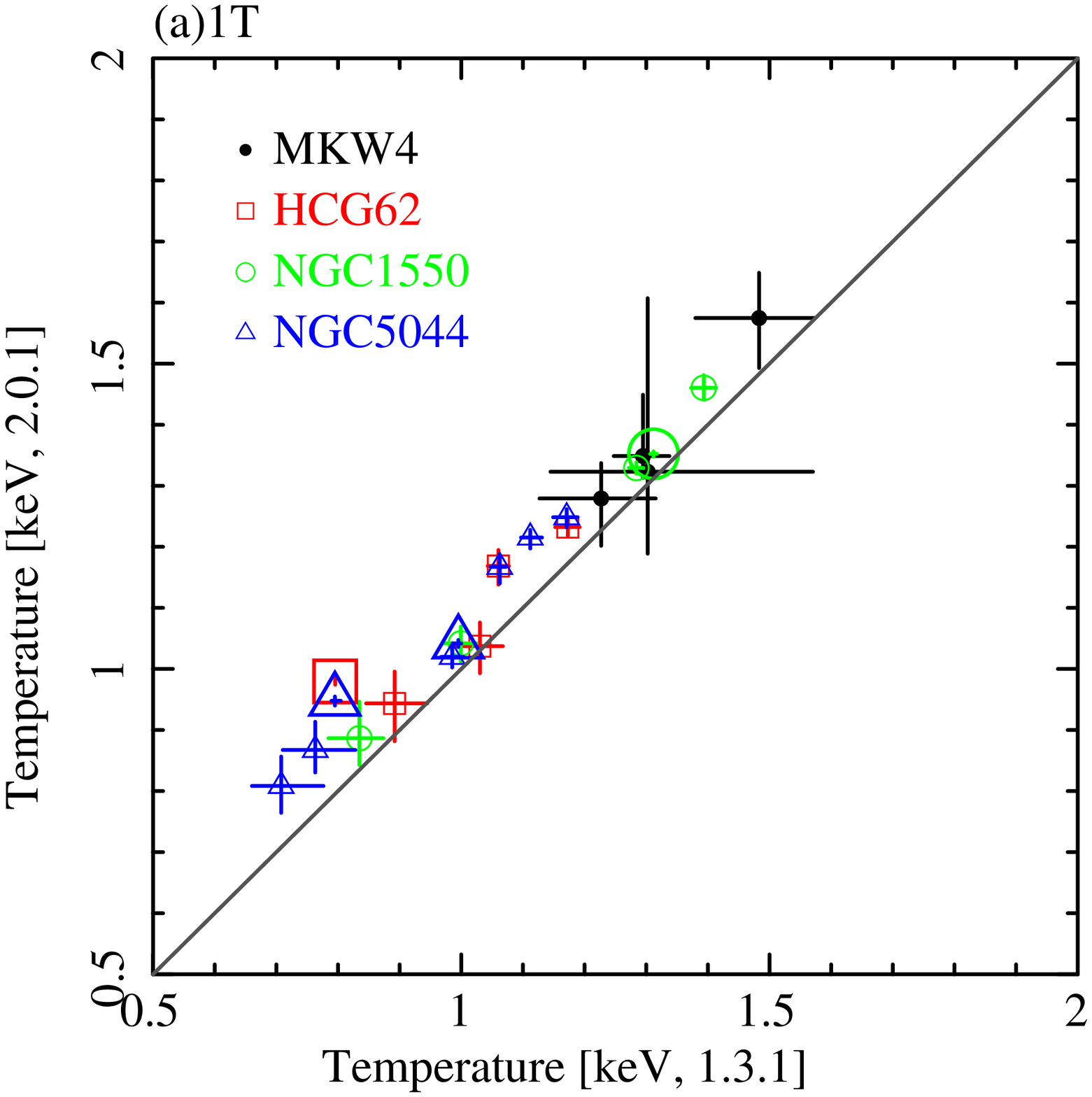}
    \includegraphics[width=0.30\textwidth,angle=0,clip]{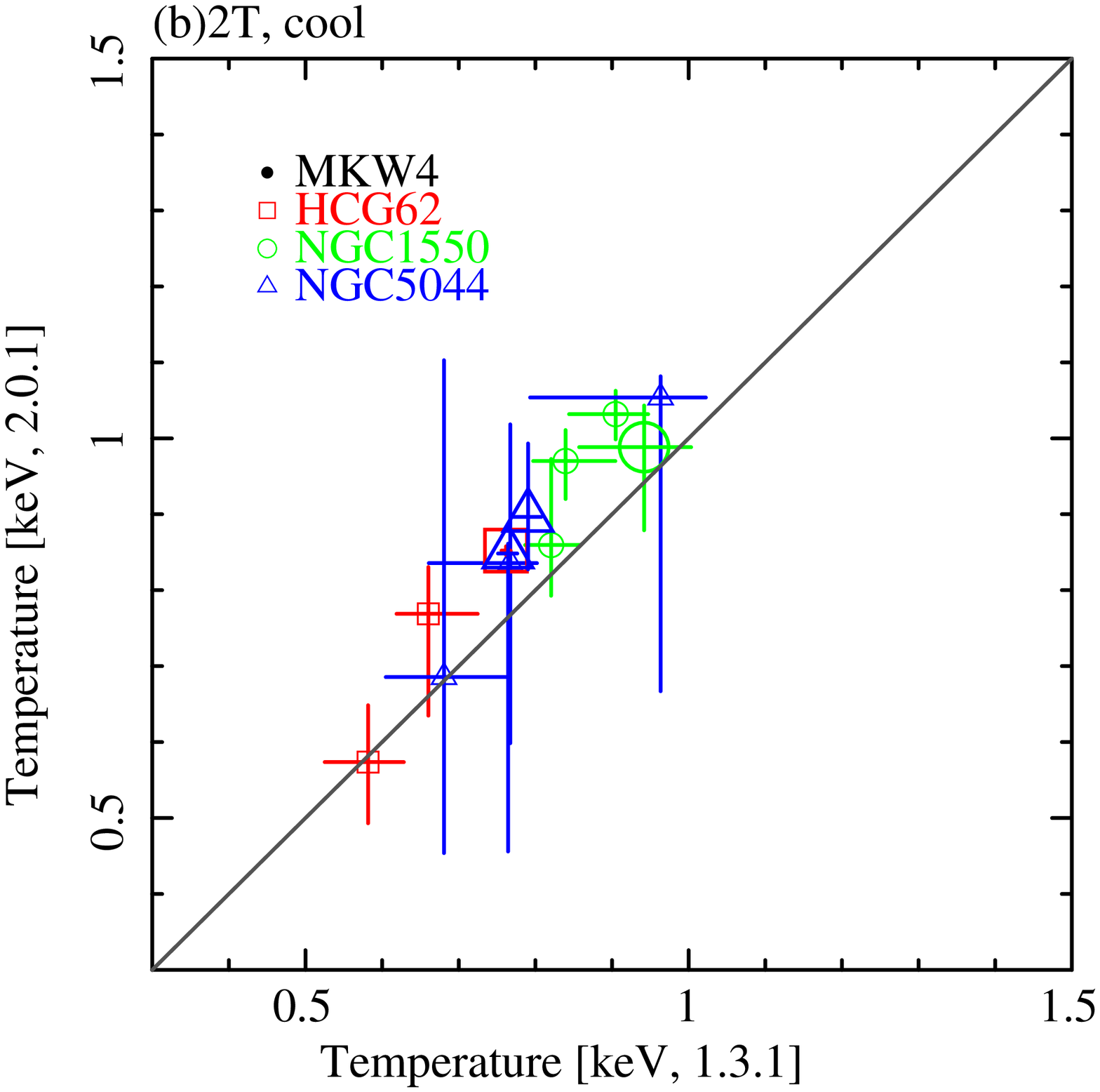} 
    \includegraphics[width=0.30\textwidth,angle=0,clip]{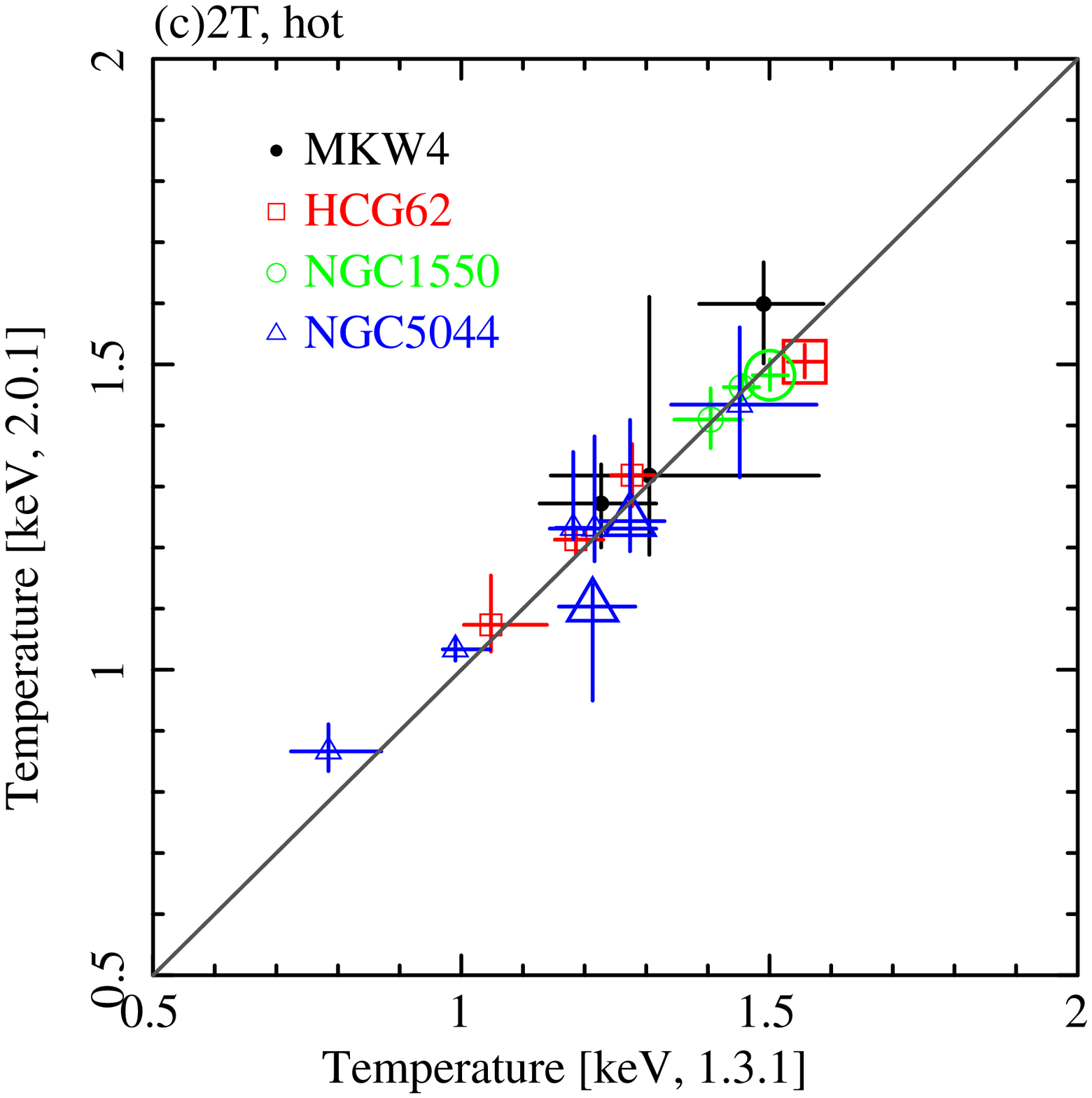} 
    \includegraphics[width=0.30\textwidth,angle=0,clip]{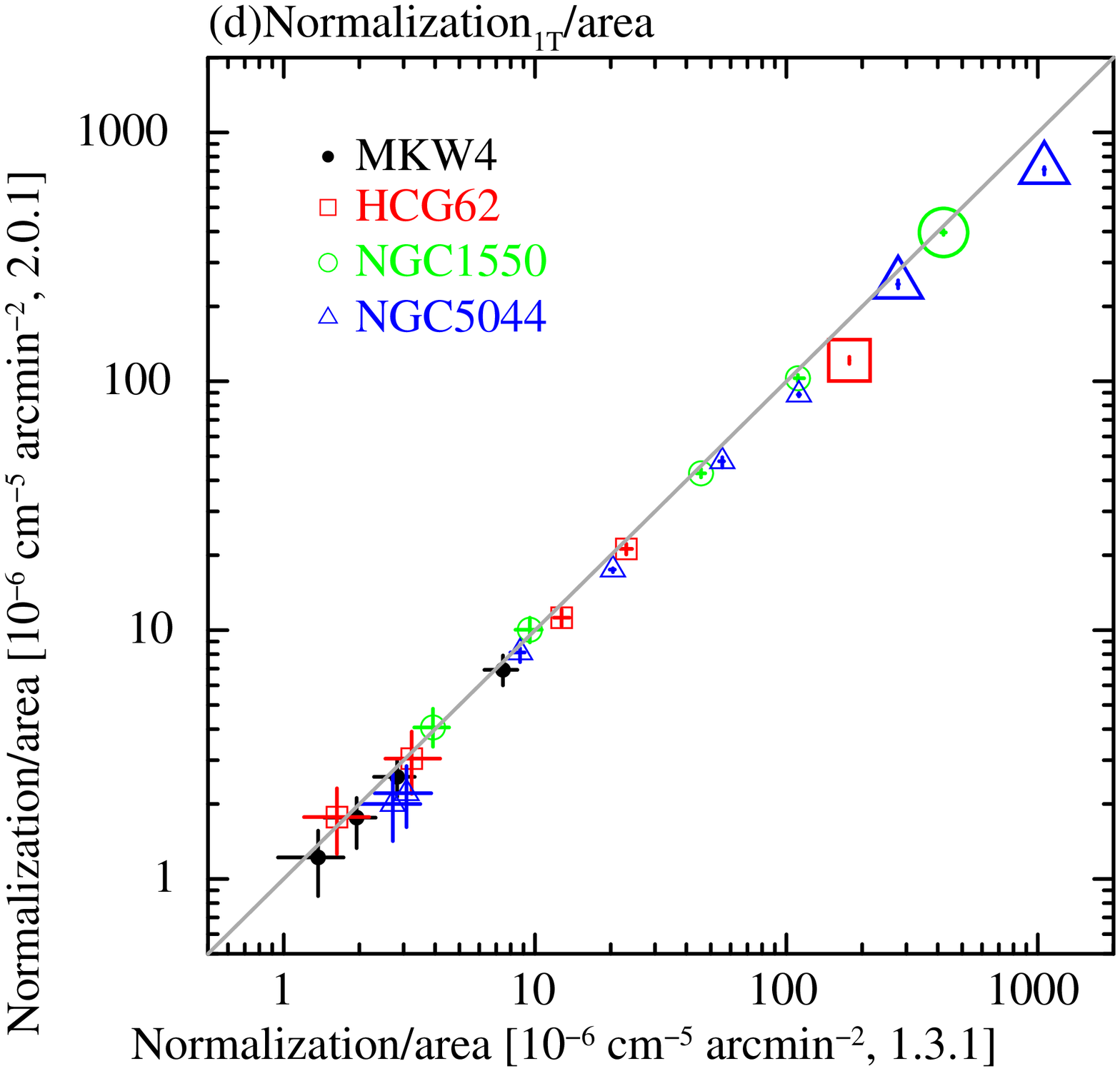}
    \includegraphics[width=0.30\textwidth,angle=0,clip]{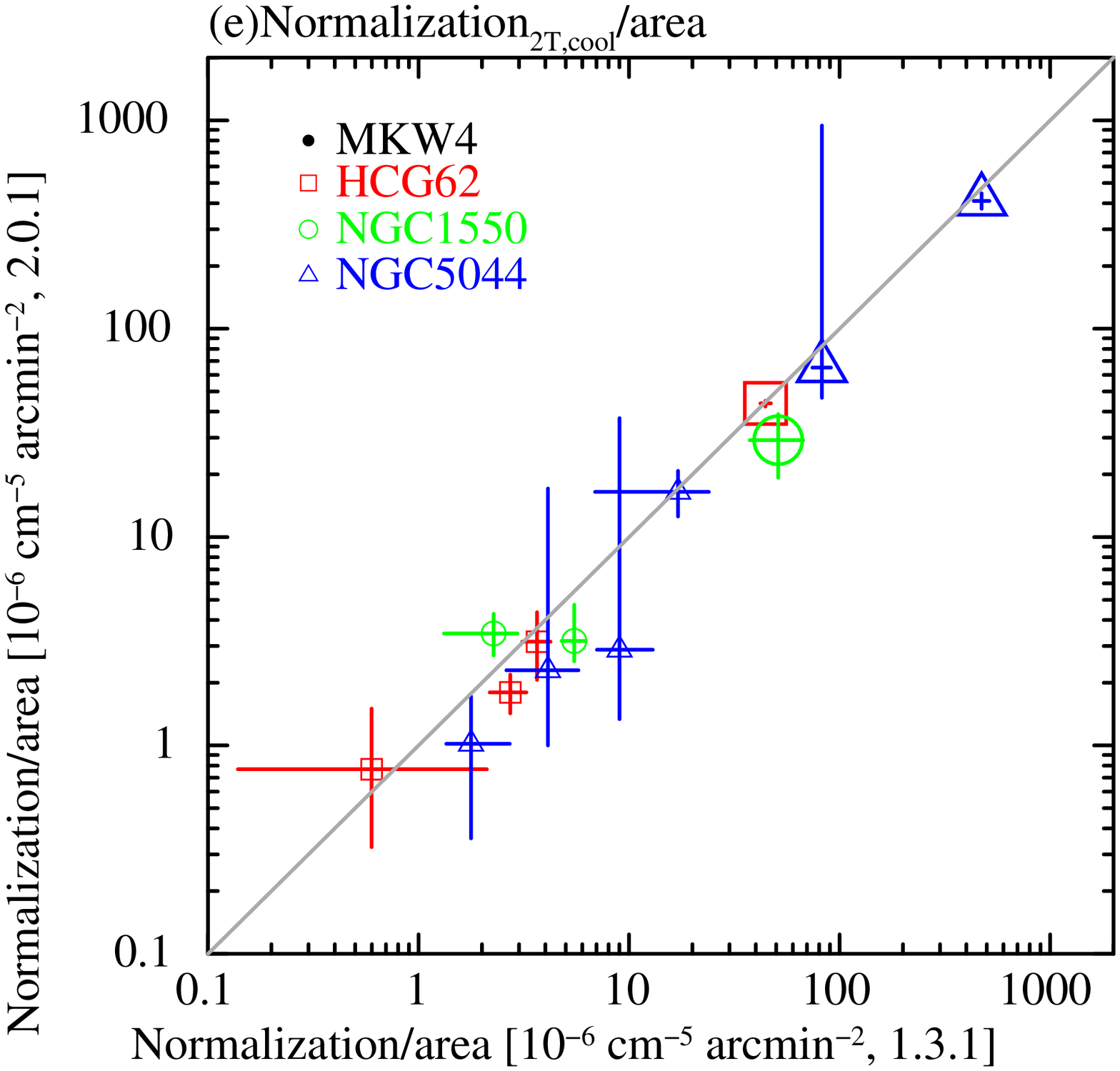} 
    \includegraphics[width=0.30\textwidth,angle=0,clip]{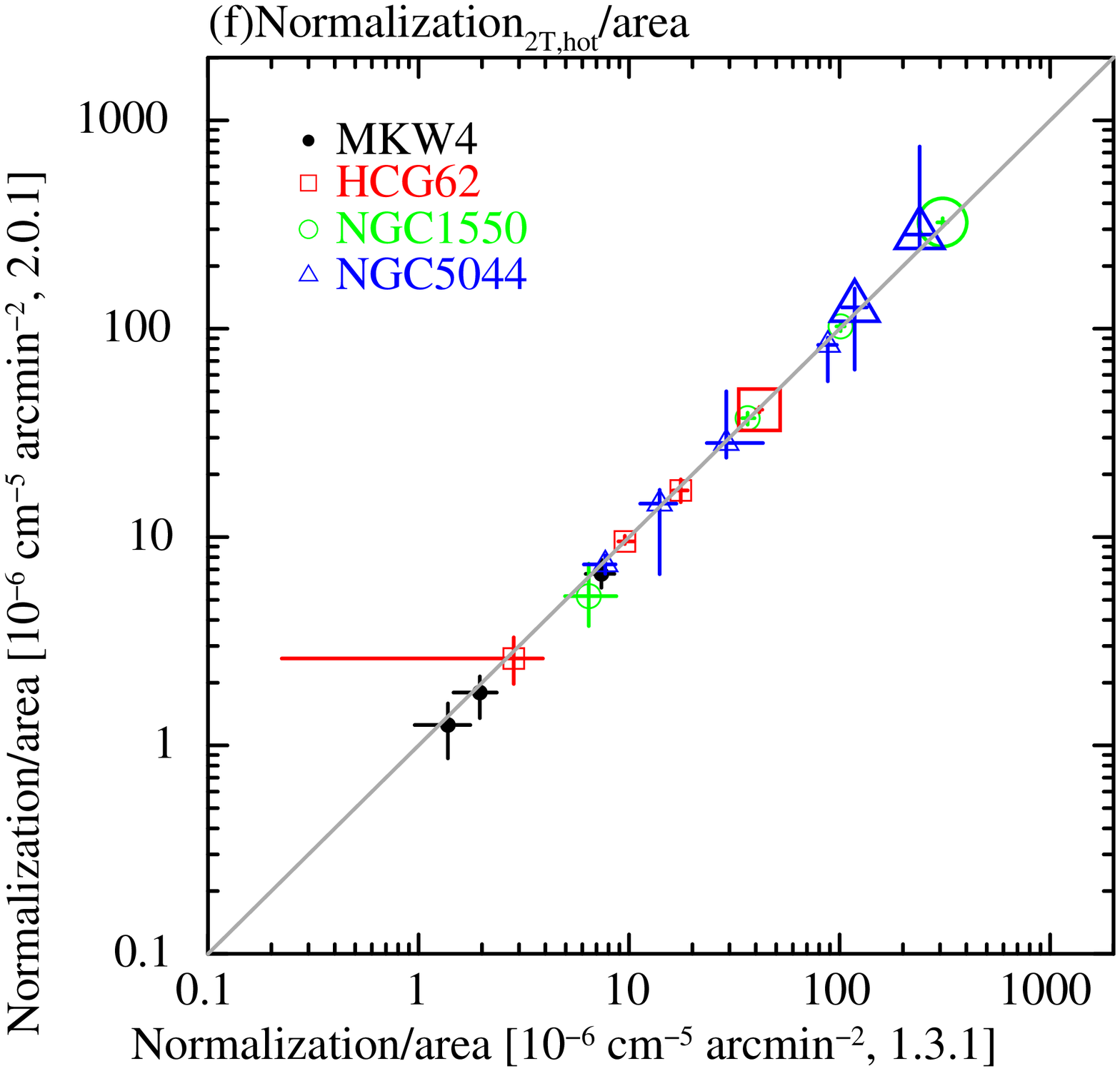} 
      \end{center}
  \caption{
Comparisons of the derived temperatures and normalizations with the 
ATOMDB version 1.3.1 and 2.0.1\@.  (a)--(c): The comparisons of the 
temperatures from the 1T model fits, and the cooler and hotter 
component temperatures from the 2T model fits for the version 1.3.1 
and 2.0.1\@.  (d)--(f):The comparisons of the normalizations divided 
by the area from the 1T model fits, and the cooler and hotter component 
temperatures from the 2T model fits for the version 1.3.1 and 2.0.1\@.  
Bigger marks indicate the temperatures or normalizations within 
$0.05~r_{180}$.
}
\label{fig:201131_temp}
\end{figure*}

\begin{figure}
  \begin{center} 
    \includegraphics[width=0.3\textwidth,angle=0,clip]{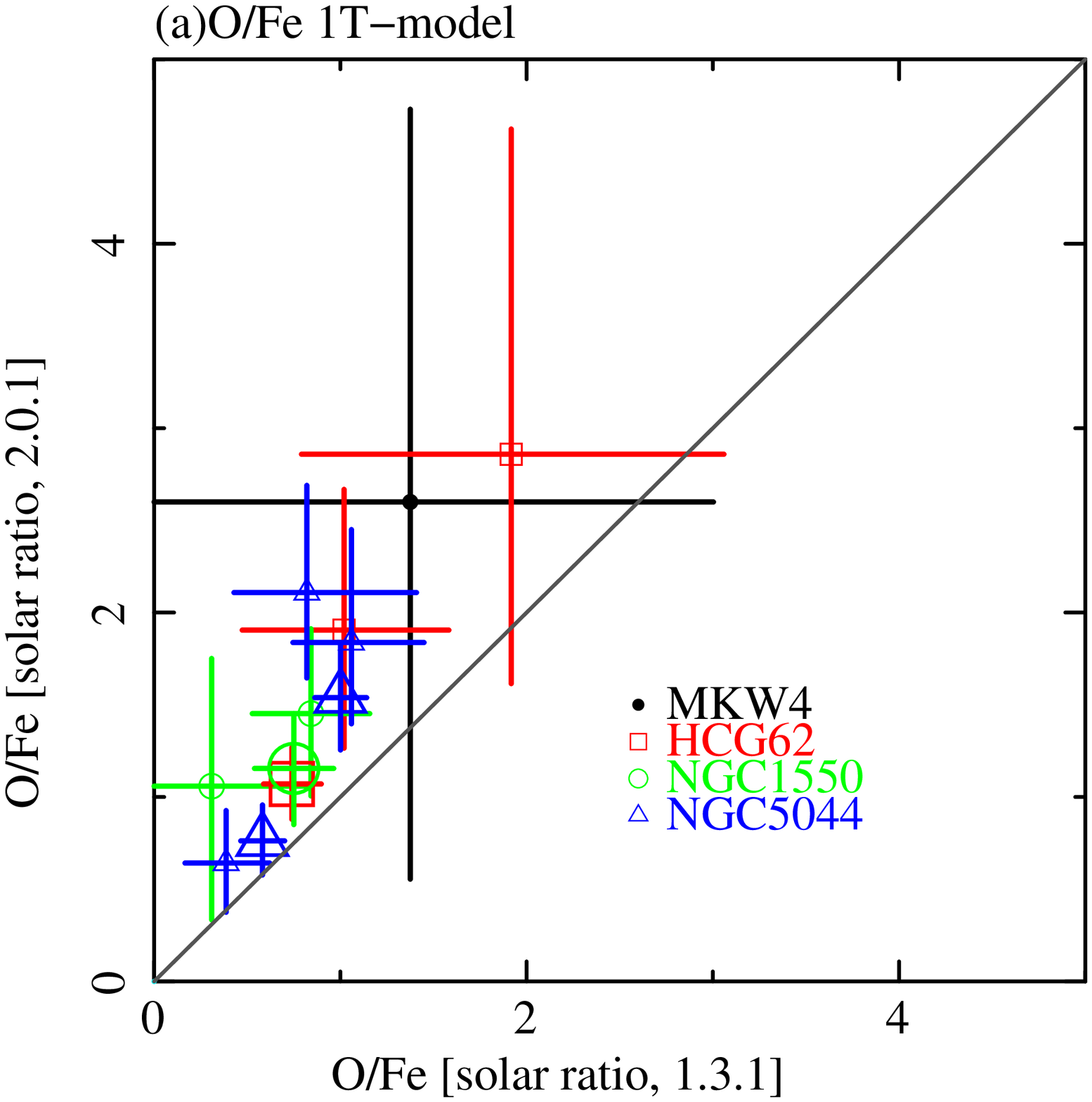} 
    \includegraphics[width=0.3\textwidth,angle=0,clip]{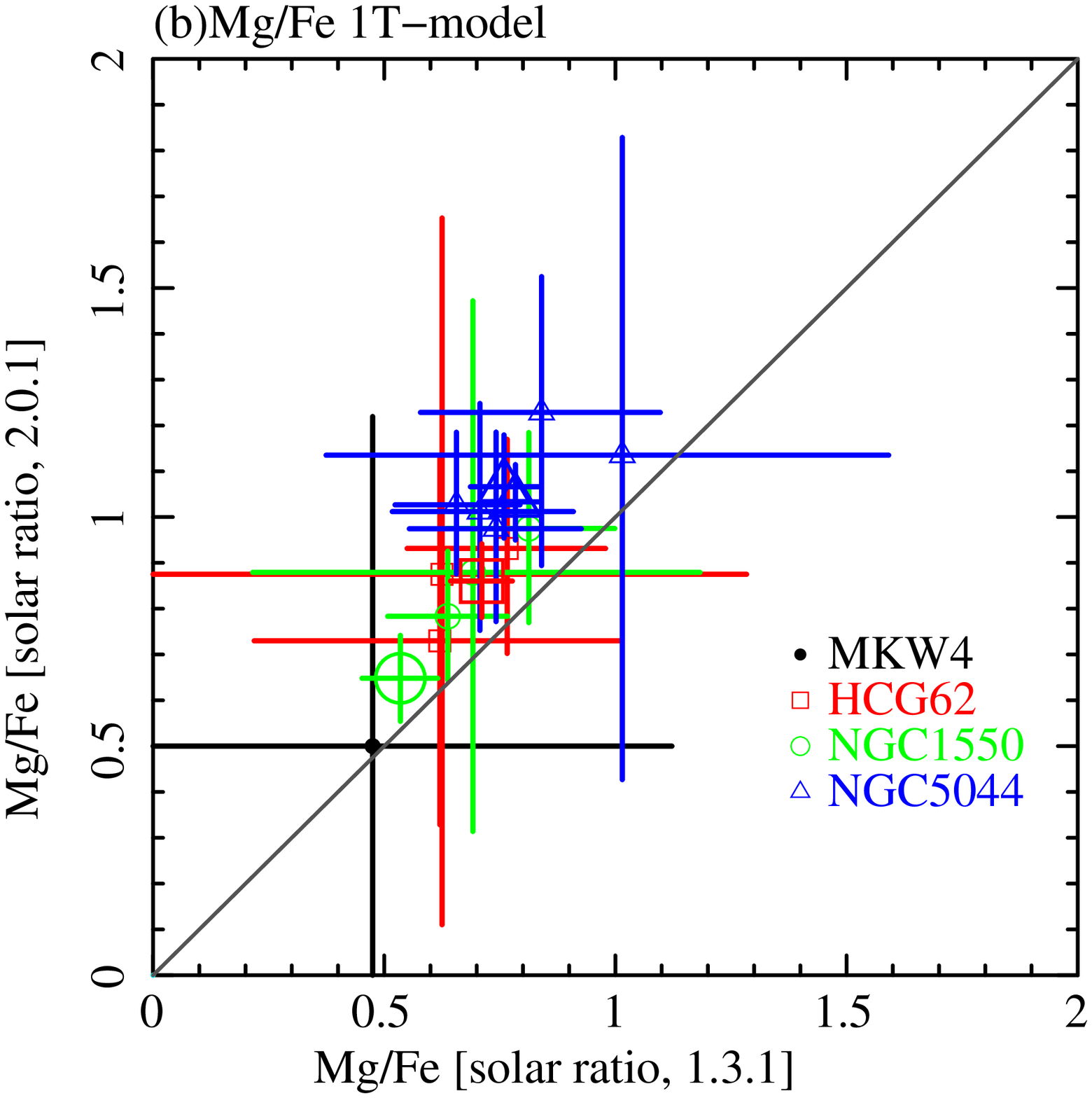}
    \includegraphics[width=0.3\textwidth,angle=0,clip]{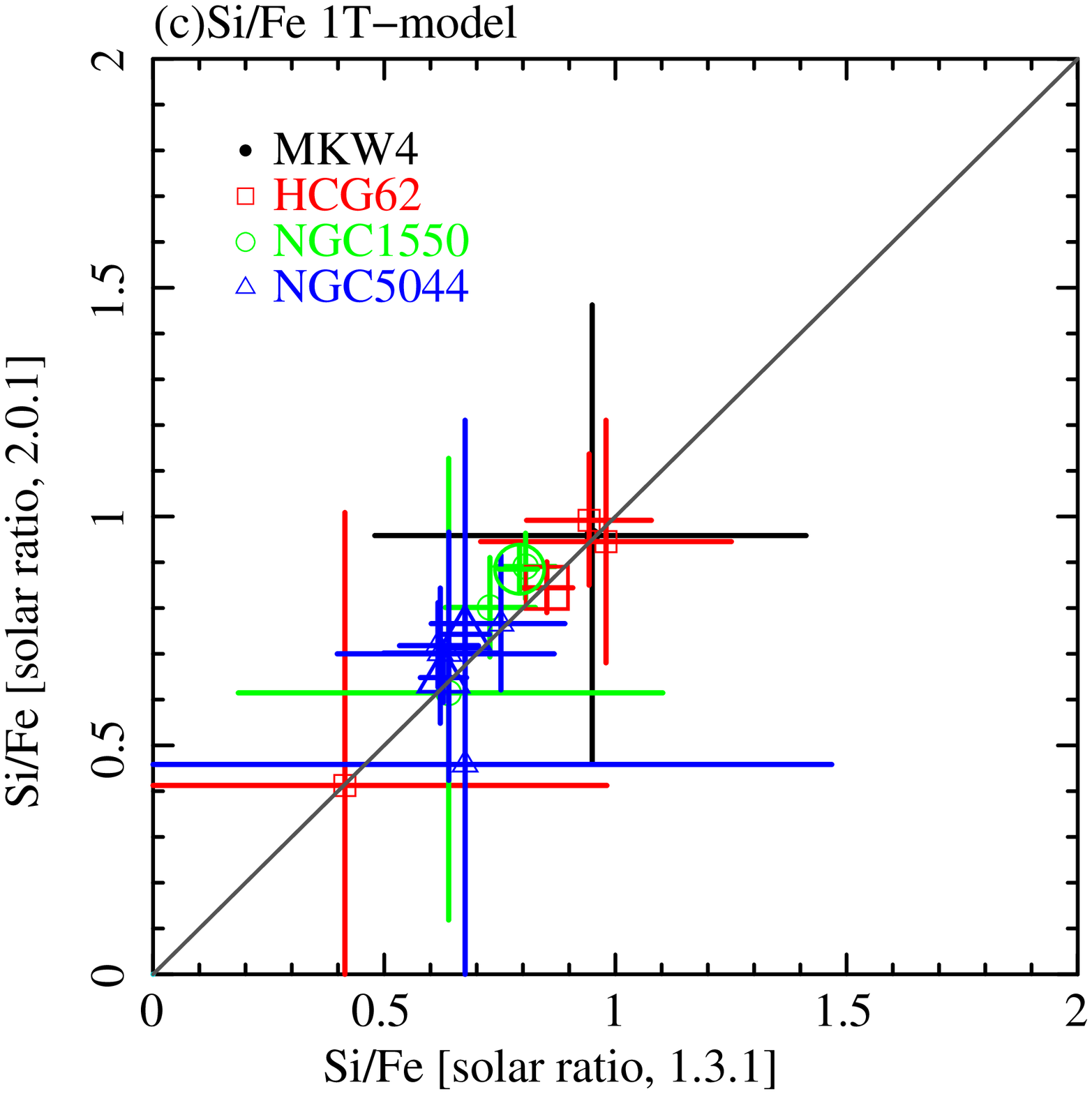}    
    \includegraphics[width=0.3\textwidth,angle=0,clip]{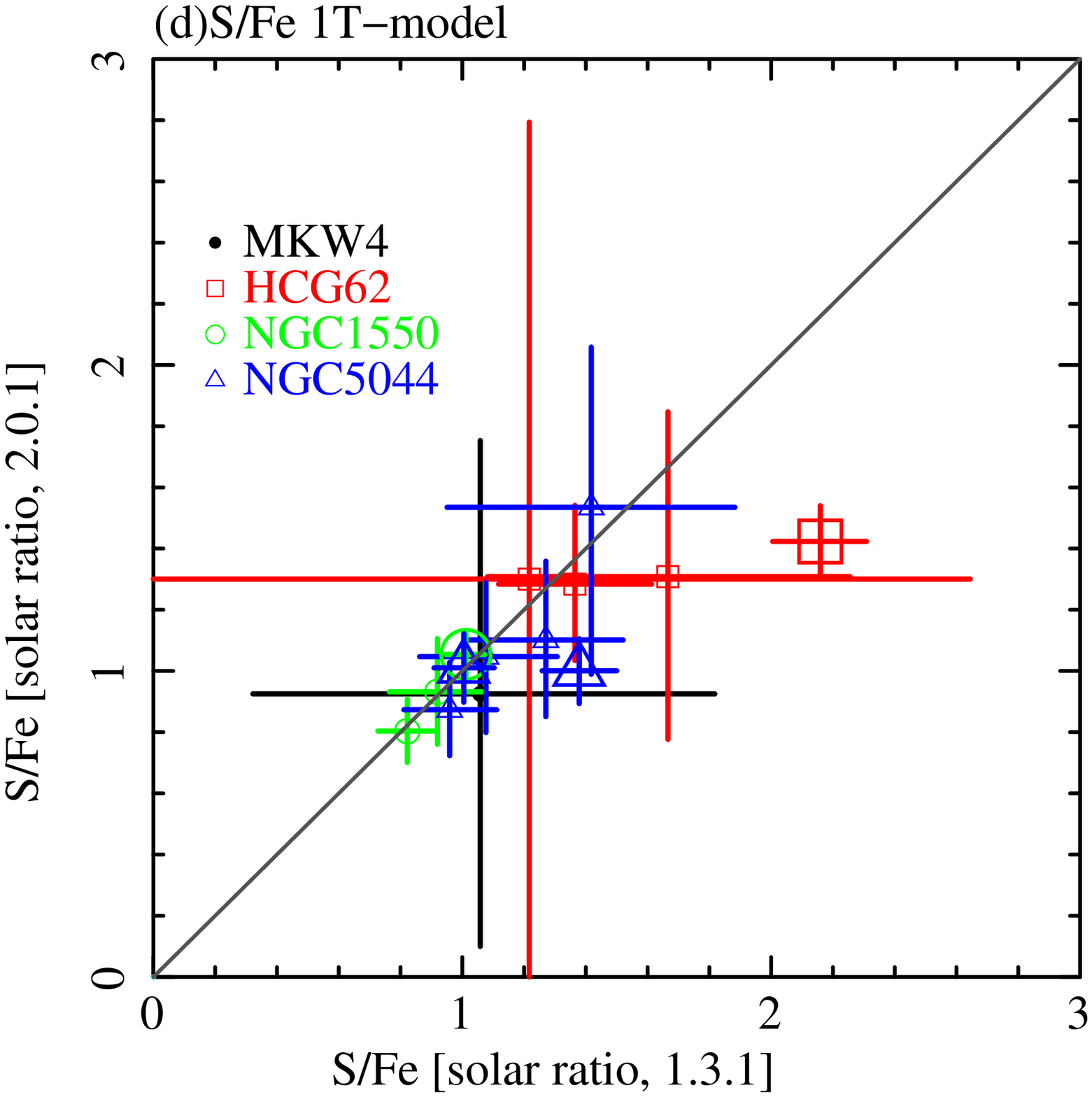}
 \end{center}
  \caption{
Comparisons of (a) O/Fe, (b)Mg/Fe, 
(c) Si/Fe, and (d) S/Fe ratios derived from the 1T model fits with 
the ATOMDB version 1.3.1 and 2.0.1\@. The colors are the same as in
figure \ref{radial_chi}.	
Bigger marks indicates the Fe abundance and the ratios within 
$0.05~r_{180}$.
}
\label{fig:201131_abund_1T}
\end{figure}

\begin{figure}
  \begin{center} 
    \includegraphics[width=0.3\textwidth,angle=0,clip]{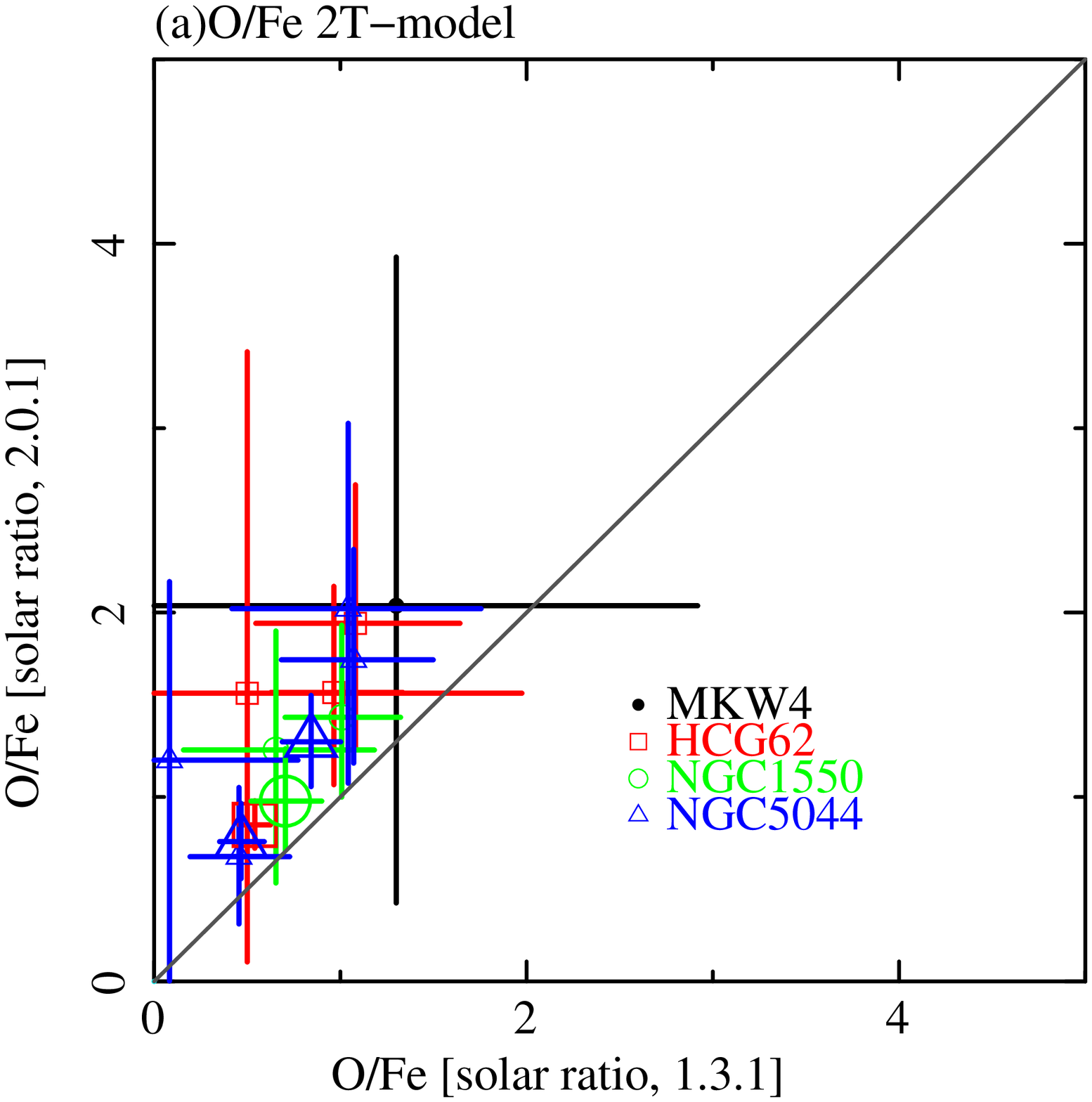} 
    \includegraphics[width=0.3\textwidth,angle=0,clip]{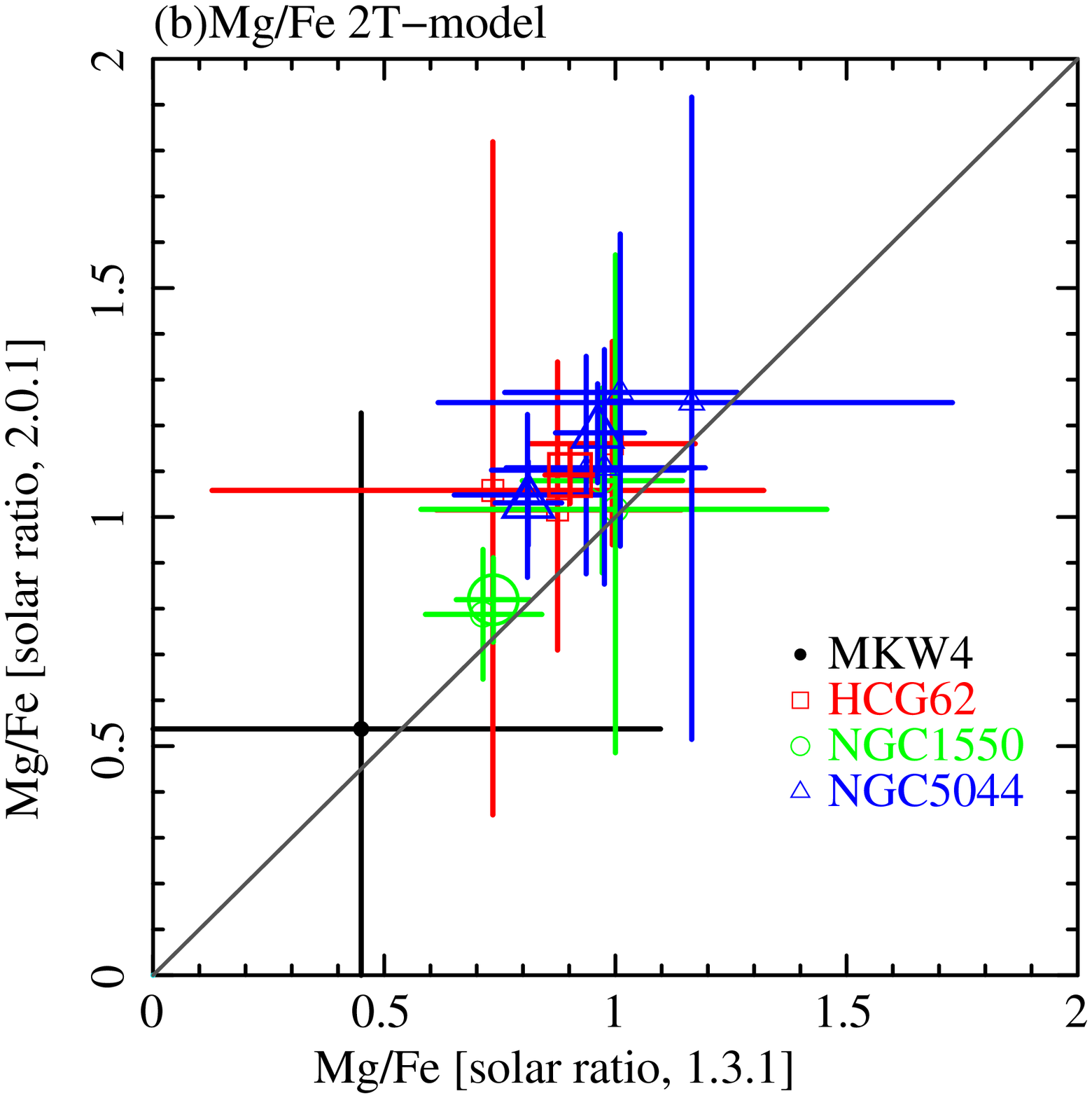}
    \includegraphics[width=0.3\textwidth,angle=0,clip]{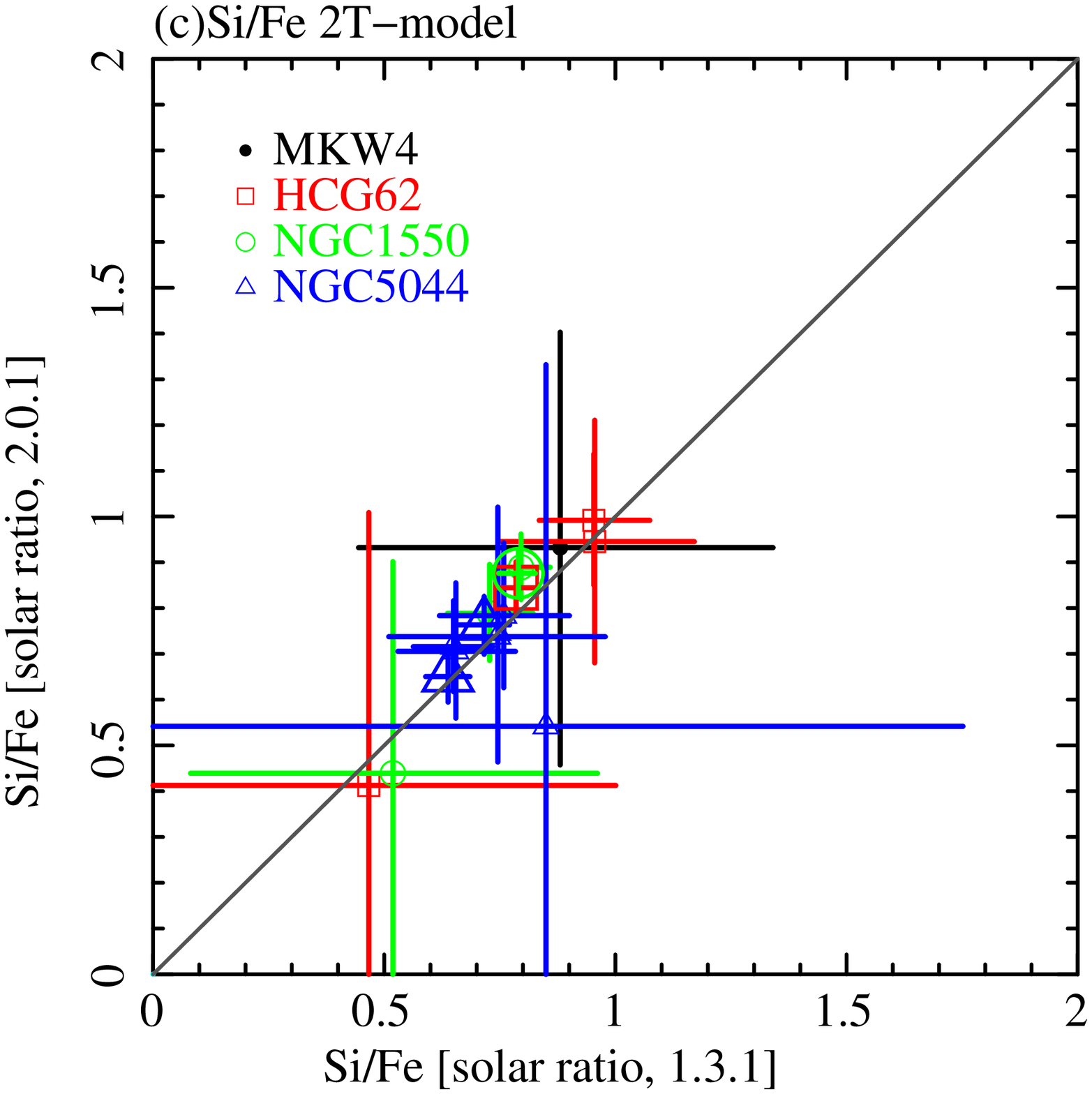}    
    \includegraphics[width=0.3\textwidth,angle=0,clip]{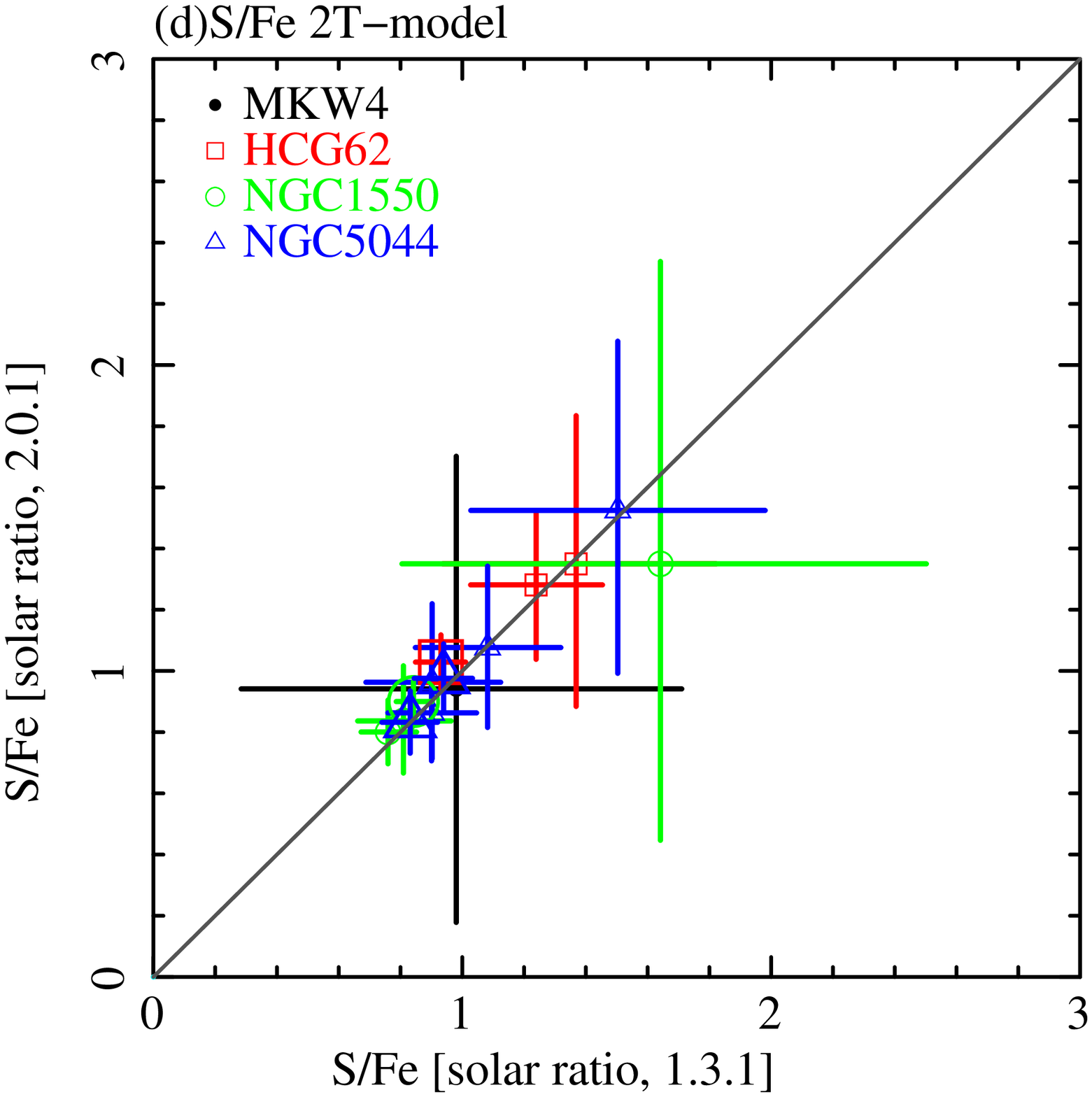}
 \end{center}
  \caption{
The same as figure \ref{fig:201131_abund_1T} derived from 2T model.
}
\label{fig:201131_abund_2T}
\end{figure}

In this section, we summarize comparisons of the results by changing
the ATOMDB versions. The reduced $\chi^2$ for the spectral fits for 
both the ATOMDB versions are shown figure \ref{fig:chi_201131}.
We compared the temperatures and normalizations derived from 
the spectral fits with both of the ATOMDB versions in figure \ref{fig:201131_temp}.
Figure \ref{fig:201131_abund_1T} and \ref{fig:201131_abund_2T} summarize 
comparisons of the abundance ratios of O, Mg, Si, and S
to Fe for the 1T and 2T model fitting with both of the ATOMDB versions,
in units of the solar ratios. 
The details of the results were addressed in subsection \ref{sec:atomdb}.

\begin{deluxetable}{llcccccccc}
\tabletypesize{\scriptsize}
%
\tablewidth{0pt}
\tablecaption{
Weighted averages of the abundance ratios in units of the 
solar ratios with ATOMDB version 1.3.1. \label{tb:weight}}
\tablehead{
\colhead{group} & \colhead{region} & \multicolumn{2}{c}{O/Fe [solar ratio]}  & \multicolumn{2}{c}{Mg/Fe [solar ratio]} & \multicolumn{2}{c}{Si/Fe [solar ratio]} & \multicolumn{2}{c}{S/Fe [solar ratio]}  \\
\colhead{} & \colhead{} & \colhead{1T} & \colhead{2T} & \colhead{1T} & \colhead{2T} & \colhead{1T} & \colhead{2T} & \colhead{1T} & \colhead{2T} \\ 
}
\startdata 
MKW~4      & All$^{a}$ &$1.34^{+2.35}_{-1.34}$ &$1.48^{+2.53}_{-1.48}$  &$0.13^{+0.56}_{-0.13}$ &$0.24^{+0.77}_{-0.24}$  &$0.75^{+0.67}_{-0.68}$ &$0.67^{+0.66}_{-0.63}$  &$1.19^{+1.18}_{-1.14}$ &$1.08^{+1.15}_{-1.08}$ \\
HCG~62 & $< 0.1~r_{180}$ &$0.77^{+0.16}_{-0.16}$ &$0.58^{+0.09}_{-0.09}$  &$0.72^{+0.08}_{-0.08}$ &$0.91^{+0.07}_{-0.07}$  &$0.87^{+0.08}_{-0.08}$ &$0.80^{+0.06}_{-0.06}$  &$1.94^{+0.20}_{-0.20}$ &$0.97^{+0.10}_{-0.10}$ \\
 & $> 0.1~r_{180}$  &$1.34^{+0.72}_{-0.72}$ &$0.96^{+0.41}_{-0.41}$  &$0.62^{+0.45}_{-0.46}$ &$0.85^{+0.32}_{-0.33}$  &$0.88^{+0.36}_{-0.34}$ &$0.89^{+0.29}_{-0.29}$  &$1.59^{+0.85}_{-0.83}$ &$1.36^{+0.64}_{-0.65}$ \\
  & All$^{a}$ &$0.80^{+0.16}_{-0.16}$ &$0.60^{+0.09}_{-0.09}$  &$0.72^{+0.08}_{-0.08}$ &$0.91^{+0.07}_{-0.07}$  &$0.87^{+0.08}_{-0.08}$ &$0.81^{+0.06}_{-0.06}$  &$1.92^{+0.19}_{-0.19}$ &$0.98^{+0.10}_{-0.10}$ \\
NGC~1550 & $< 0.1~r_{180}$ &$0.78^{+0.29}_{-0.29}$ &$0.79^{+0.27}_{-0.26}$  &$0.56^{+0.11}_{-0.11}$ &$0.73^{+0.11}_{-0.11}$  &$0.80^{+0.05}_{-0.06}$ &$0.79^{+0.05}_{-0.06}$  &$0.95^{+0.08}_{-0.09}$ &$0.82^{+0.08}_{-0.08}$ \\
 & $> 0.1~r_{180}$  &$0.23^{+0.74}_{-0.23}$ &$0.53^{+0.78}_{-0.53}$  &$0.80^{+0.29}_{-0.27}$ &$0.98^{+0.27}_{-0.26}$  &$0.72^{+0.16}_{-0.15}$ &$0.72^{+0.15}_{-0.15}$  &$0.93^{+0.25}_{-0.25}$ &$0.84^{+0.25}_{-0.24}$ \\
  & All$^{a}$ &$0.71^{+0.27}_{-0.25}$ &$0.76^{+0.26}_{-0.24}$  &$0.60^{+0.10}_{-0.11}$ &$0.77^{+0.10}_{-0.10}$  &$0.79^{+0.05}_{-0.06}$ &$0.78^{+0.05}_{-0.05}$  &$0.95^{+0.08}_{-0.08}$ &$0.82^{+0.08}_{-0.07}$ \\
NGC~5044 & $< 0.1~r_{180}$ &$0.72^{+0.14}_{-0.13}$ &$0.61^{+0.15}_{-0.14}$  &$0.76^{+0.07}_{-0.06}$ &$0.86^{+0.09}_{-0.08}$  &$0.64^{+0.05}_{-0.05}$ &$0.67^{+0.05}_{-0.05}$  &$1.11^{+0.11}_{-0.10}$ &$0.89^{+0.09}_{-0.09}$ \\
 & $> 0.1~r_{180}$  &$0.31^{+0.60}_{-0.31}$ &$0.50^{+0.78}_{-0.13}$  &$0.79^{+0.24}_{-0.24}$ &$0.98^{+0.26}_{-0.25}$  &$0.79^{+0.24}_{-0.24}$ &$0.98^{+0.26}_{-0.25}$  &$1.31^{+0.36}_{-0.37}$ &$1.17^{+0.35}_{-0.34}$ \\
 & All$^{a}$ &$0.73^{+0.13}_{-0.13}$ &$0.61^{+0.15}_{-0.10}$  &$0.76^{+0.06}_{-0.06}$ &$0.88^{+0.08}_{-0.08}$  &$0.65^{+0.05}_{-0.05}$ &$0.68^{+0.05}_{-0.05}$  &$1.13^{+0.10}_{-0.10}$ &$0.91^{+0.09}_{-0.09}$ \\
\enddata
\tablenotetext{a}{
All the regions observed with Suzaku.}
\end{deluxetable}

\newpage

\newpage

\end{document}